\documentclass[12pt]{article}
\usepackage{theorem}
\usepackage{amsmath,amssymb}
\usepackage{graphicx,makeidx}
\usepackage{caption,subcaption}
\usepackage[mathscr]{eucal}

\usepackage{titlesec}
\titleformat{\subsection}[runin]
 {\normalfont\normalsize\bfseries}{\thesubsection}{1em}{\!\!}
\titleformat{\subsubsection}[runin]
 {\normalfont\normalsize\bfseries}{\thesubsubsection}{1em}{\!\!}
{\theorembodyfont{\upshape} }
\newcommand{\boma}[1]{{\mbox{\boldmath $#1$} }}

\begin{document}
\def\epsii{\eta}
\def\ov{\overline}
\def\XX{\mathcal{X}}
\def\YY{\mathcal{Y}}
\def\cemb{\hookrightarrow}
\def\raL{\ra}
\def\E0{E_0}
\def\Dik{\Dir^{(\mm)}}
\def\vi{z}
\def\zi{v}
\def\vt{v}
\def\En{\mathfrak{E}}
\def\TAM{A_{(>)}}
\def\TAm{A_{(<)}}
\def\InB{\mathfrak{B}^\s}
\def\eu{\varepsilon^{\s}}
\def\eren{\varepsilon^{ren}}
\def\TT{\mathscr{T}}
\def\TTs{\TT^{\s}}
\def\an{\lambda}
\def\ET{{}_T\mathcal{E}}
\def\Ep{{}_p\mathcal{E}}
\def\b0{\textbf{0}}
\def\Ta{\mathsf{T}}
\def\Bp{{B_{\sp,n}}}
\def\Bs{{B_{\si,n}}}
\def\sp{{\bar{\si}}}
\def\Sp{{\bar{S}}}
\def\ba{{\bf a}}
\def\bap{\mathfrak{a}}
\def\sip{\mathfrak{p}}
\def\bj{\sip(j)}
\def\Op{\mathfrak{O}}
\def\fo{\mathfrak{F}}
\def\HE{H}
\def\uu{v}
\def\AM{A}
\def\Am{a}
\def\DD{D}
\def\blp{\bar{\bf l}}
\def\bkl{b_{\bk\bl}}
\def\bklp{\bar{b}_{\bk\blp}}
\def\bxp{\bar{\bx}}
\def\bkp{\bar{\bk}}
\def\CB{C}
\def\aa{\alpha}
\def\bxk{\bx_{\star}}
\def\xk{x_\star}
\def\Co{\lozenge}
\def\NCo{\blacksquare}
\def\Toro{{\bf T}}
\def\atanh{\mbox{arcth}\,}
\def\Iun{\mathfrak{J}}
\def\GaL{\mathfrak{g}}
\def\Mk{M_{\mm,k}}
\def\ri{\mathfrak{r}}
\def\oms{\omega_{*}}
\def\Fock{\mathfrak{F}}
\def\Ee{\mathfrak{E}}
\def\ee{\epsilon}
\def\Ome{\Om_\ee}
\def\bxe{\bx_\ee}
\def\FI{\boma{S}}
\def\dFI{\mathfrak{S}}
\def\ff{\mathcal{F}}
\def\hh{\mathcal{H}}
\def\Mel{\mathfrak{M}}
\def\Tr{\mbox{Tr}\,}
\def\xb{\bar{x}}
\def\rb{\bar{\rho}}
\def\rr{\bar{r}}
\def\uno{\mbox{\textbf{1}}}
\def\tez{\al}
\def\Res{\mbox{Res}}
\def\L2m0{L^2_0}
\def\Tm{\tilde{T}}
\def\II{\mathfrak{I}}
\def\Ns{\mathfrak{N}}
\def\Tti{\tilde{T}}
\def\DF{\mathscr{D}}
\def\F{{\mathcal F}}
\def\mm{\kappa}
\def\oo{\varpi}
\def\0{{\bf 0}}
\def\UU{{\mathcal U}}
\def\Hank{\mathfrak{H}}
\def\t{\mathfrak{t}}
\def\Dir{D}
\def\l{\left}
\def\r{\right}
\def\ha{\widehat{a}}
\def\ak{\ha_k}
\def\ah{\ha_h}
\def\had{\ha^{\dagger}}
\def\akd{\had_k}
\def\ahd{\had_h}
\def\GD{\mathfrak{G}}
\def\Fk{F_k}
\def\Fh{F_h}
\def\Fkc{\ov{\Fk}}
\def\Fhc{\ov{\Fh}}
\def\Fkd{\mathfrak{F}_{k_1}}
\def\fk{f_k}
\def\fh{f_h}
\def\fkc{\ov{\fk}}
\def\fhc{\ov{\fh}}
\def\bl{{\bf l}}
\def\bn{{\bf n}}
\def\bk{{\bf k}}
\def\bh{{\bf h}}
\def\bx{{\bf x}}
\def\by{{\bf y}}
\def\bz{{\bf z}}
\def\bq{{\bf q}}
\def\bp{{\bf p}}
\def\Nab{\square}
\def\Fi{\widehat{\phi}}
\def\s{u}
\def\Fis{\Fi^{\s}}
\def\Fiseps{\Fi^{\eps \s}}
\def\Ti{\widehat{T}}
\def\Tis{\Ti^{\s}}
\def\Tiseps{\Ti^{\eps \s}}
\def\Aa{\widehat{A}}
\def\Bb{\widehat{B}}
\def\al{\alpha}
\def\be{\beta}
\def\de{\delta}
\def\eps{\varepsilon}
\def\ga{\gamma}
\def\lam{\lambda}
\def\om{\omega}
\def\si{\sigma}
\def\te{\theta}
\def\Ga{\Gamma}
\def\Om{\Omega}
\def\Si{\Sigma}
\def\dd{\displaystyle}
\def\la{\langle}
\def\ra{\rangle}
\def\leqs{\leqslant}
\def\geqs{\geqslant}
\def\sc{\cdot}
\def\restriction{\upharpoonright}
\def\parn{\par\noindent}
\def\complessi{{\bf C}}
\def\reali{{\bf R}}
\def\razionali{{\bf Q}}
\def\interi{{\bf Z}}
\def\naturali{{\bf N}}
\def\AA{{\mathcal A}}
\def\BB{{\mathcal B}}
\def\FF{{\mathcal F}}
\def\EE{{\mathcal E}}
\def\GG{{\mathcal G} }
\def\HH{{\mathcal H}}
\def\JJ{{\mathcal J}}
\def\KK{{\mathcal K}}
\def\LL{{\mathcal L}}
\def\MM{{\mathcal M}}
\def\OO{{\mathcal O}}
\def\PP{{\mathcal P}}
\def\QQ{{\mathcal Q}}
\def\RR{{\mathcal R}}
\def\SS{{\mathcal S}}
\def\cir{{\scriptscriptstyle \circ}}
\def\parn{\par \noindent}
\def\salto{\vskip 0.2truecm \noindent}
\def\saltino{\vskip 0.1truecm \noindent}
\def\beq{\begin{equation}}
\def\feq{\end{equation}}
\def\barray{\begin{array}}
\def\farray{\end{array}}
\newcommand{\rref}[1]{(\ref{#1})}
\def\vain{\rightarrow}
\def\spazio{\vskip 0.5truecm \noindent}
\def\fine{\hfill $\square$ \vskip 0.2cm \noindent}
\def\ffine{\hfill $\lozenge$ \vskip 0.2cm \noindent}

\setcounter{secnumdepth}{5}

\makeatletter \@addtoreset{equation}{section}
\renewcommand{\theequation}{\thesection.\arabic{equation}}
\makeatother
\begin{titlepage}
{~}
\vspace{-2cm}
\begin{center}
{\Large \textbf{Local zeta regularization}}
\vskip 0.2cm
{~}
\hskip -0.4cm
{\Large \textbf{and the scalar Casimir effect I.}}
\vskip 0.2cm
{~}
\hskip -0.4cm
{\Large\textbf{A general approach based on integral kernels}}
\end{center}
\vspace{0.5truecm}
\begin{center}
{\large
Davide Fermi$\,{}^a$, Livio Pizzocchero$\,{}^b$({\footnote{Corresponding author}})} \\
\vspace{0.5truecm}
${}^a$ Dipartimento di Matematica, Universit\`a di Milano\\
Via C. Saldini 50, I-20133 Milano, Italy\\
e--mail: davide.fermi@unimi.it \\
\vspace{0.2truecm}
${}^b$ Dipartimento di Matematica, Universit\`a di Milano\\
Via C. Saldini 50, I-20133 Milano, Italy\\
and Istituto Nazionale di Fisica Nucleare, Sezione di Milano, Italy \\
e--mail: livio.pizzocchero@unimi.it
\end{center}
\begin{abstract}
This is the first one of a series of papers about zeta regularization
of the divergences appearing in the vacuum expectation value (VEV) of
several local and global observables in quantum field theory. More
precisely we consider a quantized, neutral scalar field on a domain in
any spatial dimension, with arbitrary boundary conditions and, possibly,
in presence of an external classical potential. We analyze, in particular,
the VEV of the stress-energy tensor, the corresponding boundary forces
and the total energy, thus taking into account both local and global
aspects of the Casimir effect. In comparison with the wide existing
literature on these subjects, we try to develop a more systematic approach,
allowing to treat specific configurations by mere application of a general
machinery. The present Part I is mainly devoted to setting up this general
framework; at the end of the paper, this is exemplified in a very simple
case. In Parts II, III and IV we will consider more engaging applications,
indicated in the Introduction of the present work.
\end{abstract}
\vspace{0.2cm} \noindent
\hfill \parn
\par \vspace{0.3truecm} \noindent \textbf{AMS Subject classifications:} 81T55, 83C47.
\par \vspace{0.3truecm} \noindent \textbf{PACS}: 03.70.+k, 11.10.Gh, 41.20.Cv~.
\end{titlepage}
\tableofcontents
\vfill \eject \noindent
\section{Introduction} \label{intro}
As well known, zeta regularization treats divergent quantities in
quantum field theory introducing a complex parameter, with the
role of a regulator, and defining the renormalized observables in
terms of analytic continuation with respect to the regulator. This
construction was first proposed by Dowker and Critchley
\cite{DowKri}, Hawking \cite{Hawk} and Wald \cite{Wal} to
renormalize local observables, such as the vacuum expectation
value (VEV) of the stress-energy tensor; the ultimate purpose was
the semiclassical treatment of quantum effects in general
relativity (e.g., using the stress-energy VEV as a source in
Einstein's equations). After the previously cited pioneers, a
number of authors championed the zeta approach to treat local
observables; let us mention, in particular, Cognola, Zerbini,
Elizalde \cite{CoZe1,CoZe2}, and Moretti \cite{MorBo,Mor1a,Mor1,Mor2,MorRec,Morall}
who worked in curved spacetimes, and Actor, Svaiter et al.
\cite{ActBox1,ActBox2,Actor,SvaBox1,SvaBox2} who worked on spatial
domains with boundaries in flat spacetime. \parn
The zeta strategy can be as well applied to global observables,
such as the VEV of the total energy; in this version, it has in fact
become more popular than its local counterpart. The literature on
global zeta regularization is enormous; here we only cite the classical
papers \cite{Zim1,Zim2} by Zimerman et al., \cite{Blau} by Blau, Visser,
and Wipf and the monographies of Elizalde et al. \cite{MorBo,AAP0,10AppZ},
of Bordag et al. \cite{BorNew,Bord} and of Kirsten \cite{KirBook}
(see also \cite{Kir1,Kir3,Kir5,Kir4,Kir6,Kir2}). \parn
Both in the local and in the global version, zeta methods provide a very
natural approach to study the effects of quantum vacuum; in comparison
with the original derivation of these effects by Casimir \cite{Casimir},
and with other methods such as point-splitting \cite{BirDav,BunDav,Mil}
(in particular, see \cite{Mor3,Mor2} for a comparison between
point-splitting and the zeta regularization approach) and the algebraic,
microlocal approach (see \cite{Dap1,DapCas,Ful,Par,Rad} and citations therein),
the zeta strategy is competitive and, perhaps, more elegant. \parn
The present series of papers (formed by this work and by the
subsequent Parts II,III,IV \cite{PII,PIII,PIV}) considers both the
local and the global zeta approach, developing the viewpoint
proposed in a special case by our previous work \cite{ptp}. We are
mainly interested in the theory of a quantized neutral scalar
field on flat Minkowski spacetime, of arbitrary dimension $d+1$.
The field is confined within a $d$-dimensional spatial domain
$\Om$ and arbitrary boundary conditions are prescribed for it on
the boundary $\partial \Om$; besides, we also admit the presence
of a classical external potential, which could be meant to
describe, in some sense, an effective theory of interacting
fields. (We will only occasionally mention the possibility of
replacing the flat domain $\Om$ with a Riemannian manifold, or an
open subset of it with prescribed boundary conditions; this
amounts to replace Minkowski spacetime with a non-flat ultrastatic
spacetime \cite{MorBo}.) \parn Our attention is mainly focused on
the VEV of the stress-energy tensor, of which we consider both the
conformal and the non-conformal parts; however, we also determine
the total energy and boundary forces. \parn Most of the works on
local zeta regularization cited above consider Euclidean quantum
fields; this makes a difference with respect to our formalism,
based on canonical quantization in a genuinely Lorentzian
framework. Moreover, each one of the previously mentioned works
\cite{ActBox1,ActBox2,Actor,ptp,SvaBox1,SvaBox2} about local zeta
regularization for flat domains deals with a specific spatial
configuration (e.g. a strip between parallel planes,
configurations with perpendicular planes, a rectangular wave guide
or a rectangular box). In the present series of papers we are
trying to develop a more systematic approach to the zeta strategy,
to be applied nearly automatically in each specific case. \parn In
order to set up the general formalism we use with more generality
the following fact, emerging from the previous literature: the
analytic continuations involved in the zeta approach are closely
related to certain integral kernels determined by the basic
elliptic operator $\AA$ which governs the spatial part of the
field equation; among these kernels one could mention, in
particular, the Dirichlet and heat kernels corresponding,
respectively, to complex powers of $\AA$ and to the exponential
$e^{-\t \AA}$. Our papers I-IV emphasize as far as possible the
basic role of these and other kernels in view of zeta
regularization. Our aim is to write down few very general rules to
construct the required analytic continuation via integral kernels;
these prescriptions can be applied in an almost mechanical way to
treat specific configurations, as shown by several applications.
\parn The present Part I is mainly devoted to the general
formulation, and in particular to the set of rules mentioned
before; at the end of this paper we discuss, as a first
application, the very simple case of a massless field on a segment
(i.e., in spatial dimension $d=1$) for several types of boundary
conditions. In the subsequent Part II \cite{PII} we will show how
our general schemes work in a number of explicitly solvable cases
(the field between parallel or perpendicular planes, or inside a
wedge, with arbitrary boundary conditions and arbitrary choices of
the spatial dimension and of the conformal parameter). In Parts
III \cite{PIII} and IV \cite{PIV} we will consider two cases in
which the implementation of our general rules requires a mixture
of analytic and numerical calculations; more precisely, Part III
will discuss a field in precence of a background harmonic
potential and Part IV a field confined within a rectangular box.
In all these cases we will illustrate connections with the past
literature including, when they exist, previous computations based
on
alternative approaches such as point-splitting. \vspace{0.14cm}\\
Let us describe in more detail the contents of the present Part I.
In Section \ref{back} and in the related Appendix \ref{appetmunu}, we
introduce the general framework for a neutral scalar field on a $d$-dimensional
spatial domain $\Om$, including a brief discussion on the stress-energy
tensor with its conformal and non-conformal parts. This is an occasion
to fix the attention on the basic elliptic operator $\AA := - \Delta + V(\bx)$
on $\Om$, where $\Delta$ is the Laplacian and $V(\bx)$ the external potential.
Always in Section \ref{back}, we introduce zeta regularization in the
formulation already used in our previous work \cite{ptp}; the basic idea
is to replace the quantized field $\Fi$ with its regularized version
$\Fis := (\AA/\mm^2)^{-\s/4} \Fi$, where $\s \in \complessi$ is the
regulator and $\mm > 0$ is a mass parameter. Formally, $\Fis$ becomes $\Fi$
for $\s=0$; the zeta approach implements this idea in terms of analytic
continuation. This means the following: for any local or global observable
(say, the VEV of either the stress-energy tensor or of the total energy),
after introducing a regularized version of the observable based on $\Fis$,
we define the renormalized value as the analytic continuation at $\s=0$.
We distinguish between two versions of this prescription: the restricted
zeta approach, in which the observable is analytic at $\s=0$, and the
extended approach, where a singularity appears at $\s=0$ and is
eliminated removing the negative powers of $\s$ in the Laurent expansion. \parn
In Section \ref{Secker} (and in the related Appendices \ref{appeg},
\ref{apperes} and \ref{appeslab}) we present a number of integral kernels
associated to the operator $\AA := - \Delta + V(\bx)$. The Dirichlet kernel
$\Dir_{s}(\bx,\by) := \la\de_{\bx}|\AA^{-s}\de_{\by}\raL$ ($s$ in a
complex domain, $\bx,\by \in \Om$) is closely related to the stress-energy
VEV. More precisely, the regularized stress-energy VEV, built from $\Fis$,
is determined by the Dirichlet kernel $\Dir_s(\bx,\by)$ (and by its spatial
derivatives) at $\by = \bx$, $s = (\s \pm 1)/2$; so, the renormalized
version of this VEV is determined by the analytic continuation of $\Dir_s$
near $s = \pm {1/2}$. As shown in the
cited section, the continuation of the Dirichlet kernel can be determined algorithmically
relating it to the heat kernel $K(\t\,;\bx,\by) := \la \de_{\bx}|e^{-\t\AA}\de_{\by}\raL$
or to other kernels (among them the so-called cylinder kernel $T(\t\,;\bx,\by)
:= \la \de_{\bx}|e^{-\t\sqrt{\AA}}\de_{\by} \raL$, considered by Fulling
\cite{FullEst,FulGus}). This precedure ultimately gives the set of mechanical
rules for zeta regularization mentioned previously in this Introduction. \parn
In Section \ref{SecEn} and in Appendix \ref{apperell} we relate to the
previous framework the total energy, the pressure and the total forces on
the boundary. We also take the occasion to prove (at the regularized level)
the equivalence between two alternative definitions of the pressure,
often assumed without proof in the literature: pressure as the action
of the stress-energy VEV on the normal to the boundary, and pressure as
the functional derivative of the bulk energy with respect to deformations
of the spatial domain $\Om$. \parn
In Section \ref{secVar} and in Appendix \ref{appeEps} we consider some
variations of the general framework of Sections \ref{back} and \ref{Secker}
concerning the following situations: i) the case where $0$ is either an
isolated or non-isolated point of the spectrum of the fundamental operator
$\AA := - \Delta + V(\bx)$; ii) the case where the flat spatial domain
$\Om$ is described via curvilinear coordinates or, more generally, the case
where $\Om$ is a (possibly non-flat) Riemannian manifold equipped with arbitrary
coordinates (or an open subset of it, with prescribed boundary conditions).
This suggests the possibility to apply our formalism to the case of a non-flat
ultrastatic spacetime where the line element reads $ds^2 = -dt^2 + d\ell^2$,
with $d \ell^2$ the Riemannian line element of $\Om$. \parn
The final Section \ref{SegSec} presents a first application of our formalism;
this concerns the case in which $d=1$ and $\Om$ is the interval $(0,a)$,
with suitable boundary conditions. This configuration is very simple:
we take it in consideration just to show in action our mechanical procedures
for analytic continuation. Comparison with the existing literature on
this simple case is performed. As already mentioned, more sophisticated applications
of our approach will appear in Parts II-IV.
\section{Zeta regularization for a scalar field} \label{back}
\subsection{General setting.}
In this section we summarize the zeta regularization method for the
propagator and the vacuum expectation value (VEV) of the stress-energy
tensor of a quantized scalar field, in the formulation presented in
\cite{ptp} (see also \cite{tes}). Here the scheme of \cite{ptp} is slighlty generalized,
admitting the presence of a classical background potential $V$ and
arbitrary spacetime dimensions. \parn
Throughout the paper we use natural units, so that
\beq c = 1 ~, \qquad \hbar = 1 ~. \feq
We work in $(d+1)$-dimensional Minkowski spacetime; this is identified
with $\reali^{d+1}$ using a set of inertial coordinates
\beq x = (x^\mu)_{\mu=0,1,...,d} \equiv (x^0,\bx) \equiv (t,\bx) ~, \feq
in which the Minkowski metric has coefficients $(\eta_{\mu \nu}) = \mbox{diag}
(-1,1,...\,,1)$.
Let us fix a spatial domain $\Om \subset \reali^d$ where we consider a
quantized neutral, scalar field $\Fi$ in presence of a classical
background static potential $V$; so, we have $V: \Om \to \reali$,
$\bx \mapsto V(\bx)$ and
\beq \Fi : \reali \times \Om \to \LL_{s a}(\Fock)~; \qquad
0 = (-\partial_{tt}+\Delta-V(\bx)) \Fi(\bx,t) ~, \label{daquan} \feq
(analogous settings are considered, e.g., in \cite{BorNew,Bord,MorBo,Ful})
({\footnote{\label{footdist} Of course the notation
$\Fi : \reali \times \Om \to \LL_{s a}(\Fock)$,
$(\bx,t) \mapsto \Fi(\bx,t)$ is used here in connection
with a \textsl{generalized} operator valued function; in fact,
as well known, $\Fi$ is an operator valued \textsl{distribution}.
In the papers of this series we use
the classical functional notations even for
generalized functions.}}).
Here we are referring to the space $\LL(\Fock)$ of linear operators
on the Fock space $\Fock$, and to the subset $\LL_{s a}(\Fock)$ of
selfadjoint operators; $\Delta := \sum_{i=1}^d \partial_{ii}$ is
the $d$-dimensional Laplacian. We assume $V$ to be smooth, and
prescribe appropriate boundary conditions (e.g., the Dirichlet conditions $\Fi(t,\bx)=0$
for $\bx \in \partial\Om$). From here to the end of the paper we put
\beq \AA := - \Delta + V ~, \label{defaa} \feq
intending that the boundary conditions are accounted for in the above
definition; we assume this framework to grant that
$\AA$ is a selfadjoiont operator in the Hilbert space $L^2(\Om)$, with
the inner product
\beq \la f|g \raL := \int_{\Om}d\bx\, \ov{f}(\bx)
\,g(\bx) \feq
($d\bx$ denoting the standard Lebesgue measure on $\Om$)
({\footnote{In passing, we recall that the Hilbert space $\Fock$ can
be realized as the direct sum of all symmetrized tensor powers of $L^2(\Om)$.}}).
Moreover, we assume $\AA$ to be \textsl{strictly positive}, by which
we mean that the \textsl{spectrum} $\si(\AA)$ is contained in
$[\eps^2,+\infty)$ for some $\eps > 0$\,. Let us mention that $\AA$
may have continuous spectrum, a fact typically occurring when $\Om$
is unbounded; therefore, when we speak of the eigenvectors of $\AA$
we always intend them in a generalized sense, including improper
eigenfunctions. \parn
Let us consider a complete orthonormal set $(\Fk)_{k \in \KK}$ of
(generalized) eigenfunctions of $\AA$
({\footnote{For a fully rigorous discussion of generalized eigenfunctions,
see Chapter IV of \cite{GelSh}. In the sequel, following the usual
terminology, when speaking of functions (or distributions) on $\Om$
we use the adjectives ``proper'' or ``improper'', to distinguish
between objects which actually belong to $L^2(\Om)$ or not. In this
spirit we speak of proper or improper eigenfunctions, and use the same
terminology for the corresponding eigenvalues. In the sequel, the adjective
``generalized'' in relation to eigenfunctions is sometimes omitted.}}),
indexed by an unspecified set of labels $\KK$, and let us write
the corresponding eigenvalues in the form $(\om^2_k)_{k \in \KK}$
($\om_k \geqs \eps$ for all $k \in \KK$). Thus
\begin{equation}\begin{split}
& \Fk : \Om \to \complessi; \qquad \AA\Fk= \om^2_k \Fk ~; \\
& \hspace{-0.7cm} \la \Fk | \Fh \raL = \de(k, h) \quad
\mbox{for all $k,h \in \KK$} ~.
\end{split}\end{equation}
Any eigenfunction label $k \in \KK$ can include different parameters,
both discrete and continuous. Besides, we generically write $\int_\KK dk$
to indicate summation over all labels (i.e., literal summation over
discrete parameters and integration over continuous parameters, with
respect to a suitable measure); $\de(h,k) = \de(k,h)$ is the Dirac
delta function for the label space $\KK$ (this reduces to the
Kr\"onecker symbol in the case of discrete parameters). Note that
the condition $\om_k \geqs \eps >0$ excludes the
presence of infrared divergences from all the sums over $k$
appearing in the sequel. \parn
The functions
\beq \fk : \reali \times \Om \to \complessi~, \qquad
x = (t, \bx) \mapsto \fk(x) := e^{- i \om_k t} \Fk(\bx) \feq
fulfill $(- \partial_{tt} - \AA) \fk = 0$\,; they allow us to infer
for the quantized field a normal modes expansion of the form
\beq \Fi(x) = \int_\KK {d k \over \sqrt{2 \om_k}}
\l[\ak\,\fk(x) + \akd\,\fkc(x) \r] \label{espan} \feq
\vfill \eject \noindent
{~}
\vskip -1cm \noindent
(with $^\dag$ indicating the adjoint operator, and
$\ov{\phantom{I}}$ the complex conjugate). In the above we are
considering the destruction and creation operators $\ak,\akd\in\LL(\Fock)$,
which fulfill the canonical commutation relations
\beq [\ak , \ah] = 0 ~, \quad [\ak , \ahd] = \de(h,k) ~,
\qquad \ak |0 \ra = 0 ~, \feq
where $|0\ra \in \Fock$ is the vacuum state (of unit norm).
A relevant character for the sequel of our analysis will be the
\textsl{propagator}, i.e., the vacuum expectation value (VEV)
\beq \la 0|\Fi(x)\Fi(y)|0\ra \qquad
(x,y\in\reali\times\Om)~. \label{propag} \feq
Let us pass to the stress-energy tensor operator; this depends on
a parameter $\xi \in \reali$, and its components $\Ti_{\mu\nu}:
\reali\times\Om \to \LL_{sa}(\Fock)$, for $\mu,\nu \in\{0,1,...,d\}$,
are given by
\beq \Ti_{\mu \nu} := \l(1 - 2\xi \r) \partial_\mu \Fi \circ \partial_\nu \Fi
- \!\l({1\over 2} - 2\xi \r)\eta_{\mu\nu}(\partial^\lam\Fi \, \partial_\lam \Fi + V\Fi^2)
- 2 \xi \, \Fi \circ \partial_{\mu \nu} \Fi \label{tiquan} \feq
where we use the symmetrized operator product $\Aa \circ \Bb :=
(1/2) (\Aa \Bb + \Bb \Aa)$ and all the bilinear terms in the field
are evaluated on the diagonal (e.g., $\partial_\mu \Fi \circ
\partial_\nu \Fi$ indicates the map $x \mapsto \partial_\mu \Fi(x)
\circ \partial_\nu \Fi(x)$). \parn
Eq. \rref{tiquan} provides a natural quantization of what is often
called the ``improved'' stress-energy tensor; this is a well-known
modification of the canonical stress-energy tensor with an additive
term proportional to the parameter $\xi$, that does not alter its
divergence. For certain boundary conditions, e.g. of Dirichlet type,
this addition does not even alter the corresponding momentum vector.
We refer to Appendix \ref{appetmunu} for further details on this topic.
Here we only recall that the improved stress-energy tensor was first
proposed by Callan, Coleman and Jackiw \cite{Call} in order to deal
with some pathologies appearing in perturbation theory; later on, this
tensor was reinterpreted in terms of the Minkowskian limit for a scalar
field coupled to gravity via the curvature scalar \cite{BirDav,MorBo,DowKri,Oha,Par}. \parn
Needless to say, the vacuum expectation value of the stress-energy tensor
\rref{tiquan} is the main character in the theory of the Casimir effect.
It is evident from Eq. \rref{tiquan} that $\la 0|\Ti_{\mu\nu}(x)|0\ra$
can be expressed formally in terms of the propagator \rref{propag} (and
of its derivatives) evaluated on the diagonal $y = x$. On the other hand,
the propagator is known to be plagued with ultraviolet divergences along
the diagonal, so that $\la 0|\Ti_{\mu\nu}(x)|0\ra$ is a merely formal
expression for a divergent quantity; our purpose is to redefine the
propagator and the stress-energy VEV via a suitable regularization,
ultimately yielding finite values for these quantities.
\vspace{-0.4cm}
\subsection{Zeta regularization.}
Let $\mm > 0$ denote a parameter, to which we
attribute the dimension of a mass (or of an inverse length, since $\hbar=1$).
$\mm$ will be called the mass scale; it is introduced for dimensional reasons
and plays the role of a normalization scale. See \cite{Blau,MorBo,AAP0,Mor1a,Mor1}
for further comments regarding this parameter and its presence or absence
in the renormalized observables of the field.
We will check that the final, renormalized results depend on $\mm$ only
when singularities appear in the analytic continuations involved in the
following construction. \parn
The zeta strategy, in the version proposed in \cite{ptp} to give  meaning
to the VEV of $\Ti_{\mu \nu}$, relies on the powers
\beq (\mm^{-2}\AA)^{-\s/4} = \mm^{\s/2} \AA^{-\s/4} \feq
where $\AA$ is the operator \rref{defaa} and $\s \in \complessi$\,;
these are employed to define the \textsl{smeared}, or
\textsl{zeta-regularized}, \textsl{field operator}
\beq \Fis := (\mm^{-2} \AA)^{-\s/4}\, \Fi ~, \label{Fis} \feq
depending on the complex parameter $\s$ and coinciding with the usual
field operator $\Fi$ for $\s = 0$. If $(\Fk)_{k \in \KK}$ is a
complete orthonormal set of eigenfunctions of $\AA$ with eigenvalues
$(\om^2_k)_{k\in \KK}$, we have $(\mm^{-2}\AA)^{-\s/4}\Fk = \mm^{\s/2}
\om_k^{-\s/2} \Fk$, for any $k \in \KK$; so, the functions $\fk(t,\bx) =
e^{-i\om_k t}\Fk(\bx)$ fulfill $(\mm^{-2}\AA)^{-\s/4}\fk = \mm^{\s/2}
\om_k^{-\s/2} \fk$, and the expansion \rref{espan} for $\Fi$ becomes,
after application of $(\mm^{-2}\AA)^{-\s/4}$,
\beq \Fis(x) = \mm^{\s/2}\!\int_{\KK} {d k \over \sqrt{2}\,\om_k^{1/2+\s/2}}
\l[\ak\,\fk(x) + \akd\,\fkc(x) \r] \label{espans} . \feq
Note that, in the limit $\om_k \to + \infty$, the term $1/\om_k^{1/2+\s/2}$
in the above integral vanishes rapidly if $\Re \s$ is large; this is a
manifestation of the regularizing effect of the operator
$(\mm^{-2}\AA)^{-\s/4}$ for large $\Re \s$, a fact we will describe
much more precisely in the sequel. \parn
Using $\Fis$, we can define a \textsl{regularized propagator}
\beq \la 0|\Fis(x)\Fis(y)|0\ra \qquad
(x,y\in\reali\times\Om)  \label{proprel} \feq
and a \textsl{zeta regularized stress-energy tensor}
\beq \Tis_{\mu \nu} \!:= (1\!-\!2\xi)\partial_\mu \Fis\!\circ\partial_\nu \Fis\!
- \!\l({1\over 2}\!-\!2\xi\!\r)\!\eta_{\mu\nu}\!\l(\!\partial^\lam\Fis\partial_\lam \Fis\!+\!V(\Fis)^2\r)
- 2 \xi \, \Fis\!\circ \partial_{\mu \nu} \Fis \,, \label{tiquans} \feq
where, as in Eq. \rref{tiquan}, all the bilinear terms in the field
are evaluated on the diagonal. \parn
We are interested in the VEV of this regularized stress-energy tensor,
which formally gives $\la 0|\Ti_{\mu\nu}(x)|0\ra$ in the limit $\s \to 0$\,.
Of course, we can relate the VEV of $\Tis_{\mu\nu}(x)$ to the
regularized propagator \rref{proprel} in the following way:
$$ \la 0 | \Tis_{\mu \nu}(x) | 0 \ra = $$
$$ = \! \l. \l({1 \over 2}\!-\!\xi\!\r)\!(\partial_{x^\mu y^\nu}\!
+ \partial_{x^\nu y^\mu}\!)\! -\!\l({1\over 2}\!-\!2\xi\!\r)\!
\eta_{\mu\nu}\!\l(\partial^{x^\lam}\!\partial_{y^\lam}\!+\! V(\bx) \r)\!
- \xi (\partial_{x^ \mu x^\nu}\!+ \partial_{y^\mu y^\nu}\!) \r|_{y=x}\!\!\cdot $$
\beq \cdot ~ \la 0 | \Fis(x) \Fis(y) | 0 \ra ~. \label{tisprop} \feq
We will return later on this equation and on its use for the actual
computation of the above VEV. \parn
Typically, the regularized propagator and the VEV of $\Tis_{\mu \nu}(x)$
are analytic functions of $\s$, for $\Re\s$ sufficiently large; the same
can be said of many related observables (including global object, such
as the total energy, which is related to the space integral of the $(0,0)$
component of the stress-energy tensor). Let us consider any one of these
(local or global) observables, and denote with $\F(\s)$ its zeta-regularized
version, based on Eq. \rref{Fis} (see, e.g., Eq. \rref{proprel} or Eq.
\rref{tiquans}); we assume that the function $\s \mapsto \FF(\s)$ is well
defined and analytic for $\s$ in a suitable domain $\UU_0$ of the complex
plane. The zeta approach to renormalization can be formulated in either a
``restricted'' or an ``extended'' version, both described hereafter. \salto
i) \textsl{Zeta approach, restricted version}. Assume that the function
$\UU_0 \to \complessi, \s \mapsto \F(\s)$ can be analytically continued
to a larger open subset $\UU$ of $\complessi$ such that $0 \in \UU$\,;
let us use the notation $\s \mapsto \F(\s)$ even for this extension
({\footnote{In the style of our previous work \cite{ptp} we should
write $\F : \UU_0 \to \complessi$ for the initially given function and
$AC\,\F: \UU \to \complessi$ for its analytic continuation; here, we
choose to simplify the notation, writing $\F$ for both functions. Note
that the analogue of Eq. \rref{ren} in the style of \cite{ptp} would be
$\F_{ren} := (AC\,\F)(0)$.}}).
In this case, making reference to the analytic continuation, we define
the renormalized value of the observable under consideration as
\beq \F_{ren} := \F(0) ~. \label{ren} \feq
ii) \textsl{Zeta approach, extended version}. Assume that there is an
open subset $\UU$ of $\complessi$, larger than $\UU_0$, such that
$0 \in \UU$ and the function $\s \in \UU_0 \mapsto \F(\s)$ has an
analytic continuation to $\UU \setminus \{0\}$, still indicated with $\F$.
In this case, since there is an isolated singularity at $\s=0$, in a
neighborhood of this point we have the Laurent expansion $\F(\s) =
\sum_{k = -\infty}^{+\infty}\F_k \s^k$\,. Let us consider the
\textsl{regular part}
\beq (RP\,\F)(\s) := \sum_{k=0}^{+\infty} \F_k \s^k~; \feq
we define the renormalized value of the given observable as
\beq \F_{ren} := (RP\,\F)(0) \label{renest} \feq
(i.e., $\F_{ren} = \F_0$). In most applications $\F$ is \textsl{meromorphic}
close to $\s=0$, which means that it has a \textsl{pole} at this point; in
this case the previous Laurent expansion has the form $\F(\s) = \sum_{k=-N}^{+\infty}
\F_k \s^k$, where $N \in \{1,2,3,...\}$ is the order of the pole.
Let us stress that the prescription \rref{renest} is a quite
straightforward generalization of the approach considered in
\cite{Blau,AAP0}, where $\F$ was assumed to posses a simple pole
in $\s=0$ (i.e., it was assumed that $N=1$). \salto
Of course, the restricted zeta approach of item (i) is equivalent to a
special case of the extended approach, in which $\F(\s)$ has a removable
singularity at $\s=0$ and the Laurent expansion at this point is the usual
power series expansion. \salto
A large part of our subsequent work will be devoted to the application
of the previous scheme to the VEV of the stress-energy tensor. In general,
we define the renormalized version of the latter as
\beq \la 0 | \Ti_{\mu \nu}(x) | 0 \ra_{ren} :=
RP \Big|_{\s=0} \la 0 | \Tis_{\mu \nu}(x) | 0 \ra~; \label{prii} \feq
when no singularity appears at $\s=0$, according to the restricted
approach (i) the above definition reduces to
\beq \la 0 | \Ti_{\mu \nu}(x) | 0 \ra_{ren} := \la 0 | \Tis_{\mu \nu}(x) | 0 \ra
\Big|_{\s=0} ~. \label{pri} \feq
In \cite{ptp}, we only considered the prescription \rref{pri} in the
special case of a Dirichlet field (with $V = 0$) between two parallel
planes, i.e., in the configuration corresponding to the standard theory
of the Casimir effect. In that case the approach \rref{pri} was
implemented via a direct computation of the analytic continuation appearing
therein; as already stressed, here we aim to much more generality.
\vspace{-0.4cm}
\subsection{A remark.} \label{aremark}
In the sequel, while performing zeta regularization and the consequent
renormalization, it is sometimes natural to consider, in place of $\s$,
some complex parameter $s$ related to $\s$ by a simple transformation.
In view of such situations, it is convenient to generalize some notations
of the previous subsection in the following way: \parn
i) Consider an analytic function $\SS_0 \to \complessi$, $s \mapsto \F(s)$,
where $\SS_0$ is an open subset of $\complessi$; if this admits an analytic
continuation to a larger open subset $\SS$, the latter will be still
denoted with $s \mapsto \F(s)$. \parn
ii) Suppose the analytic function $\SS_0 \to \complessi$, $s \mapsto \F(s)$
has an analytic extension to $\SS \setminus \{s_0\}$, where $\SS$ is an
open subset of $\complessi$ and $s_0 \in \SS$\,. Then, the Laurent expansion
$\F(s) = \sum_{k=-\infty}^{+\infty} \F_k (s\!-\!s_0)^k$ will be used to
define the regular part (near $s_0$) of this analytic continuation as
$(RP\,\F)(s) := \sum_{k=0}^{+\infty} \F_k (s\!-\!s_0)^k$\,; of course,
this implies $(RP\,\F)(s_0) = \F_0$\,.
\vspace{-0.4cm}
\subsection{Staticity features of the VEV of $\boma{\Ti_{\mu\nu}}$.}
\label{statSubsec}
Let us return to the regularized stress-energy tensor; for $x = (t, \bx) \in \reali
\times \Om$, we claim that
\begin{equation}\begin{split}
& \hspace{1.8cm} \la 0|\Tis_{\mu\nu}(x)|0\ra~~\mbox{is independent of $t$} ~, \\
& \la 0|\Tis_{0i}(x)|0\ra = \la 0|\Tis_{i0}(x)|0\ra = 0
\qquad \mbox{for $i \in \{1,...,d\}$} ~. \label{eqstatics}
\end{split}\end{equation}
These statements are not surprising, due to the staticity of the
general framework considered in the present paper; a formal proof
will be given in subsection \ref{subsecrel}
(see Eq.s (\ref{Tidir00}-\ref{Tidirij}) and the considerations which
follow them). Of course the features of Eq. \rref{eqstatics} are
preserved by analytic continuation, so that an analogue of this equation
holds for the renormalized VEV $\la 0|\Ti_{\mu\nu}(x)|0\ra_{ren}$ as well.
\subsection{Conformal and non-conformal parts of the stress-energy tensor.}
\label{ConfSubsec}
In the literature (see, e.g., \cite{BirDav,MorBo,WaldRG}) it is customary
to write the stress-energy tensor (here to be intended as one of the
operators $\Ti_{\mu\nu}$, $\Tis_{\mu\nu}$, or either one of the VEVs
$\la 0|\Tis_{\mu\nu}|0\ra$, $\la 0|\Ti_{\mu\nu}|0\ra_{ren}$) as the
sum of a \textsl{conformal} and a \textsl{non-conformal} part. In order
to define these quantities, let us consider for $\xi$ the critical value
\beq \xi_d := {d\!-\!1 \over 4d} ~; \label{xic} \feq
it is known that, when coupling of the scalar field to gravity is taken
into account, the theory is invariant (for $V=0$) under conformal
transformations of the spacetime line element if $\xi$ has the above
critical value (see, e.g., \cite{WaldRG}, p.447). In the sequel we
adopt the notations
\beq \mbox{$\Co \equiv$ conformal}~, \qquad \mbox{$\NCo \equiv $ non-conformal}
\label{CoNCo} \feq
and we define, for example, the conformal and non-conformal parts of the
renormalized stress-energy VEV $\la 0|\Ti_{\mu\nu}|0\ra_{ren}$ in the
following way:
\beq \la 0|\Ti^{(\Co)}_{\mu\nu}|0\ra_{ren} :=
\la 0|\Ti_{\mu\nu}|0\ra_{ren}\Big|_{\xi = \xi_d} ~, \label{TCo} \feq
\beq \la 0|\Ti^{(\NCo)}_{\mu\nu}|0\ra_{ren} := {1 \over \xi\!-\!\xi_d}\,
\Big(\la 0|\Ti_{\mu\nu}|0\ra_{ren} - \la 0|\Ti^{(\Co)}_{\mu\nu}|0\ra_{ren}\Big) ~.
\label{TNCo} \feq
Of course, this implies
\beq \la 0|\Ti_{\mu\nu}|0\ra_{ren} = \la 0|\Ti^{(\Co)}_{\mu\nu}|0\ra_{ren}
+ (\xi\!-\!\xi_d)\,\la 0|\Ti^{(\NCo)}_{\mu\nu}|0\ra_{ren} ~. \label{TRinCo}\feq
In the applications to be considered in Section \ref{SegSec} and in
the subsequent Parts II,III and IV, when presenting our final results
for the renormalized stress-energy VEV, we will either write them in
the form \rref{TRinCo} or give separately the conformal and
non-conformal parts \rref{TCo} \rref{TNCo}.
\vspace{-0.4cm}
\subsection{Total energy and pressure on the boundary.}
\label{TotEnSubsec}
The \textsl{total energy} is, by definition, the integral of
$\la 0|\Ti_{00}(x)|0\ra$ over the spatial domain $\Om$\,. We defer the
discussion of this topic to Section \ref{SecEn}; therein we will describe
the representation of the total energy as the sum of a bulk term and
a boundary term, in the framework of zeta regularization. \parn
In the same section we will use zeta regularization to treat the
\textsl{pressure} on the boundary $\partial\Om$ of the
spatial domain; this quantity can be defined in terms of the VEV
of the spatial components $\Ti_{i j}$\,. There is an alternative
characterization of the pressure in terms of the variation
of the bulk energy (see Eq. \rref{defEs} for the definition) with
respect to deformations of the spatial domain $\Om$\,; the equivalence
of this definition with the previous one has often been assumed
uncritically in the literature, so we think it can be useful to
produce a formal proof (see Section \ref{SecEn}).
\section{Expressions of the zeta regularized stress-energy VEV in terms of integral kernels}
\label{Secker}
\setcounter{subsection}{1}
In this section $\Om$ always denotes a spatial domain in $\reali^d$,
and we often consider the Hilbert space $L^2(\Om)$ of the square
integrable complex-valued functions on $\Om$.
\vspace{-0.4cm}
\subsection{Basics on integral kernels.}
Let us consider a linear operator $\BB$ acting on $L^2(\Om)$. The
\textsl{integral kernel} of $\BB$ is the (generalized) function
\beq \BB(~,~) : \Om \times \Om \to \complessi \,, \qquad (\bx,\by) \mapsto
\BB(\bx, \by) := \la \de_\bx | \BB \, \de_{\by} \raL \label{kerb} \feq
where $\de_\bx$ and $\de_{\by}$ are the Dirac delta functions centered
at $\bx$ and $\by$, respectively, here viewed as improper vectors of the
Hilbert space $L^2(\Om)$
({\footnote{Let us repeat what we have declared in the footnote of page
\pageref{footdist}: in these papers we use functional notations even for
objects which are not ordinary functions. Making reference to the theory
of Schwartz distributions, we can understand the statement
``$\BB(~,~) : \Om \times \Om \to \complessi$'' as a way to indicate that
we are considering a (complex) distribution on $\Om \times \Om$; let us
describe how to intend Eq. \rref{kerb} from this viewpoint. We refer to the
space of test functions $D(\Om)$, formed by the $C^\infty$, compactly supported
functions $\varphi: \Om \to \complessi$ and equipped with its standard inductive
limit topology; $D(\Om \times \Om)$ has a similar meaning. We wish to define rigorously
$(\bx,\by) \mapsto \la \de_\bx | \BB \, \de_{\by} \raL$ as a distribution on
$\Om \times \Om$, i.e., as a continuous linear form on $D(\Om \times \Om)$.
To this purpose we consider $\varphi, \psi \in D(\Om)$ and note that the identities
$\varphi = \int_{\Om} d \bx \,\varphi(\bx) \delta_{\bx}$ and $\psi =
\int_{\Om} d \bx \, \psi(\bx) \delta_{\bx}$ give formally
$$ \int_{\Om \times \Om} d \bx d \by~
\overline{\varphi}(\bx) \la \delta_{\bx} | \BB \delta_{\by} \raL \psi(\by) = \la \varphi | \BB \psi \raL~. $$
On the other hand, if the domain of the operator $\BB$
contains $D(\Om)$ and the sesquilinear
map $D(\Om) \times D(\Om) \to \complessi$,
$(\psi, \varphi) \mapsto \la \varphi | \BB \psi \raL$
is continuous, using the nuclear theorem of Schwartz
(see \cite{Schwa}, Thm. II)
one proves rigorously the existence of
a unique distribution on $\Om \times \Om$,
denoted with $(\bx,\by) \mapsto \la \de_\bx | \BB \, \de_{\by} \raL$,
such that the above relation holds for all
$\varphi, \psi \in D(\Om)$, intending the integral in
the left hand side as the action of
this distribution on the test function
$(\bx,\by) \mapsto \overline{\varphi}(\bx) \psi(\by)$.
One can use similar considerations to give a
distributional meaning to Eq. \rref{compsuch} (see again
\cite{Schwa}).
Throughout the papers of this series, whenever we speak
of the integral kernel of an operator $\BB$ we implicitly
assume $\BB$ to possess the regularity features mentioned
before, so that $\la \delta_{\bx} | \BB \delta_{\by} \raL$
makes sense at least as a distribution. In some
cases described explicitly in the sequel, stronger assumptions
on $\BB$ ensure that $\la \delta_{\bx} | \BB \delta_{\by} \raL$
can be understood as an ordinary function, continuous
with its derivatives up to a certain order; this is
the situation outlined in Appendix \ref{AppKer}, often
mentioned in the sequel in relation to several
kernels of interest for us.}}).
Equivalently, the integral
kernel of the operator $\BB$ can be defined as the unique
(generalized) function $\BB(~,~):\Om \times \Om \to \complessi$
such that
\beq (\BB \psi)(\bx) = \int_\Om d\by \;\BB(\bx,\by) \, \psi(\by) ~,
\qquad \label{compsuch} \feq
for all sufficiently regular $\psi : \Om \to \complessi$\,.
If $\BB$ possesses a complete orthonormal set of (generalized) eigenfunctions
$(\Fk)_{k \in \KK}$ with corresponding eigenvalues $\be_k \in \complessi$
($\BB \Fk = \be_k \Fk$), then
\beq \BB(\bx, \by) = \int_\KK dk \,\be_k\,\Fk(\bx) \Fkc(\by)  \label{eqker} \feq
(since the function in the right-hand side fulfills equation
\rref{compsuch} for all $\psi$). The precise sense in which the eigenfunction
expansion \rref{eqker} converges depends on the specific features of the operator
$\BB$; we generally assume distributional convergence
({\footnote{Meaning that $\int_{\Om \times \Om} d\bx d\by\,\BB(\bx,\by) \varphi(\bx,\by)
= \int_\KK dk \,\be_k\, \int_{\Om \times \Om} d\bx d\by\,\Fk(\bx)\Fkc(\by) \varphi(\bx,\by)$
for all test functions $\varphi$ on $\Om \times \Om$.}}),
leaving to future subsections the consideration of special cases where convergence
can be intended in a stronger sense. \parn
Incidentally, let us mention the relation existing between the
kernel $\BB(~,~)$ and the \textsl{trace} of $\BB$; the latter,
if it exists, is the number $\Tr \BB := \int_\KK dk\,
\la \Fk| \BB \Fk \raL \in \complessi$, where $(\Fk)_{k \in \KK}$ is any
complete orthonormal set of $L^2(\Om)$. The right-hand side does not depend
on the choice of $(\Fk)_{k\in \KK}$; in particular, if $\BB$ has purely
discrete spectrum, $(\Fk)_{k\in \KK}$ is a complete orthormal set of proper
eigenfunctions labelled by a countable set $\KK$ and $(\be_k)_{k \in \KK}$
are the corresponding eigenvalues, we have $\Tr \BB = \sum_{k \in \KK} \be_k$,
if this series converges. Returning to definition \rref{kerb} of the kernel
$\BB(~,~)$, we see that
\beq \Tr \BB = \int_\Om d\bx\; \BB(\bx,\bx) \label{trb} \feq
(since $(\de_\bx)_{\bx \in \Om}$ is a generalized complete orthonormal set)
({\footnote{In the sequel, the adjective ``generalized'' in relation to
complete orthonormal sets is sometimes omitted.}}). \salto
Let us move on and note that the boundary conditions possibly involved
in the definition of $\BB$ have implications for the kernel $\BB(~,~)$\,;
for example, if boundary conditions of the Dirichlet type are involved,
the eigenfunctions $(\Fk)_{k\in\KK}$ in Eq. \rref{eqker} vanish on
$\partial\Om$, thus yielding $\BB(\bx,\by) = 0$ for $\bx \in \partial\Om$
or $\by \in \partial\Om$\,. \parn
Let us also mention that from Eq. \rref{kerb} one infers
\beq \BB^{\dagger}(\bx,\by) = \ov{\BB(\by,\bx)}  ~, \qquad
\ov{\BB}(\bx, \by) = \ov{\BB(\bx, \by)} \feq
where $\BB^{\dagger}$ is the adjoint operator of $\BB$ with respect to
the inner product of $L^2(\Om)$ while $\ov{\BB}$ is the complex
conjugate operator, such that $ \ov{{\BB}\psi} =
\ov{\BB}\,\ov{\psi}$ for all $\psi$. These facts imply
\beq \BB(\by, \bx) = \BB(\bx, \by) \qquad
\mbox{if \;$\BB^\dagger = \ov{\BB}$} ~. \label{imp} \feq
\vspace{-0.8cm}
\subsection{The operator $\boma{\AA}$.} \label{subA} Most of the integral kernels
considered in the sequel will be related to a selfadjoint operator $\AA$ acting
in $L^2(\Om)$; for example, they will be the kernels associated to some function
of the operator $\AA$ (say, a power $\AA^s$). To treat these kernels, precise
assumptions on $\AA$ will be specified whenever necessary; in any case, we will
typically consider three situations of descreasing generality, described by the
forthcoming Eq.s \rref{i}\rref{ii}\rref{iii}. \parn
In the first situation, we will simply assume that
\beq \mbox{$\AA$ is a strictly positive, selfadjoint operator in $L^2(\Om)$} \label{i} \feq
(let us recall that strict positivity means $\si(A) \subset [\eps^2,+\infty)$ for some $\eps >0$). \parn
In the second situation, the assumptions are the following:
\beq \barray{c} \mbox{$\AA = - \Delta+V$ in $\Om$, with $V$
a real $C^\infty$ potential on $\Om$ and} \\
\mbox{boundary conditions such that $\AA$ be selfadjoint and stricly positive.}
\label{ii} \farray \feq
In the third situation, the assumptions are:
\beq \hspace{-0.12cm} \barray{c} \mbox{$\AA = - \Delta + V$ on a bounded domain
$\Om$ with $\partial\Om$ of class $C^\infty$ and Dirichlet} \\
\mbox{boundary conditions,\! $V$\! a real $C^\infty$\! potential on $\ov{\Om}$
and $V(\bx)\!\geqs\!0$ for all $\bx\!\in\!\ov{\Om}$}\,.\farray\!\!\! \label{iii} \feq
Here and in the sequel  $\ov{\Om} = \Om \cup \partial \Om$ is the closure of $\Om$.
The assumptions \rref{iii} grant selfadjointness
and strict positivity of $\AA$; moreover, they imply that
$\AA$ has a purely discrete spectrum and one can build a complete
orthonormal system of proper eigenfunctions $\Fk$ with eigenvalues $\om^2_k$
labelled by $\KK = \{1,2,3,....\}$
in such a way that $0 < \om_1 \leqs \om_2 \leqs \om_3 \leqs ...$ (with the
possibility that some of these inequalities are equalities, to deal with the
case of degenerate eigenvalues). It is well-known that the eigenvalues, when
ordered in this manner, fulfill the Weyl asymptotic relation
\beq \om_k \sim C \,k^{1/d} \qquad \mbox{for $k \to + \infty$} ~, \label{weyl} \feq
where $C := 2 \sqrt{\pi}\, \Ga(d/2\!+\!1)^{1/d}\,\mbox{Vol}(\Om)^{-1/d}$ (see \cite{Mikh},
Thm.5, page 189 and \cite{Shu}, $\S$ 8.2, pages 99-101 for elementary derivations).
Appendix \ref{AppKer} contains a number of technical results concerning cases \rref{i}
\rref{ii} \rref{iii}, that will be mentioned whenever necessary in the sequel of this section. \parn
These results often refer to the spaces $C^j(\Om)$ and $C^j(\Om \times \Om)$ or
(in the case \rref{iii}) $C^j(\ov{\Om})$ and \hbox{$C^j(\ov{\Om} \times \ov{\Om})$}, for
$j \in \naturali$ or $j = \infty$; these are formed by the complex functions
on $\Omega, \Omega \times \Omega$ and so on, which are continuous along with
their partial derivatives of all orders $\leqs j$. In particular, in the rest
of the section we will present situations connected to cases \rref{i} \rref{ii}
\rref{iii}, in which certain integral kernels $(\bx,\by) \mapsto \BB(\bx,\by)$
related to $\AA$ are of class $C^j(\Om \times \Om)$ or $C^j(\ov{\Om} \times \ov{\Om})$,
for suitable $j$. \parn
The condition of strict positivity for $\AA$ in Eq.s \rref{i} \rref{ii} will
be occasionally relaxed, assuming only that the eigenvalues of $\AA$ are non-negative
(and declaring this explicitly, to avoid misunderstandings); see, e.g., subsection \ref{nonNeg}.
\vspace{-0.4cm}
\subsection{The Green function.}\label{GreenSubSec}
Let $\AA$ be a strictly positive selfadjoint operator in $L^2(\Om)$; then,
we can introduce the inverse operator $\AA^{-1}$ and the corresponding kernel
\beq G(\bx, \by) := \AA^{-1}(\bx, \by) ~, \feq
which is called the \textsl{Green function} of $\AA$\,. In terms of this
kernel, the identity $\AA\,\AA^{-1} = \boma{1}$ can be re-expressed as
\beq \AA_\bx\,G(\bx,\by) = \de(\bx-\by) ~, \label{ag} \feq
where $\AA_\bx$ indicates the operator $\AA$ acting on $G(\bx,\by)$ as a
function of $\bx$\,. Using a complete orthonormal system $(\Fk)_{k \in \KK}$
of eigenfunctions of $\AA$ with corresponding eigenvalues $(\om_k^2)_{k \in \KK}$,
we can express the Green function as
\beq G(\bx, \by) = \!\int_\KK {d k \over \om_k^2}\;\Fk(\bx)\Fkc(\by) ~. \feq
The Green function is among the most familiar integral kernels, especially
when $\AA = - \Delta\! + V$ with suitable boundary conditions (of the Dirichlet,
Neumann or Robin type); in this case, uniqueness results are available for
the Poisson equation (perturbed with an external potential), allowing to
characterize the Green function $G(\bx,\by)$ as the unique solution of
Eq. \rref{ag} fulfilling the prescribed boundary conditions for
$\bx \in \partial\Om$ or $\by \in \partial\Om$\,. The literature on this
topic is enormous, and here we only mention some well-known monographies:
Berezanskii \cite{Bere}, Krylov \cite{Kry}, Sauvigny \cite{Sau} and
Shimakura \cite{Shi} give abstract and rigorous analyses, while Duffy \cite{Duffy},
Kythe \cite{Kyt}, Sommerfeld \cite{Somm} and Stakgold and Holst \cite{Stack}
present more practical and explicit discussions.
\vspace{-0.4cm}
\subsection{A digression on complex powers.}
Throughout the present paper (and in Parts II-IV), the following
conventions are employed: \parn
i) $\ln : (0,+\infty) \vain \reali$ is the elementary logarithm; \parn
ii) for any $\al \in \complessi$, we systematically refer to the standard
definition
\beq x^\al := e^{\al\ln x} \qquad \mbox{for all $x \in (0,+\infty)$} ~ ;
\label{stand} \feq
iii) for $\al \in \complessi$ and $z$ in a convenient subset
$\complessi^{\times}$ of the complex plane, we define
\beq z^\al := e^{\al\ln |z| + i \al\arg z} ~, \label{power} \feq
where $\arg : \complessi^{\times} \to \reali$ is some determination of
the argument; this determination depends on the domain $\complessi^{\times}$
and must be specified in each case of interest. In most applications
considered hereafter we set
\begin{equation}\begin{split}
& \hspace{5.3cm} \complessi^{\times} := \complessi \setminus [0,+\infty)~; \\
& \arg := \mbox{the unique determination of the argument with values
in $(0,2 \pi)$} ~. \label{arg}
\end{split}\end{equation}
\vspace{-0.8cm}
\subsection{The Dirichlet kernel.}
Let again $\AA$ be a strictly positive selfadjoint operator in $L^2(\Om)$.
The power $\AA^{-s}$ can be defined through the standard functional calculus
for each $s \in \complessi$; the corresponding integral kernel
\beq \Dir_s(\bx, \by) := \AA^{-s}(\bx, \by) \label{Dirdef}\feq
is called the $s$-th \textsl{Dirichlet kernel}. In passing, let us note that
$D_{-1}(\bx,\by)$ coincides with the Green function $G(\bx,\by)$ considered
in subsection \ref{GreenSubSec}. \parn
If $(\Fk)_{k \in \KK}$ is a complete orthonormal set of eigenfunctions of $\AA$
with corresponding eigenvalues $(\om_k^2)_{k \in \KK}$ we have $\AA^{-s} \Fk =
\om_k^{-2s}\Fk$, so that (by Eq. \rref{eqker})
\beq \Dir_s(\bx, \by) = \int_\KK {dk \over \om_k^{2s}}\;
\Fk(\bx) \Fkc(\by) ~. \label{eqkerdi} \feq
The denomination of ``Dirichlet kernel'' employed for $\Dir_s$ is suggested
by the similarity between the above expansion and the Dirichlet series,
considered in \cite{HarDir,Minak1,Minak2,Seel}. \parn
For our purposes, it is important to give sufficient conditions under which
$\Dir_s(\bx,\by)$ is a regular function of $(\bx,\by)$ (even on the diagonal
$\by = \bx$, of special interest in the sequel); we are also interested in
cases in which the expansion \rref{eqkerdi} converges in a stronger sense,
in comparison with the distributional sense that we are generically ascribing
to kernel expansions. Let us present some results of this kind, which are
proved in Appendix \ref{AppKer}. \parn
Let $\AA = - \Delta + V$ be as in Eq. \rref{ii}; then
\beq \Dir_s \in C^j(\Om \times \Om) \qquad \mbox{for $s \in \complessi$, $j \in
\naturali$ such that $\dd{\Re s > {d \over 2} + {j \over 2}}$}~. \label{then} \feq
With the stronger assumptions \rref{iii}, we have
\beq \Dir_s \in C^j(\ov{\Om} \times \ov{\Om}) \qquad \mbox{for $s \in \complessi$,
$j \in \naturali$ such that $\dd{\Re s > {d \over 2} + {j \over 2}}$} ~. \label{thenn} \feq
Again with the assumptions \rref{iii}, $\AA$ possesses a complete orthonormal
set of proper eigenfunctions $(\Fk)_{k =1,2,3,...}$ (see the comments
after the cited equation), and the expansion \rref{eqkerdi} reads
\beq \Dir_s(\bx, \by) = \sum_{k=1}^{+\infty} {1 \over \om_k^{2s}}\;
\Fk(\bx) \Fkc(\by) ~. \label{eqkerdiSer} \feq
In this case one has $\Fk \in C^\infty(\ov{\Om})$ for each $k$; moreover,
for $\Re s > d + j/2$ the expansion \rref{eqkerdiSer} is absolutely and
uniformly convergent on $\ov{\Om} \times \ov{\Om}$, with  all its derivatives
up to order $j$: see Appendix \ref{AppKer} (especially Eq. \rref{absconv})
for more details. \parn
To conclude the present subsection let us mention that the general statement
of Eq. \rref{trb}, here applied with $\BB = \AA^{-s}$, yields
\beq \int_\Om d\bx \; \Dir_s(\bx, \bx) = \Tr \AA^{-s} ~, \label{TrAs} \feq
provided that the above trace exists. With the assumptions \rref{iii} for
$\AA = - \Delta + V$, using the Weyl estimates \rref{weyl} for $\om_1 \leqs
\om_2 \leqs ... $ we find
\beq \Tr \AA^{-s} = \sum_{k=1}^{+\infty} {1 \over \om^{2 s}_k}
\quad \mbox{is finite for $\dd{\Re s > {d \over 2}}$} ~. \label{case2} \feq
\vspace{-0.7cm}
\subsection{Some remarks concerning the Dirichlet kernel and its derivatives.}
Let us consider again a strictly positive selfadjoint operator $\AA$ in
$L^2(\Om)$; moreover, assume this operator to be \textsl{real}, in the
sense that $\ov{\AA} = \AA$ (i.e., $\ov{A \psi} = A \ov{\psi}$
for all $\psi$)\,. \parn
If $(\Fk)_{k\in \KK}$ is a complete orthonormal set of eigenfunctions of
$\AA$ with related eigenvalues $(\om_k^2)_{k \in \KK}$, then the conjugate
system $(\Fkc)_{k\in \KK}$ is as well a complete orthonormal set of
eigenfunctions of $\AA$ with the same eigenvalues; so, besides Eq.\!
\rref{eqkerdi} we have an alternative representation for the Dirichlet
kernel $\Dir_s(\bx,\by)$, based on this conjugate system. From here, we
easily infer that, for $s \in \complessi$ with complex conjugate $\bar{s}$,
\beq \Dir_s(\bx,\by) = \Dir_s(\by,\bx) \qquad \mbox{and} \qquad
\ov{\Dir_s(\bx,\by)} = \Dir_{\bar{s}}(\bx,\by) ~. \label{dirsym}\feq
To go on we claim that, for any pair of multi-indexes $\al,\be$,
\beq \l.\partial_{\bx}^\al \partial_{\by}^\be \Dir_s(\bx,\by)\r|_{\by = \bx} =
\l.\partial_{\bx}^\be \partial_{\by}^\al \Dir_s(\bx,\by)\r|_{\by = \bx} ~.
\label{derdirdia}\feq
Indeed, due to the first identity in Eq. \rref{dirsym}, we have
$\partial_{\bx}^\al \partial_{\by}^\be \Dir_s(\bx,\by) =
\partial_{\bx}^\al \partial_{\by}^\be \Dir_s(\by,\bx)$; when evaluating
the right-hand side of this equality on the diagonal $\by = \bx$, the
variables can be relabeled to yield
$\l.\partial_{\bx}^\al \partial_{\by}^\be \Dir_s(\by,\bx)\r|_{\by = \bx} =
\l.\partial_{\by}^\al \partial_{\bx}^\be \Dir_s(\bx,\by)\r|_{\bx = \by}$,
thus proving Eq. \rref{derdirdia}. \salto
All the above results can be applied to the (real) operator $\AA := -\Delta+V(\bx)$;
the symmetry properties outlined here for the corresponding
Dirichlet kernel will be relevant in connection with the results of the
next section.
\vspace{-0.4cm}
\subsection{The regularized propagator and stress-energy VEV: connections with the Dirichlet kernel.}
\label{subsecrel}
Let us refer to the framework of the previous section, where the
operator $A = - \Delta + V$ in $L^2(\Om)$ has been considered in
connection with a quantized scalar field. In the sequel
$x = (x^0, \bx), y = (y^0, \by) \in \reali \times \Om$;
if we use the expansion \rref{espans} for the regularized field
$\Fis$ in terms of creation and destruction operators we obtain
for the regularized propagator the expression
\begin{equation}\begin{split}
& \hspace{4.7cm} \la 0 | \Fis(x) \Fis(y) 0 \ra = \\
& = \mm^{\s}\!\int_{\!\KK \times \KK}{dk dh \over 2 (\om_k \om_h)^{\s+ 1 \over 2}}\,
\la 0|\!\l[\ak \fk(x)\!+\akd \fkc(x)\r]\!\l[\ah\fh(y)\!+ \ahd \fhc(y) \r]\! |0 \ra ~.
\end{split}\end{equation}
This relation, along with the identities $\la 0|\ak\ah|0\ra =
\la 0 |\akd\ahd| 0 \ra = \la 0|\akd\ah|0\ra = 0$ and
$\la 0|\ak\ahd|0\ra = \de(k,h)$, gives
\begin{equation}\begin{split}
& \la 0 | \Fis(x) \Fis(y) | 0 \ra =
\mm^{\s} \int_{\KK} {d k \over 2\,{\om_k}^{1 + \s}}\;\fk(x) \fkc(y) = \\
& \hspace{0.5cm}
= \mm^{\s} \int_{\KK} {d k \over 2\,{\om_k}^{1 + \s}}\;\Fk(\bx) \Fkc(\by)\;
e^{- i \om_k(x^0 -y^0)} ~.  \label{319}
\end{split}\end{equation}
From here we can easily obtain the derivatives of the propagator;
for example, for $j \in \{1,...d\}$, we have
\beq \partial_{x^0 y^j} \la 0 | \Fis(x) \Fis(y) | 0 \ra =
- i \,\mm^{\s}\! \int_{\KK} {d k \over 2 \, {\om_k}^{\s}}\;\Fk(\bx)
(\partial_{y^j}\Fkc)(\by)\;e^{- i \om_k(x^0 -y^0)} ~. \label{320} \feq
In particular, if we apply Eq.s \rref{319} \rref{320} with
$y=x$ and compare with the eigenfunction expansion \rref{eqkerdi} of
the Dirichlet kernel, we get
\beq \l. \la 0 | \Fis(x) \Fis(y) | 0 \ra \r|_{y=x} = {\mm^{\s} \over 2}
\l. D_{{\s + 1 \over 2}}(\bx, \by) \r|_{\by=\bx} ~; \label{regProp}\feq
\beq \l. \partial_{x^0 y^j}\la 0 | \Fis(x) \Fis(y) | 0 \ra \r|_{y=x} =
- i \mm^{\s} \l. \partial_{y^j} D_{{\s \over 2}}(\bx, \by) \r|_{\by=\bx} ~.\feq
One can express similarly all the derivatives with $y=x$ appearing in
Eq. \rref{tisprop} for the regularized stress-energy VEV.
In this way (and using as well the identity \rref{derdirdia}) we obtain
the following results, where $i,j,\ell$ are spatial indexes ranging in
$\{1,...,d\}$
({\footnote{To prove Eq. \rref{Tidiri0}, note that
$$ \la 0 | \Tis_{0 j}(t,\bx) | 0 \ra = - \la 0 | \Tis_{0 j}(t,\bx) | 0 \ra
= -{i \mm^\s \over 2}\l.\Big(\partial_{y^j}\Dir_{{\s \over 2}}(\bx,\by)
- \partial_{x^j}\Dir_{{\s \over 2}}(\bx,\by)\Big)\r|_{\by = \bx} $$
and that the last expression vanishes due to identity \rref{derdirdia}.
Besides, note that Eq. \rref{Tidirij} is equivalent to the more
symmetric expression
$$ \la 0 | \Tis_{i j}(t,\bx) | 0 \ra = \la 0 | \Tis_{j i}(t,\bx) | 0 \ra
= \mm^\s \!\l[\!\Big({1\over 4} - \xi\Big) \de_{i j}
\Big(\!\Dir_{{\s - 1 \over 2}}(\bx,\by) -
(\partial^{\,x^\ell}\!\partial_{y^\ell}\!+\!V(\bx))
\Dir_{{\s + 1 \over 2}}(\bx,\by) \Big) \, + \r. $$
$$ \l. + \l(\!\Big({1\over 4} - {\xi\over 2}\Big)
(\partial_{x^i y^j}\! + \partial_{x^j y^i})
- {\xi\over 2}\,(\partial_{x^i x^j}\! + \partial_{y^i y^j})\!\r)\!
\Dir_{{\s + 1 \over 2}}(\bx,\by) \r]_{\by = \bx} . $$}}):
\beq {~}\hspace{-0.5cm} \la 0 | \Tis_{0 0}(t,\bx) | 0 \ra \! = \!\mm^\s
\!\!\l[\!\l(\!\frac{1}{4}\!+\!\xi\!\r)\!\!\Dir_{{\s - 1\over 2}}(\bx,\by)
\!+\!\l(\!\frac{1}{4}\!-\!\xi\!\r)\!\!
(\partial^{x^\ell}\!\partial_{y^\ell}\!+\!V(\bx))
\Dir_{{\s + 1 \over 2}}(\bx,\by)\r]_{\by = \bx}\!, \label{Tidir00} \feq
\beq \la 0 | \Tis_{0 j}(t,\bx) | 0 \ra =
\la 0 | \Tis_{j 0}(t,\bx) | 0 \ra = 0 ~, \label{Tidiri0} \feq
\begin{equation}\begin{split}
& {~}\hspace{3.8cm} \la 0 | \Tis_{i j}(t,\bx) | 0 \ra =
\la 0 | \Tis_{j i}(t,\bx) | 0 \ra = \\
& = \mm^\s \!\l[\!\Big({1\over 4} - \xi\Big) \de_{i j}
\Big(\!\Dir_{{\s - 1 \over 2}}(\bx,\by) -
(\partial^{\,x^\ell}\!\partial_{y^\ell}\!+\!V(\bx))
\Dir_{{\s + 1 \over 2}}(\bx,\by) \Big) \, + \r. \\
& \hspace{5cm} \l. + \l(\!\Big({1\over 2} - \xi\Big)\partial_{x^i y^j}
- \xi\,\partial_{x^i x^j}\!\r)\!
\Dir_{{\s + 1 \over 2}}(\bx,\by) \r]_{\by = \bx} .
\label{Tidirij}
\end{split}\end{equation}
The above equations indicate, amongst else, that $\la 0|\Tis_{\mu\nu}(t,\bx)|0\ra$
does not depend on the time variable $t$; this comes as no surprise
at all, since our general framework is itself static (indeed, the
spatial domain $\Om$ and the potential $V$ are time independent).
These features of the regularized stress-energy VEV had been
anticipated in subsection \ref{statSubsec}; due to them, in
the rest of the paper we will use the notation
({\footnote{This is slightly abusive, since staticity occurs
only \textsl{after} taking the VEV; the alternative notation
$\la 0|\Tis_{\mu\nu}|0\ra (\bx)$ is more precise, but graphically
disturbing and will not be employed in the sequel.}})
\beq \la 0 | \Tis_{\mu \nu}(\bx) | 0 \ra \equiv
\la 0 | \Tis_{\mu \nu}(t,\bx) | 0 \ra ~. \label{notab} \feq
Up to now we have been working rather loosely, but it is not difficult
to indicate the precise conditions for the validity of our manipulations.
Indeed, Eq.s (\ref{Tidir00}-\ref{Tidirij}) for $\la 0|\Tis_{\mu\nu}(\bx)|0\ra$
involve the Dirichlet kernels $\Dir_{\s-1 \over 2}$ and $\Dir_{\s+1 \over 2}$,
with its second order derivatives, along the diagonal $\by=\bx$. So, for these
equations to be meaningful it is sufficient that $\Dir_{\s-1 \over 2} \in C^0(\Om \times \Om)$
and $\Dir_{\s+1 \over 2} \in C^2(\Om \times \Om)$; with the assumptions \rref{ii}
on $\AA = - \Delta + V$, recalling Eq. \rref{then} we see that both conditions
are fulfilled if
\beq \Re \s > d + 1 ~. \label{esse} \feq
For $\AA$, $\s$ as in Eq.s \rref{ii} \rref{esse}, the map $\bx \mapsto
\la 0|\Tis_{\mu\nu}(\bx)|0\ra$ is in $C^0(\Om)$ for all $\mu,\nu$. With the
stronger assumptions \rref{iii} on $\AA$, and again with $\s$ as in \rref{esse},
we infer from \rref{thenn} that $\bx \mapsto \la 0|\Tis_{\mu\nu}(\bx)|0\ra$ is
in $C^0(\ov{\Om})$ for all $\mu, \nu \in \{0,...,d\}$ (i.e., the regularized
stress-energy VEV is continuous up to the boundary)
({\footnote{Generalizing the previous considerations, one proves the following
for any $\ell \in \naturali$: with the assumptions \rref{ii} (resp., \rref{iii}),
if $\Re \s > d + \ell + 1$ the function $\bx \mapsto \la 0|\Tis_{\mu\nu}(\bx)|0\ra$
is in $C^\ell(\Om)$ (resp., in $C^\ell(\ov{\Om})$).}}).
\parn
Now, let us fix a point $\bx \in \Om$ and, making the assumptions \rref{ii},
consider the functions $\s \mapsto \la 0|\Tis_{\mu\nu}(\bx)|0\ra$; these are
not only well defined, but even analytic on the half plane $\{\s \in \complessi
~|~\Re \s > d + 1 \}$
({\footnote{This fact can be established proving analyticity with respect to $\s$
of $\Dir_{\s-1 \over 2}(\bx,\bx)$, of $\Dir_{\s+1 \over 2}(\bx,\bx)$ and of the second
order derivatives of $\Dir_{\s+1 \over 2}$ at $(\bx,\bx)$. We do not go into the details
of the proofs, that will be given elsewhere \cite{InPre} using rather obvious
analyticity results for the operator functions $\s \mapsto \AA^{-(\s \mp 1)/2}$. }}).\parn
Let us recall that, according to Eq.s \rref{ren} \rref{renest}, the analytic
continuation of $\la 0|\Tis_{\mu \nu}(\bx)|0\ra$ at $\s=0$ determines
the zeta renormalized VEV of the stress-energy tensor; of course, the
latter does not depend on $t$ as well and we will write
\beq \la 0 | \Ti_{\mu \nu}(\bx) | 0 \ra_{ren} \equiv
\la 0 | \Ti_{\mu \nu}(t,\bx) | 0 \ra_{ren} ~. \label{notabre} \feq
Of course, due to Eq.s (\ref{Tidir00}-\ref{Tidirij}), the renormalized
stress-energy VEV is determined by the ``renormalized'' functions
\beq \Dik_{\pm{1 \over 2}}(\bx,\by) := RP\Big|_{\s= 0}
\l(\mm^\s \Dir_{\s \pm 1 \over 2}(\bx,\by)\r) \,, \label{Dik} \feq
\beq \partial_{z w} \Dik_{{1 \over 2}}(\bx,\by) := RP\Big|_{\s= 0}
\l(\mm^\s \partial_{z w} \Dir_{\s + 1 \over 2}(\bx,\by)\r) \label{Dikzw} \feq
(with $z, w$ any two spatial variables), to be evaluated along the
diagonal $\by = \bx$\,. More precisely, we have
\beq \la 0 | \Ti_{0 0}(\bx)|0\ra_{ren} =
\!\l[\!\l(\!\frac{1}{4}\!+\!\xi\!\r)\!\Dik_{\!-{1\over 2}}(\bx,\by)
\!+\!\l(\!\frac{1}{4}\!-\!\xi\!\r)\!
(\partial^{x^\ell}\!\partial_{y^\ell}\!+\!V(\bx))
\Dik_{\!+{1 \over 2}}(\bx,\by)\r]_{\by = \bx} \!, \label{Tidir00R} \feq
\beq \la 0 | \Ti_{0 j}(\bx) | 0 \ra_{ren} =
\la 0 | \Ti_{j 0}(\bx) | 0 \ra_{ren} = 0 ~, \label{Tidiri0R} \feq
\begin{equation}\begin{split}
& \hspace{4.1cm} \la 0 | \Ti_{i j}(\bx) | 0 \ra_{ren} =
\la 0 | \Ti_{j i}(\bx) | 0 \ra_{ren} = \\
& = \!\l[\!\Big({1\over 4} - \xi\Big) \de_{i j}
\Big(\!\Dik_{\!-{1 \over 2}}(\bx,\by) -
(\partial^{\,x^\ell}\!\partial_{y^\ell}\!+\!V(\bx))
\Dik_{\!+{1 \over 2}}(\bx,\by) \Big)\, + \r. \\
& \hspace{5.6cm} \l. + \l(\!\Big({1\over 2} - \xi\Big)\partial_{x^i y^j}
- \xi\,\partial_{x^i x^j}\!\r)\!
\Dik_{\!+{1 \over 2}}(\bx,\by) \r]_{\by = \bx} . \label{TidirijR}
\end{split}\end{equation}
Let us remark that, if $\Dir_{\s \pm 1 \over 2}(\bx,\by)$ and
$\partial_{z w} \Dir_{{\s + 1 \over 2}}(\bx,\by)$ have analytic
continuations regular at $\s=0$, indicated hereafter with
$\Dir_{\pm {1 \over 2}}(\bx,\by)$ and
$\partial_{z w} \Dir_{{1 \over 2}}(\bx,\by)$, one has
\begin{equation}\begin{split}
& \hspace{0.34cm} \Dik_{\pm{1 \over 2}}(\bx,\by) =
\Dir_{\pm {1 \over 2}}(\bx,\by) ~, \\
& \partial_{z w} \Dik_{{1 \over 2}}(\bx,\by) =
\partial_{z w} \Dir_{{1 \over 2}}(\bx,\by) \label{DikAC}
\end{split}\end{equation}
for any choice of the mass scale $\mm$; clearly, in this case the
renormalized stress-energy VEV is independent of $\mm$. On the contrary,
an explicit dependence on $\mm$ appears if the analytic continuations
of $\Dir_{\s \pm 1 \over 2}(\bx,\bx)$ or
$\partial_{z w} \Dir_{{\s + 1 \over 2}}(\bx,\bx)$ (or both) have a
singularity at $\s = 0$; this will occur in some specific examples,
to be considered in the subsequent Parts II and III. \salto
A connection between a regularized stress-energy VEV and a Dirichlet-like
kernel is mentioned by Cognola, Vanzo and Zerbini \cite{CogVanZer0} in a
slightly different framework, where the field theory becomes Euclidean
after Wick rotation of the time coordinate, and our operator $\AA$ (in
the spatial variables $\bx$) is replaced by the (spacetime) differential
operator $ -\partial_{tt}-\Delta+V$. \parn
A variant of the previous results about the Dirichlet kernel and the
regularized VEV $\la 0|\Tis_{\mu\nu}(\bx)|0 \ra$ can be formulated in
the case of a \textsl{slab}. In this case $\Om = \Om_1 \times \reali^{d_2}$,
with $\Om_1$ a domain in $\reali^{d_1}$ and $d_1\!+\!d_2 = d$\,;
moreover, the potential $V$ depends only on the coordinates $\bx_1\in\Om_1$.
In this situation we can express $\la0|\Tis_{\mu\nu}(\bx)|0\ra$ in
terms of the Dirichlet kernel associated to the operator $\AA_1 :=
-\Delta_1 + V(\bx_1)$ (we defer to subsection \ref{slabSubsec} a
comprehensive discussion of slabs configurations). \parn
In the following subsections we return to the case where $\Om$ is an
arbitrary domain in $\reali^d$ and we connect the Dirichlet kernel $\Dir_s$
to other integral kernels, in order to shed light on the analytic
continuation of $\Dir_s$ (of course, these connections will also be
useful, in their $d_1$-dimensional formulation, in the case of a slab).
\vspace{-0.4cm}
\subsection{The heat kernel, the cylinder kernel and some variations.}\label{subsecKT}
Let us consider again a strictly positive selfadjoint operator $\AA$ in
$L^2(\Om)$ (that will be $-\Delta+V$ in the subsequent applications).
For all $\t\in [0,+\infty)$, using the standard functional calculus
we can define the operators
\beq e^{-\t \AA}\,, \qquad e^{- \t\sqrt{\AA}} ~. \feq
These fulfill the following conditions:
\beq \l({d \over d \t} + \AA \r)e^{-\t\AA} = 0 ~,
\qquad \l. e^{- \t \AA} \r|_{\t=0} = \mathbf{1}~,
\label{conh} \feq
\beq \l({d^2 \over d \t^2} - \AA \r)e^{- \t \sqrt{\AA}} = 0 ~, \qquad \l.
e^{- \t \sqrt{\AA}} \r|_{\t=0} = \mathbf{1} ~; \quad
\label{cont} \feq
moreover, due to the strict positivity of $\AA$\,, $e^{-\t\AA}$ and
$e^{-\t\sqrt{\AA}}$ are expected to vanish for $\t\to+\infty$, in some
sense that can be made more precise in terms of integral kernels. Let
us now pass to the kernels
\beq K(\t\,; \bx, \by) := e^{- \t \AA}(\bx, \by)~, \qquad
T(\t\,; \bx, \by) = e^{- \t \sqrt{\AA}}(\bx, \by)~, \label{deft} \feq
which can be expressed as follows in terms of a complete orthonormal
set $(\Fk)_{k \in \KK}$ of eigenfunctions of $\AA$ and of the corresponding
eigenvalues $(\om^2_k)_{k \in \KK}$:
\beq K(\t\,; \bx, \by) = \int_\KK dk \; e^{- \t\,\om_k^2} \,
\Fk(\bx) \Fkc(\by) ~, \label{eqheat} \feq
\beq T(\t\,; \bx, \by) = \int_\KK dk \; e^{- \t\,\om_k} \,
\Fk(\bx) \Fkc(\by) ~. \label{eqcyl} \feq
We note that
\beq \l(\partial_\t + \AA_\bx \r) K(\t\,; \bx, \by) = 0 ~,
\qquad K(0; \bx, \by) = \de(\bx - \by) ~; \label{eqk} \feq
\beq \l(\partial_{\t \t} - \AA_\bx \r) T(\t\,; \bx, \by) = 0 ~,
\qquad T(0; \bx, \by) = \de(\bx - \by) ~. \label{eqt} \feq
In the above $\AA_\bx$ indicates the operator $\AA$ acting on
$K(\t\,;\bx,\by)$ and $T(\t\,;\bx,\by)$ as functions of the $\bx$
variable; Eq.s \rref{eqk}\rref{eqt} follow, respectively, from
Eq.s \rref{conh} \rref{cont}. Besides, under minimal supplementary
conditions one can prove that $K(\t\,;\bx,\by)$ and $T(\t\,;\bx,\by)$
vanish exponentially for fixed $\bx,\by \in \Om$ and $\t\to+\infty$;
we shall return on this in subsection \ref{slKT}. \parn
In case \rref{ii} (i.e., $\AA = -\Delta+V$ with $V$ smooth) we have $\partial_\t + \AA_\bx = \partial_\t
- \Delta_\bx + V(\bx)$ and $\partial_{\t\t} - \AA_\bx = \partial_{\t\t}
+ \Delta_\bx - V(\bx)$, so Eq. \rref{eqk} contains a heat equation and
Eq. \rref{eqt} a $(d+1)$-dimensional Laplace equation (with an external
potential); note as well that $K(\t\,;\bx,\by)$ and $T(\t\,;\bx,\by)$
fulfill the boundary conditions in the definition of $\AA$ for $\bx$
or $\by$ in $\partial \Om$. For obvious reasons, $K$ is called the
\textsl{heat kernel} of $\AA$ (even in cases where $\AA$ is not of the
form $-\Delta\!+\!V$); $T$ is called by Fulling \cite{FullAs3} the
\textsl{cylinder kernel} of $\AA$. It should be mentioned that,
under the assumptions \rref{ii}, there are rigorous proofs that
\beq (\t, \bx, \by) \mapsto K(\t,\bx,\by),
(\t, \bx, \by) \mapsto T(\t,\bx,\by)~\mbox{are in $C^\infty((0,+\infty) \times
\Omega \times \Omega)$}; \label{similar} \feq
with the stronger assumptions \rref{iii} there is a result
similar to \rref{similar}, with $\Om$ replaced by $\ov{\Om}$.
(These results can be proved by a slight generalization of
Thm. 5.2.1 in \cite{Dav}. For some related statements,
see also Appendix \ref{AppKer} and \cite{InPre}.) \parn
Again with the assumptions \rref{iii}, $\AA$ has pure point spectrum and
possesses a complete orthonormal set of proper eigenfunctions $(\Fk)_{k = 1,2,3,...}$,
for which the expansions \rref{eqheat} \rref{eqcyl} hold with
$\int_{\KK} d k = \sum_{k=1}^{+\infty}$. One can prove that, for
each $\t >0$, these expansions converge absolutely and uniformly on $\ov{\Om}
\times \ov{\Om}$ with their derivatives of any order (see again Appendix \ref{AppKer},
especially Eq. \rref{absconvheat}).
\salto
With the more general assumptions \rref{ii} we have $\AA^\dagger = \ov{\AA} = \AA$,
which in turn implies similar relations for $e^{-\t \AA}$,
$e^{-\t \sqrt{\AA}}$; due to Eq. \rref{imp}, this gives
\beq K(\t\,; \by, \bx) = K(\t\,; \bx, \by)~, \qquad
T(\t\,; \by, \bx) = T(\t\,; \bx, \by) ~. \feq
Needless to say, the heat kernel has been the object of intensive and
detailed studies, even in a much more general framework than the one
considered in the present paper; exhaustive analyses have been given,
for example, by Berline et al. \cite{Berl}, Calin et al. \cite{Calin},
Chavel \cite{Chav}, Davies \cite{Dav}, Gilkey \cite{Gil} and Grigor'yan
\cite{Gryg}\label{Monog}. On the contrary, the cylinder kernel is a less
popular object; it has mainly been investigated by Fulling and co-authors
\cite{FulGus,FullAs3,FullAs}. Some considerations of Fulling (see, e.g.,
\cite{FulWed}) also involve the operator $\sqrt{\AA}^{\;-1} e^{-\t\sqrt{\AA}}$
and the associated kernel
\beq \Tm(\t\,; \bx, \by) := (\sqrt{\AA}^{\;-1}
e^{- \t \sqrt{\AA}})(\bx, \by) = \int_\KK {dk \over \om_k} \;
e^{- \t\,\om_k} \, \Fk(\bx) \Fkc(\by) ~,  \label{SKer} \feq
which we will refer to as the \textsl{modified cylinder kernel}, for
reasons which become apparent hereafter (see Eq. \rref{STKer}).
Let us observe that the trivial relation $e^{-\t \sqrt{\!\AA}}\!
=\!- {d \over d\t}(\sqrt{\!\AA}^{\;-1}\! e^{- \t \sqrt{\!\AA}})$
can be reformulated in terms of integral kernels as
\beq T(\t\,;\bx,\by) = - \partial_\t \Tm (\t\,;\bx,\by) ~; \label{STKer} \feq
conversely, $\Tm$ can be determined as the primitive of $-T$ which
vanishes for $\t \to +\infty$, that is
\beq \Tm(\t\,;\bx,\by) = \int_{\t}^{+\infty}\! d\t'\;T (\t'\,;\bx,\by) ~.
\label{STKerPrim} \feq
In some cases $\Tm$ is easier to compute than $T$, and some identities
relating the cylinder kernel $T$ to the Dirichlet kernel $\Dir_s$
can be applied more efficiently if they are rephrased in terms of
$\Tm$ (this situation will be exemplified in the case of a wedge
domain, to be discussed in the subsequent Part II; see Section
5 therein). \parn
With the assumptions \rref{ii} or \rref{iii}, one could give for
$\Tm$ some regularity results very similar to the ones
illustrated previously for $T$ (e.g.: statement \rref{similar}
holds as well for $\Tm$ under the conditions \rref{ii}).
\salto
Before moving on, let us consider the \textsl{heat} and \textsl{cylinder
traces}; these are respectively defined, for $\t \in (0,+\infty)$, as
\beq K(\t) := \Tr e^{-\t\AA} ~, \qquad T(\t) := \Tr e^{-\t\sqrt{\AA}} ~.
\label{KTTr} \feq
Assuming the above traces to exist, the general identity \rref{trb} for
the trace of an operator $\BB$ (here applied with either $\BB = e^{-\t\AA}$
or $\BB = e^{-\t\sqrt{\AA}}$) yields respectively
\beq K(\t) = \int_\Om d\bx \; K(\t\,;\bx,\bx) ~, \qquad
T(\t) = \int_\Om d\bx \; T(\t\,;\bx,\bx) ~. \label{KTInt} \feq
In particular, in the case \rref{iii} where the eigenvalues of $\AA$
are labelled by $\KK = \{1,2,3,...\}$ and the Weyl estimate \rref{weyl}
holds, we have:
\beq K(\t) = \sum_{k=1}^{+\infty} e^{-\t\,\om_k^2} < + \infty~, \quad
T(\t) = \sum_{k=1}^{+\infty} e^{-\t\,\om_k} < + \infty \qquad
\mbox{for all $\t > 0$} ~. \label{case2KT} \feq
Clearly enough, an analogous discussion could be made for the space integral
of the diagonal, modified cylinder kernel $\Tm(\t\,;\bx,\bx)$; we
omit this discussion for brevity.
\vspace{-0.4cm}
\subsection{The case of a non-negative $\boma{\AA}$. The heat and cylinder kernels.}\label{nonNeg}
Let us remark that the heat and cylinder kernels can both be defined
even in case $\AA$ is \textsl{non-negative}, without requiring strict
positivity; by this we mean that $\si(\AA) \subset [0,+\infty)$, and
that we are not assuming $\si(\AA) \subset [\eps^2, + \infty)$ for
any $\eps >0$\,. The non-negativity of $\AA$ is equivalent to the
existence of a complete orthonormal system $(\Fk)_{k \in \KK}$ of
(either proper or improper) eigenfunctions with corresponding
non-negative eigenvalues $(\om^2_k)_{k \in \KK}$ ($\om_k \geqs 0$).
Most of the considerations of the previous subsection still hold,
in particular Eq.s \rref{eqheat} \rref{eqcyl}. \parn
For example, if $\AA = -\Delta$ and $\Om = \reali^d$, then
$\AA$ is non-negative with eigenfunctions
$F_{\bk}(\bx) = (2 \pi)^{-d/2} e^{i \bk \cdot \bx}$ and eigenvalues
$\om^2_{\bk} = |\bk|^2$, labelled by $\bk \in \reali^d$;
the measure $d \bk$ on the set of labels is the usual Lebesgue
measure of $\reali^d$. The eigenfunction expansion
\rref{eqheat} of the heat kernel yields in this case the familiar result
\beq K(\t\,;\bx,\by) = {1 \over (4\pi\t)^{d/2}}\;
e^{-{|\bx-\by|^2\! \over 4\t}} ~; \label{eq1} \feq
moreover, the expansion \rref{eqcyl} of the cylinder kernels gives
the result
\beq T(\t\,;\bx,\by) = {\Ga({d+1 \over 2})\,\t \over
\pi^{d+1 \over 2}(\t^2 + |\bx-\by|^2)^{{d+1 \over 2}}} ~, \label{eq2} \feq
which is a bit less popular and appears, e.g., in \cite{FulMass}. \parn
In some subcases with non-negative
spectrum we can speak as well of the modified cylinder kernel
$\Tm(\t\,;\bx,\by)$. In fact, if $0$ has zero spectral measure
(a fact holding when $0$ belongs to the continuous spectrum,
but not holding when $0$ is a proper eigenvalue),
$\sqrt{\AA\,}^{\,-1}e^{-\t \sqrt{\AA}}$ can be defined through
the standard functional calculus for selfadjoint operators.
With minimal supplementary assumptions of regularity,
the kernel $\Tm(\t\,;\bx,\by) := \la\de_\bx |\sqrt{\AA}^{\;-1}e^{-\t\sqrt{\AA}}\de_\by\raL$
makes sense, as well as its expansion
\rref{SKer} in terms of a complete orthonormal set of (generalized)
eigenfunctions $\Fk$ with eigenvalues $\om^2_k$ ($k\!\in\!\KK$);
note that the assumption of zero spectral
measure for $0$ is equivalent to the requirement that
$\om_k = 0$ only on a zero-measure
subset of $\KK$. In these situations we have again Eq. \rref{STKerPrim},
describing the cylinder kernel $T$ as the primitive of $-\Tm$\,. \parn
For example, let us return to the case where $\AA = -\Delta$ and
$\Om = \reali^d$, in which the spectrum $\si(\AA) = [0,+\infty)$ is
purely continuous. We have mentioned previously the eigenfunctions
$F_{\bk}$ and the eigenvalues $\om^2_{\bk}$ (labelled by $\bk\!\in\!\reali^d$),
where $F_{\bk}(\bx) = (2\pi)^{-d/2} e^{i \bk\cdot\bx}$ and $\om_k = |\bk|$;
of course $\om_k = 0$ only on a set of zero Lebesgue
measure (consisting of the unique point $\bk = 0$). In this case,
the expansion \rref{SKer} (involving an integral in the Lebesgue measure
$d \bk$) gives the following for $d \geqs 2$:
({\footnote{For $d = 1$ the right hand side of Eq.\rref{SKer} for $\Tm$ does not
converge (not even distributionally); we take this as an indication
that the modified cylinder kernel is ill-defined.}})
\beq \Tm(\t\,;\bx,\by) = {\Ga({d-1 \over 2}) \over
2\pi^{d+1 \over 2}(\t^2 + |\bx-\by|^2)^{{d-1 \over 2}}} ~. \label{eq3}\feq
\vspace{-0.8cm}
\subsection{Connections between the cylinder kernel and a $\boma{(d\!+\!1)}$-dimensional Green function.}
\label{ment}
Due to the limited popularity of $T$ it can be useful to connect
this kernel to a more familiar
object, namely a Green function, even though this requires to pass
to $d+1$ dimensions. All details of this construction are given in
Appendix \ref{appeg} where, as an example, this approach is used
for a novel derivation of Eq. \rref{eq2} not relying on the eigenfunction
expansion \rref{eqcyl}.
\subsection{Behaviour of the heat and cylinder kernels for small and large $\boma{\t}$.}\label{slKT}
{~}
\parn
\textsl{The small $\t$ limit.}
The asymptotic expansion of the heat
kernel of an operator $\AA = - \Delta + V$
on an open set $\Om \subset \reali^d$
has been extensively studied (see, e.g, the work of Minakshisundaram and
Pleijel \cite{Minak2}, or the already cited monographies
\cite{Berl,Calin,Chav,Dav,Gil,Gryg} on the heat kernel).
From here to the end of this paragraph, we discuss
the $\t \vain 0^{+}$ behavior of the heat and
cylinder kernels of $\AA = - \Delta + V$
under the assumptions \rref{iii}
(these could be generalized to get the same results, but we are not going
to discuss this subject). As known from the previosly cited references,
there is a unique sequence of real functions
$a_n:\Om\times\Om\to\reali$ ($n=1,2,3....$), usually referred to as HMDS
(Hadamard-Minakshisundaram-DeWitt-Seeley) coefficients, such that for any
$N \in \{1,2,3,...\}$ one has
\beq K(\t\,;\bx,\by) = {1 \over (4\pi\t)^{d/2}}\;e^{-{|\bx-\by|^2\! \over 4\t}}\!
\l[1 + \sum_{n = 1}^{N}\,a_n(\bx,\by)\, \t^n + O(\t^{N+1}) \r]
\quad\! \mbox{for $\t \to 0^+$}. \label{asinK} \feq
In the above equation notice the factor
$K_0(\t\,;\bx,\by) := {1 \over (4\pi\t)^{d/2}}\;e^{-{|\bx-\by|^2\!\over 4\t}}$,
which is just the heat kernel associated to $-\Delta$ on $\reali^d$.
In case $V = 0$, we have $a_n = 0$ for all $n$; thus, $K(\t\,;\bx,\by)
= K_0(\t\,;\bx,\by)[1 + O(t^\infty)]$ (where the last term indicates a
remainder which is $O(\t^N)$ for each $N \in \{1,2,3,...\}$). \parn
Along the diagonal $\by = \bx$ (for any $V$) Eq. \rref{asinK} reduces to
\beq K(\t\,;\bx,\bx) = {1 \over (4\pi\t)^{d/2}} \l[1 + \sum_{n = 1}^{N}\,
a_n(\bx)\,\t^n + O(\t^{N+1}) \r] \quad \mbox{for $\t \to 0^+$} \label{asinKD}\feq
where $a_n(\bx)$ is shorthand for $a_n(\bx,\bx)$. The small $\t$
analysis of the cylinder kernel is more involved; however, Fulling
proved (see, e.g., \cite{FulGus}) that its asymptotic behaviour along
the diagonal $\by = \bx$ is as follows: there exist functions
$e_n,f_n : \Om \to \reali$ ($n=0,1,2,...$) such that, for any
$N\in\{0,1,2,...\}$,
\beq T(\t\,;\bx,\bx) = {1 \over \t^d}\l[\sum_{n = 0}^{N} e_n(\bx)\,\t^n +
\hspace{-0.2cm} \sum_{\substack{n = d+1 \\ n - d \mbox{ \scriptsize{odd}}}}^{N}
\hspace{-0.2cm} f_n(\bx)\,\t^n \ln \t + O(\t^{N+1}\!\ln\t)\r]\;\,
\mbox{for $\t \to 0^+$}.\! \label{asinTD} \feq
As pointed out in \cite{FulGus}, some of the functions $e_n, f_n$
(but not all of them) can be expressed in terms of the diagonal
HMDS coefficients $\bx \mapsto a_n(\bx)$ mentioned before. \salto
Before proceeding, let us remark that the heat and cylinder traces
$K(\t),T(\t)$ (see Eq.s \rref{KTTr} \rref{KTInt}) are well-known to
admit small $\t$ expansions analogous to those in Eq.s \rref{asinKD}
\rref{asinTD}; see once more the references cited above. In particular,
assuming again $\Om$ to be compact and $V$ to be smooth and bounded below,
for $\t \to 0^+$ there hold
\beq K(\t) = {1 \over (4\pi\t)^{d/2}} \l[Vol(\Om) + \sum_{n = 1}^{N}
A_n\,\t^{n/2} + O(\t^{N+1 \over 2}) \r] ~, \label{asinKTr} \feq
\beq T(\t) = {1 \over \t^d}\l[\sum_{n = 0}^{N} E_n\,\t^n +
\hspace{-0.2cm} \sum_{\substack{n = d+1 \\ n - d
\mbox{ \scriptsize{odd}}}}^{N} \hspace{-0.2cm}
F_n\,\t^n \ln \t + O(\t^{N+1}\!\ln\t)\r] \label{asinTTr} \feq
($Vol(\Om)$ denotes the volume of the spatial domain $\Om$). Notice,
in particular, that the expansion \rref{asinKTr} for $K(\t)$ involves
half-integer powers of $\t$, whereas in expansions \rref{asinK} \rref{asinKD}
for the local heat kernel $K(\t\,;\bx,\by)$ only integer powers of $\t$
appear; besides, let us stress that the real coefficients $A_n,E_n,F_n$
in Eq.s \rref{asinKTr} \rref{asinTTr} are not just the integrals over
the spatial domain $\Om$ of the functions $a_n(\bx),e_n(\bx),f_n(\bx)$
of Eq.s \rref{asinKD} \rref{asinTD}, since boundary contributions arise
as well. \salto
\textsl{The large $\t$ behavior.}
Let us move on and note that, as anticipated in subsection \ref{subsecKT},
both the heat and the cylinder kernel (along with their traces) vanish
exponentially for large $\t$, under minimal regularity conditions.
As an example, we will prove the exponential decay of the cylinder kernel
with the following assumptions: \parn
a) $\AA$ is a strictly positive, selfadjoint operator in $L^2(\Om)$,
possessing a generalized complete orthonormal set $(\Fk)_{k \in \KK}$
of eigenfunctions with eigenvalues $\om_k^2$ ($\om_k\!>\!\eps\!>\!0$);
the $\Fk$'s are continuous functions on $\Om$. \parn
b) for all points $\bx,\by \in \Om$ and $\t > 0$, one has
\beq \hat{T}(\t\,;\bx,\by) := \int_\KK dk \;
e^{- \om_k \t}\, |\Fk(\bx)|\,|\Fk(\by)| < + \infty ~. \label{hatt} \feq
In this case, starting from the eigenfunction expansion \rref{eqkerdi} we find
\beq |T(\t\,;\bx,\by)| \leqs \hat{T}(\t\,;\bx,\by) \label{368} \feq
at all points $\bx,\by \in \Om$. On the other hand, after fixing $\tau > 0$
we see that
\beq \hat{T}(\t\,;\bx,\by) \leqs e^{-\eps (\t - \tau)}\,\hat{T}(\tau\,\;\bx,\by)
\quad \mbox{for all $\t \geqs \tau$} \label{369} \feq
(this follows from the definition \rref{hatt} of $\hat{T}$, noting that
the inequalities $\om_k \geqs \eps >0$ and $\t - \tau \geqs 0$ imply
$e^{-\om_k \t} = e^{-\om_k (\t - \tau)} e^{-\om_k \tau} \leqs
e^{-\eps (\t - \tau)} e^{-\om_k \tau}$). The relations \rref{368} and
\rref{369} give the desired result of exponential decay for $T$, since
at all points $\bx, \by \in \Om$ they imply
\beq |T(\t\,;\bx,\by)| \leqs e^{-\eps(\t-\tau)} \,
\hat{T}(\tau\,;\bx,\by)\qquad \mbox{for all $\t \geqs \tau$} ~. \label{bound} \feq
In the above argument, condition \rref{hatt} plays a crucial role. This
condition is fulfilled, for example, if $\AA = - \Delta + V$ and the assumptions
\rref{iii} are satisfied. In this case we have a uniform bound $\hat{T}(\t\,;\bx,\by)
\leqs \check{T}(\t) < + \infty$ for all $\bx, \by \in \ov{\Om}$ and $\t >0$
(see Appendix \ref{AppKer}, Eq. \rref{ub}), so that Eq. \rref{bound} implies
\beq |T(\t\,;\bx,\by)| \leqs e^{-\eps(\t-\tau)} \,
\check{T}(\tau)\qquad \mbox{for all $\bx, \by \in \ov{\Om}$ and $\t \geqs \tau$}~.
\label{boundiii} \feq
The exponential decay of the heat kernel and of the traces of both the
heat and cylinder kernels can be derived by similar considerations.
For alternative approaches, see \cite{Dav} or \cite{Gryg}.
\subsection{The Dirichlet kernel as the Mellin transform of the heat or cylinder kernel.}\label{MelDir}
The results reported in this subsection are well-known; they were
derived by Dowker and Critchley \cite{DowKri}, Hawking \cite{Hawk}
and Wald \cite{Wal} and later reconsidered by Moretti et al. (see
\cite{MorBo,MorRec} and citations therein). \parn
The representations of $\Dir_s$ mentioned in the title are useful
in view of the analytic continuation with respect to $s$; they
can be derived starting from the well-known relation (see \cite{NIST},
p.139, Eq.5.9.1)
\beq {1 \over z^s} = {1 \over \Ga(s)} \int_0^{+\infty} \!d\t \;
\t^{s-1}\,e^{-z\t} \quad \mbox{for all $z \in (0,+\infty)$,
$s \in \complessi$ with $\Re s > 0$} ~. \label{uzs} \feq
This identity, along with the eigenfunction expansions
\rref{eqkerdi}, \rref{eqheat} and \rref{eqcyl} of the Dirichlet,
heat and cylinder kernels, yields
\beq \Dir_s(\bx,\by) = {1 \over \Ga(s)} \int_0^{+\infty}
\!d\t \; \t^{s-1}\,K(\t\,;\bx,\by) ~, \label{DirHeat} \feq
\beq \Dir_s(\bx,\by) = {1 \over \Ga(2s)}\int_0^{+\infty}
\!d\t \; \t^{2s-1}\,T(\t\,;\bx,\by)~.  \label{DirCyl} \feq
For example, Eq. \rref{DirHeat} is derived via the following
chain of equalities:
\begin{equation}\begin{split}
& \Dir_s(\bx,\by) = \!\int_\KK {dk \over \om_k^{2s}}\,
\Fk(\bx) \Fkc(\by) = \!\int_\KK \!dk \, {1 \over \Ga(s)}\!
\int_0^{+\infty}\!\! d\t \; \t^{s-1}\,e^{-\om_k^2\t}
\Fk(\bx) \Fkc(\by) = \\
& \quad = {1 \over \Ga(s)}\!\int_0^{+\infty}\!\! d\t\;
\t^{s-1}\!\int_\KK \!dk\,e^{-\om_k^2\t} \Fk(\bx)\Fkc(\by)
= {1 \over \Ga(s)}\!\int_0^{+\infty}\!\! d\t\;\t^{s-1}\,
K(\t\,;\bx,\by) ~.
\end{split}\end{equation}
In the second passage above, we used Eq. \rref{uzs} with $z = \om_k^2$\,;
in the third passage, the exchange in the order of integration is
justifed with arguments similar to those in \cite{Minak1,Minak2}.
The derivation of Eq. \rref{DirCyl} is similar; in this case one has
to resort to Eq. \rref{uzs} with $z = \om_k$ and $s$ replaced by $2s$. \parn
Eq.s \rref{DirHeat} \rref{DirCyl} state that the Dirichlet kernel
$\Dir_s$ can be represented as the \textsl{Mellin transform} of
either the heat or the cylinder kernel; there are analogous relations
for the derivatives of the Dirichlet kernel, involving the corresponding
derivatives of the heat and cylinder kernels. \parn
In the sequel we are especially interested in the case
in which the integral representations \rref{DirHeat} and
\rref{DirCyl} hold \textsl{pointwisely}, at a given pair
of points $\bx,\by \in \Om$ (including the case $\by = \bx$).
This occurs under the following three conditions: \parn
a) $\Dir_s(\bx,\by)$ is well defined at the given points $\bx, \by$; \parn
b) $K(\t\,;\bx,\by)$ and $T(\t\,;\bx,\by)$ are also well-defined
at these points, for all $\t > 0$. \parn
c) the integrals in Eq.s \rref{DirHeat} \rref{DirCyl} converge. \parn
With the assumptions \rref{iii}, a) b) c) hold at all points
$\bx,\by \in \Om$ if $\Re s > d/2$. In fact: for $\Re s > d/2$,
$\Dir_s$ is continuous on $\ov{\Om} \times \ov{\Om}$
(see Eq. \rref{thenn}), so there is no problem with its pointwise
evaluation; for each $\t >0$, the functions $K(\t\,;~,~)$ and $T(\t\,;~,~)$
are continuous (in fact $C^\infty$) on $\ov{\Om} \times \ov{\Om}$, and
again their pointwise evaluation is not a problem; due to the exponential
decay of $K(\t\,;\bx,\by)$ and $T(\t\,;\bx,\by)$, the integrals \rref{DirHeat}
\rref{DirCyl} have no convergence problems for large $\t$; for any
$\bx,\by \in \Om$, due to the $\t \to 0^{+}$ asymptotic expansions
\rref{asinK} \rref{asinTD}, the same integrals are both convergent for
$\t$ close to zero if $\Re s > d/2$
({\footnote{For example, from \rref{asinK} it is clear that,
in the limit $\t \vain 0^{+}$, one has
$K(\t,\bx,\by) = O(1)$ for $\by \neq \bx$ and
$K(\t,\bx,\bx) = O(1/t^{d/2})$; so the integral
in \rref{DirHeat} converges for all $\bx, \by$ if $\Re s > d/2$.}}). \parn
The pointwise representations \rref{DirHeat} \rref{DirCyl} can be used
as a starting point to build the analytic continuation of the function
$s \mapsto \Dir_s(\bx,\by)$ to a larger domain than the one where they
are initially granted to hold (e.g., larger than the half plane $\{s \in \complessi
~|~ \Re s > d/2 \}$ of case \rref{iii}); we return to this point in the
next two subsections. \parn
One could make similar considerations for the derivatives
of $\Dir_s$. For example, Eq. \rref{DirHeat} gives
\beq \partial^\alpha_{\bx} \partial^{\beta}_{\by} \Dir_s(\bx,\by) = {1 \over \Ga(s)} \int_0^{+\infty}
\!d\t \; \t^{s-1}\, \partial^\alpha_{\bx} \partial^{\beta}_{\by} K(\t\,;\bx,\by) ~; \label{DirHeatder} \feq
for any pair of multi-indices $\alpha,\beta$. If we make the assumptions \rref{iii} on
$\AA = -\Delta + V$ and admit that the $\t \vain 0^{+}$ asymptotic expansion
\rref{asinK} can be derived term by term,
Eq. \rref{DirHeatder} holds pointwisely at all $\bx,\by \in \Om$ if $\Re s > d/2 + j/2$,
where $j := |\alpha | + |\beta|$
({\footnote{In fact: for $\Re s > d/2 + j/2$,
$\Dir_s$ is $C^j$ on $\ov{\Om} \times \ov{\Om}$
(see Eq. \rref{thenn}), so there is no problem with pointwise
evaluation of its derivatives up to order $j$; for each $\t >0$, the function $K(\t\,;~,~)$
is $C^\infty$ on $\ov{\Om} \times \ov{\Om}$, and
again the pointwise evaluation of its derivatives is not a problem; the derivatives
of $\partial^{\alpha}_{\bx} \partial^{\beta}_{\by} K(\t\,;\bx,\by)$
are easily seen to decay exponentially, so the integral \rref{DirHeatder}
has no convergence problem for large $\t$; deriving
term by term the asymptotic expansion \rref{asinK} we see
that, for $\t \vain 0^{+}$, $\partial^\alpha_{\bx} \partial^{\beta}_{\by} K(\t\,;\bx,\by)
= O(1)$ for $\by \neq \bx$ and
$\partial^\alpha_{\bx} \partial^{\beta}_{\by} K(\t\,;\bx,\bx) = O(1/\t^{{d/2} + [j/2]})$
which implies that, for $\Re s > d/2 + j/2$, the integral
\rref{DirHeatder} is in any case convergent.}}).
\salto
To conclude this subsection let us notice that, setting $\by = \bx$ in Eq.s \rref{DirHeat}
\rref{DirCyl} and integrating over the spatial domain $\Om$
({\footnote{Assuming that the order of integration can be interchanged
for $s$ is a suitable complex domain.}}),
the relations \rref{TrAs} for the trace $\Tr \AA^{-s}$ and \rref{KTTr}
\rref{KTInt} for the heat and cylinder traces $K(\t),T(\t)$ allow us to infer
\beq \Tr \AA^{-s} = {1 \over \Ga(s)} \int_0^{+\infty}
\!d\t \; \t^{s-1}\,K(\t) ~; \label{DirHeatTr} \feq
\beq \Tr \AA^{-s} = {1 \over \Ga(2s)}\int_0^{+\infty}
\!d\t \; \t^{2s-1}\,T(\t) ~. \label{DirCylTr} \feq
Eq.s \rref{DirHeatTr} \rref{DirCylTr} can be used to
continue analytically the function $s \mapsto \Tr(\AA^{-s})$; the situation
is similar to the one outlined previously for the local counterparts of
these equations, and will be reconsidered in the next two subsections.
\subsection{Analytic continuation of Mellin transforms via integration by parts.}
\label{contiparts}
In the first part of this subsection, the analytic continuation via
integration by parts will be presented for an arbitrary Mellin transform;
in the second part, we will connect this general construction to the
representation of the Dirichlet kernel (resp., of $\Tr\AA^{-s}$) as the
Mellin transform of either the heat or the cylinder kernel (resp., of
their traces). \parn
Let $\ff:(0,+\infty) \to \complessi$ be a function of the form
\beq \ff(\t) = {1 \over \t^\rho}\,\hh(\t) \label{ff} \feq
for some $\rho \in \complessi$ and some smooth function $\hh:[0,+\infty) \to
\complessi$, vanishing exponentially for $\t \to +\infty$; consider the
Mellin transform of $\ff$, i.e., the function
\beq \Mel(\si) := \int_0^{+\infty}\!\!d\t\;\t^{\si-1}\,\ff(\t) ~, \label{Mel}\feq
defined for appropriate $\si \in \complessi$. Due to the hypotheses on
$\ff$, the integral in Eq. \rref{Mel} converges only for $\si \in \complessi$
with $\Re \si > \Re \rho$\, and gives an analytic function of $\si$ in
this region. However, integrating by parts $n$ times (for any
$n \in \{1,2,3,...\}$) and noting that the boundary terms vanish (for
$\Re \si > \Re \rho$), we obtain
\beq \Mel(\si) = {(-1)^n \over (\si\!-\!\rho) ... (\si\!-\!\rho\!+\!n\!-\!1)}
\int_0^{+\infty} \!\!d\t \; \t^{\si-\rho+n-1}\,{d^n \hh \over d \t^n}(\t) ~.
\label{MelCon} \feq
In consequence of the features of the function $\hh$, the above integral
converges for $\Re \si > \Re \rho - n$; thus, Eq. \rref{MelCon} yields the
analytic continuation of the Mellin transform $\Mel(\si)$ to the region
\beq \{\si \in \complessi ~|~ \Re \si > \Re \rho - n\} \feq
from which the zeros of the denoninator in \rref{MelCon} must be removed;
this gives a meromorphic function with (possibly) simple poles at the
above zeros, which are the points
\beq \si \in \{\rho\,,\,\rho - 1\,,\,...\,,\,\rho - n + 1\} ~. \feq
Moreover, since the above results hold for any given $n \in \{1,2,3,...\}$,
they actually allow to determine the analytic continuation of $\Mel(\si)$
to the whole complex plane with simple poles at the points $\si \in \{\rho,
\rho - 1,\rho - 2,...$\}\,. \salto
As mentioned before, the above results can be employed to obtain
the sought-for analytic continuation of the Dirichlet kernel $\Dir_s$
(treating $\bx,\by\in\Om$ as fixed parameters) starting from
its representations \rref{DirHeat} \rref{DirCyl} in terms of the
heat and cylinder kernel, respectively. \parn
More precisely, consider the case in which the heat or the cylinder
kernel is given by a smooth function of $\t$ rapidly vanishing at
infinity, divided by a power of $\t$; by this we mean that
\beq K(\t\,;\bx,\by) = {1 \over \t^p}\;H(\t\,;\bx,\by) \quad
\mbox{or} \quad T(\t\,;\bx,\by) = {1 \over \t^q}\;J(\t\,;\bx,\by) ~, \label{espk} \feq
where $p,q \in \reali$, $H,J : [0,+\infty)\times \Om \times \Om \to
\reali$, and it is assumed that (for fixed $\bx, \by \in \Om$)
the function $\t \in [0,+\infty) \mapsto H(\t\,;\bx,\by)$ or
$J(\t\,;\bx,\by)$ is smooth and rapidly vanishing for $\t \to +\infty$\,.
In these cases the integrals in the right-hand sides of Eq.s \rref{DirHeat}
and \rref{DirCyl} converge for $\Re s > p$ and $\Re s > q/2$, respectively.
Let us mention that, with the assumptions \rref{iii},
the heat kernel of $\AA= - \Delta + V$ is as in Eq. \rref{espk} with $p = {d/2}$
(recall the asymptotics \rref{asinK}); again with the assumptions \rref{iii},
the cylinder kernel of $\AA$ is
as in \rref{espk} with $q = d$, provided that no logarithmic terms appear in the asymptotic expansion
\rref{asinTD}. \parn
Under the previous assumptions, Eq. \rref{MelCon}, along with
Eq.s \rref{DirHeat} \rref{DirCyl}, gives the following for any
$n \in \{1,2,3,...\}$
({\footnote{\label{foot1}To obtain Eq. \rref{HeatCon} one uses Eq.s
(\ref{ff}-\ref{MelCon}) with $\ff(\t) = K(\t\,;\bx,\by)$, $\rho = p$,
$\hh(\t) = H(\t\,;\bx,\by)$) and $\si = s$\,. To obtain Eq. \rref{CylCon}
one uses Eq.s (\ref{ff}-\ref{MelCon}) with $\ff(\t) = T(\t\,;\bx,\by)$,
$\rho = q$, $\hh(\t) = J(\t\,;\bx,\by)$ and $\si = 2s$\,.}}):
\beq \Dir_s(\bx,\by) = {(-1)^n \over \Ga(s)(s\!-\!p)...(s\!-\!p\!+\!n\!-\!1)}
\int_0^{+\infty} \!\!d\t \;\t^{s-p+n-1}\, \partial_\t^n H(\t\,;\bx,\by) ~;
\label{HeatCon} \feq
\beq \Dir_s(\bx,\by) = {(-1)^n \over \Ga(2s)(2s\!-\!q) ... (2s\!-\!q\!+\!n\!-\!1)}
\int_0^{+\infty} \!\!d\t \; \t^{2s - q + n - 1}\,\partial_\t^n J(\t\,;\bx,\by) ~.
\label{CylCon} \feq
Comments analogous to the ones below Eq. \rref{MelCon} can be done for the
above representations. More in detail, on the one hand Eq. \rref{HeatCon}
gives the analytic continuation of the Dirichlet kernel $\Dir_s$ in the
region $\{s \in \complessi ~|~ \Re s > p-n\}$ to a meromorphic function
with simple poles at $s \in \{p,\,p-1, ...,\,p-n+1\}$\,; on the other hand,
Eq. \rref{CylCon} gives the analytic continuation of $\Dir_s$ to the
region $\{s \in \complessi ~|~ \Re s > {(q-n)/ 2}\}$, with (possibly)
simple poles at $s \in\{{q/2},\, {(q-1)/2},\,...,\,{(q-n+1)/ 2}\}$\,. \parn
Of course, relations analogous to \rref{HeatCon} and \rref{CylCon} hold
as well for the spatial derivatives of the Dirichlet kernel. \salto
In conclusion, let us stress that similar results can be deduced for
the trace $\Tr \AA^{-s}$ (see Eq. \rref{TrAs}), giving its analytic
continuation to wider regions in the complex plane. For example,
assume the heat trace has the form (compare with the first relation
in Eq. \rref{espk})
\beq K(\t) = {1 \over \t^p}\;H(\t) ~, \label{espkTr} \feq
for some $p \in \reali$ and some smooth function $H: [0,+\infty) \to
\reali$, rapidly vanishing for $\t \to +\infty$; then, starting with
Eq. \rref{DirHeatTr} and using the relations (\ref{ff}-\ref{MelCon}),
we obtain the following, for $n \in \{1,2,3,...\}$:
\beq \Tr \AA^{-s} = {(-1)^n \over \Ga(s)(s\!-\!p)...(s\!-\!p\!+\!n\!-\!1)}
\int_0^{+\infty} \!\!d\t \;\t^{s-p+n-1}\, {d^n H \over d \t^n}(\t) ~. \label{HeatConTr} \feq
The above relation gives the analytic continuation of $\Tr \AA^{-s}$
to the region $\{s \in \complessi ~|~ \Re s > p-n\}$ to a meromorphic
function with simple poles at $s \in \{p,\,p-1, ...,\,p-n+1\}$\,. A similar
result can be derived using the cylinder trace $T(\t)$\,.
\vspace{-0.4cm}
\subsection{Analytic continuation of Mellin transforms via complex integration.}
\label{anacont} Another way to obtain the analytic continuation of
the Mellin transform of a given function is available (assuming
the latter to fulfill suitable conditions). Consider again the
framework of the previous subsection; this time the idea is to
re-express the integral in Eq. \rref{Mel} as an integral along a
suitable path in the complex plane. \parn To this purpose, first
consider the following identity concerning Mellin transforms. Let
$\t \mapsto h(\t)$ be a complex-valued function, analytic in a
complex neighborhood of $[0,+\infty)$ and exponentially vanishing
for $\Re \t \to +\infty$ in this neighborhood; then
\beq \int_0^{+\infty}\!\!d\t\; \t^{s-1}h(\t) = {e^{-i \pi s} \over 2i
\sin(\pi s)} \int_\Hank\! d\t\; \t^{s-1} h(\t) \quad \mbox{for $s
\in \complessi\!\setminus\!\{1,2,3,...\}$, $\Re s > 0$}~, \label{HankIden} \feq
where $\Hank$ denotes the \textsl{Hankel contour}, that is a simple
path in the complex plane that starts in the upper half-plane near
$+\infty$, encircles the origin counterclockwise and returns to $+\infty$
in the lower half-plane (see Fig. \ref{fig:HankFig} below).
\vskip 0cm \noindent
\begin{figure}[h] \centering
\includegraphics[width=9cm]{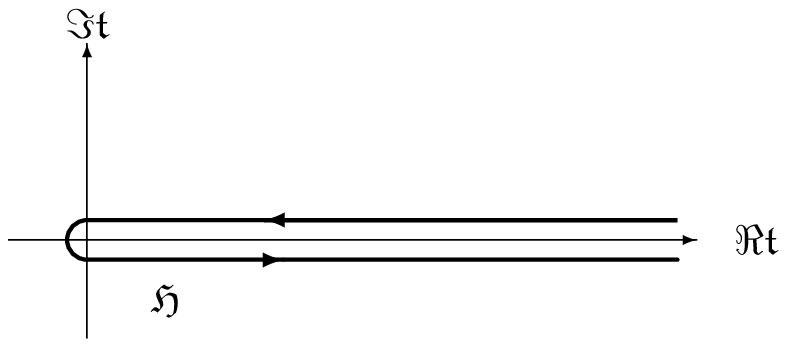}
\caption{The Hankel contour $\Hank$.} \label{fig:HankFig}
\end{figure}
\vskip 0.1cm \noindent
In the right-hand side of Eq. \rref{HankIden}, the complex power
$\t^{s-1}$ is defined making reference to Eq.s \rref{power} \rref{arg};
in the left-hand side, since $\t \in (0,+\infty)$, we define $\t^{s-1}$
according to the standard convention \rref{stand}. See Appendix \ref{apperes}
for the derivation of Eq. \rref{HankIden}. \parn
Assume now $\ff$ is as in Eq. \rref{ff} with $\hh$ a complex function,
analytic in a neighborhood of $[0,+\infty)$ and exponentially vanishing
for $\Re \t \to +\infty$; then, considering the Mellin transform $\Mel(\si)$
in Eq. \rref{Mel} and using Eq. \rref{HankIden} with $s = \si - \rho$ and
$h = \hh$\,, we obtain
\beq \Mel(\si) = {e^{-i \pi (\si-\rho)} \over 2i
\sin(\pi (\si\!-\!\rho))} \int_\Hank d\t\; \t^{\si-1} \ff(\t) ~.
\label{HankCont} \feq
In principle, Eq. \rref{HankCont} holds
under certain conditions: to ensure existence of $\Mel(\si)$ as
defined by Eq. \rref{Mel} we must require $\Re\si > \Re\rho$, and
the denominator $\sin(\pi (\si\!-\!\rho))$ must be nonzero.
However, the integral in Eq. \rref{HankCont} converges for any
$\si\in\complessi$\,, so this equation yields the analytic
continuation of the Mellin transform $\MM(\si)$ to the whole
complex plane, possibly with simple poles for
\beq \si \in \{\rho,\,\rho -1,\,\rho-2,\,...\} ~, \feq
due to the vanishing of the sine function in the denominator
({\footnote{By inspection of the denominator, it would seem that
also the points $\si \in \{\rho +1,\,\rho + 2,\,\rho+3,\,... \}$
are singular, but we know this is not the case since the original
expression \rref{Mel} for $\Mel(\si)$ is regular at these points.
The reason for the apparent contradiction lies in the fact that
the integral over the Hankel contour in Eq. \rref{HankCont} vanishes
for the above mentioned values of $\si$ (as can be easily checked via
the residue theorem recalling the properties of $\ff$), thus yielding
an indeterminate form $\infty \cdot 0$\,.}}). \salto Recall once
more that, according to Eq.s \rref{DirHeat} \rref{DirCyl}, the
Dirichlet kernel can be expressed as the Mellin transform of
either the heat or the cylinder kernel; so, the above results on
the analytic continuation via contour integration can be applied
to $\Dir_s(\bx,\by)$ (for fixed $\bx,\by \in \Om$). More
precisely, suppose that either the heat or the cylinder kernel has
the form \rref{espk}:
$$ K(\t\,;\bx,\by) = {1 \over \t^p}\;H(\t\,;\bx,\by) \quad
\mbox{or} \quad T(\t\,;\bx,\by) = {1 \over \t^q}\;J(\t\,;\bx,\by)~, $$
with $p,q \in \reali$ and suitable functions $H,J : [0,+\infty)\times
\Om \times \Om \to \reali$. Assume these functions to have
extensions $H,J : \UU([0,+\infty))\times \Om \times \Om \to \complessi$,
where $\UU([0,+\infty)) \subset \complessi$ is an open neigbourhood
of the interval $[0,+\infty)$ and, for fixed $\bx,\by \in \Om$,
the function $\t\in\UU([0,+\infty)) \mapsto H(\t\,;\bx,\by)$
or $J(\t\,;\bx,\by)$ is analytic and exponentially vanishing for
$\Re \t \to + \infty$. Making these hypoteses and using
Eq. \rref{HankCont} along with Eq.s \rref{DirHeat} \rref{DirCyl}, we obtain
({\footnote{One proceeds as in Footnote \ref{foot1},
using Eq. \rref{HankCont} in place of Eq. \rref{MelCon}.}})
\beq \Dir_s(\bx,\by) = {e^{-i \pi (s-p)} \over 2 i\,\Ga(s)\,\sin(\pi(s\!-\!p))}
\int_\Hank d\t\;\t^{s-1}\,K(\t\,;\bx,\by) ~; \label{DirHankHeatsin} \feq
\beq \Dir_s(\bx,\by) = {e^{-i \pi (2s-q)} \over 2 i\,\Ga(2s)\,\sin(\pi(2s\!-\!q))}
\int_\Hank d\t\;\t^{2s-1}\,T(\t\,;\bx,\by) ~. \label{DirHankCylsin} \feq
Due to the comments after Eq. \rref{HankCont}, both the above
identities yield the analytic continuation of the Dirichlet kernel
to a meromorphic function on the whole complex plane; more precisely,
the analytic continuations obtained via Eq.s \rref{DirHankHeatsin} and
\rref{DirHankCylsin} may have simple poles respectively for
$s \in \{ p,\,p-1,\,p-2,\,...\}$ $\setminus$ $\{0,-1,-2,...\}$ and
$s \in \{ q/2,(q-1)/2,(q-2)/2,\,...\}$ $\setminus$ $\{0,-1/2,-1,-3/2,...\}$
(as readily understood analysing the denominators in the right-hand sides
of the cited equations). \parn
In the subcases where $p,q \in \interi = \{0,\pm 1,\pm 2,...\}$,
using trivial trigonometric identities and recalling that
$\Ga(s)\Ga(1\!-\!s)\sin(\pi s) = \pi$ for any $s \in \complessi$
(see \cite{NIST}, p.138, Eq.5.5.3), Eq.s \rref{DirHankHeatsin}
\rref{DirHankCylsin} can be rephrased as
\beq \Dir_s(\bx,\by) = {e^{-i \pi s} \,\Ga(1\!-\!s) \over 2 \pi i}
\int_\Hank d\t\;\t^{s-1} K(\t\,;\bx,\by) ~;
\label{DirHankHeat}\feq \beq \Dir_s(\bx,\by) = {e^{-2 i \pi s}
\,\Ga(1\!-\!2s) \over 2 \pi i} \int_\Hank d\t\;\t^{2s-1}
T(\t\,;\bx,\by) ~. \label{DirHankCyl}\feq
In these subcases the integrals along the Hankel contour can be
computed straightforwardly for integer and half-integer values of
$s$, respectively, by means of the residue theorem; for example,
for $s = -n/2$ and $n \in \{0,1,2,...\}$, Eq. \rref{DirHankCyl} yields
\beq \Dir_{-{n \over 2}}(\bx,\by) = (-1)^n \,\Ga(n\!+\!1)\,
\Res\Big(\t^{-(n+1)}\,T(\t\,;\bx,\by)\,; 0\Big) ~.
\label{DirHankCylRes} \feq
We can obtain a variant of Eq. \rref{DirHankCyl}, giving the analytic
continuation of the Dirichlet kernel $\Dir_s$ in terms of the modified
cylinder kernel $\Tm$ (recall Eq.s \rref{SKer} \rref{STKerPrim}). To
this purpose, we assume $\Tm$ to admit a meromorphic extension in $\t$
to a neighborhood of $[0,+\infty)$, having a pole in $\t = 0$ and rapidly
vanishing for $\Re\t \to +\infty$; then, expressing the cylinder kernel
$T(\t\,;\bx,\by)$ in Eq. \rref{DirHankCyl} as $-\partial_\t \Tm(\t\,;\bx,\by)$
and integrating by parts, we obtain
\beq \Dir_s(\bx,\by) = -\,{e^{-2 i \pi s}\,\Ga(2\!-\!2s) \over 2 \pi i}
\int_\Hank d\t\;\t^{2s-2}\,\Tm(\t\,;\bx,\by) ~ \label{DirHankS}\feq
(note that no boundary contribution arises, due to the rapid vanishing
of $\Tm(\t\,;\bx,\by)$ for $\Re\t\to +\infty$).
Again, for half-integer values of $s$ we can compute explicitly the
analytic continuation \rref{DirHankS} by means of the residue theorem;
to be more precise, for $s = -n/2$ and $n \in \{-1,0,1,2,...\}$ we have
\beq \Dir_{-{n \over 2}}(\bx,\by) = (-1)^{n+1} \,\Ga(n\!+\!2)\,
\Res\Big(\t^{-(n+2)}\,\Tm(\t\,;\bx,\by)\,; 0\Big) ~.
\label{DirHankCylResTm} \feq
Relations similar to the ones obtained above hold for the spatial
derivatives of the Dirichlet kernel, allowing in turn to determine
their analytic continuations. \salto
To conclude, let us mention that similar results hold as well for
the trace $\Tr \AA^{-s}$; these are obtained using the representations
\rref{DirHeatTr} \rref{DirCylTr} in terms of the heat and cylinder
trace $K(\t),T(\t)$, and assuming suitable features for the latters.
In particular, if the map $\t \mapsto T(\t)$ admits a meromorphic
extension to a neighborhood of $[0,+\infty)$ which only has a pole
singularity at $\t = 0$ and vanishes exponentially for $\Re\t \to +\infty$,
for $n \in \{0,1,2,...\}$, we have
\beq \Tr \AA^{n/2} = (-1)^n \,\Ga(n\!+\!1)\,\Res\Big(\t^{-(n+1)}\,T(\t)\,; 0\Big) ~.
\label{DirHankCylTr} \feq
\vspace{-0.9cm}
\subsection{Other kernels, and their relations with $\boma{\Dir_s}$.}
In this section we are considering a number of integral kernels
connected with the zeta regularization of the stress-energy VEV;
attention is mainly focused on the kernels used in the subsequent
applications (including the subsequent Parts II,III and IV). However,
it would be against the spirit of this section to ignore completely
the \textsl{resolvent kernel}, i.e., the kernel of the operator
$(\AA - \lam)^{-1}$, where $\AA$ is a selfadjoint operator and
$\lam \in \complessi$ is outside the spectrum $\si(\AA)$. Under
appropriate conditions (in particular, the strict positivity of $\AA$),
the powers $\AA^{-s}$ can be related to suitable contour integrals
involving the resolvent \cite{Seel}; this fact can be restated in
terms of a relation between the corresponding kernels. This possibility
will not be considered here and in Parts II-IV, but we plan to recover
it elsewhere. \\
$\phantom{a}$ \vspace{-0.9cm}
\subsection{The case of product domains. Factorization of the heat kernel.} \label{prodDom}
Let us consider the case where $\AA = -\Delta + V$ and the spatial
domain $\Om \subset \reali^d$ has the form
\beq \Om = \Om_1 \times \Om_2 \label{omfact} \feq
with $\Om_a$, for $a \in\{1,2\}$, indicating an open subset of $\reali^{d_a}$
($d_1 + d_2 = d$); in this case, points of $\Om$ will be written as
\beq \bx = (\bx_1, \bx_2) ~, \qquad \by = (\by_1, \by_2) \feq
etc., where $\bx_a, \by_a \in \Om_a$ ($a \in \{1,2\}$). In addition to
Eq. \rref{omfact}, we assume that the external potential has the form
\beq V(\bx) = V_1(\bx_1) + V_2(\bx_2) \feq
and that the boundary conditions specified on $\partial \Om = (\partial \Om_1
\times \Om_2) \cup (\Om_1 \times \partial \Om_2)$ arise from suitable
boundary conditions prescribed separately on $\partial \Om_1$ and
$\partial \Om_2$, in such a way that, for $a \in \{1,2\}$, the operator
\beq \AA_a := - \Delta_a + V(\bx_a) \feq
($\Delta_a$ the Laplacian on $\Om_a$) is selfadjoint in $L^2(\Om_a)$.
Moreover, each one of these operators is assumed to be strictly
positive or, at least, non-negative. \parn
In the situation described above, the Hilbert space $L^2(\Om)$ and the
fundamental operator $\AA := - \Delta + V(\bx)$ acting therein can be
represented, respectively, as
\beq L^2(\Om) = L^2(\Om_1) \otimes L^2(\Om_2) ~, \qquad
\AA = \AA_1 \otimes \uno + \uno \otimes \AA_2 ~. \label{prodOmAA}\feq
Because of the assumptions we have made, each of the two operators $\AA_a$
($a \in \{1,2\}$) possesses a complete orthonormal system of eigenfunctions
$(F_{a, k_a})_{k_a \in \KK_a}$ with eigenvalues $\om^2_{a,k_a}$;
as for the fundamental operator $\AA$, we see that it has a complete
orthonormal set of eigenfunctions of the form
\beq \Fk(\bx) := F_{1, k_1}(\bx_1) F_{2, k_2}(\bx_2) \qquad
\mbox{for \;$k = (k_1, k_2) \in \KK_1 \times \KK_2$} \label{fk1k2} \feq
and that
\beq \AA \Fk = \om^2_k \Fk ~, \qquad
\om^2_k = \om^2_{1, k_1}\!+ \om^2_{2, k_2} ~. \label{omk1k2} \feq
Or course $\si(\AA) = \si(\AA_1) + \si(\AA_2)$, so that $\AA$ is non-negative;
besides, $\AA$ is strictly positive if so is one at least between $\AA_1$
and $\AA_2$. \parn
In the product case under analysis, a number of interesting facts occurs
for the integral kernels associated to $\AA$ and $\AA_1$, $\AA_2$.
The most elementary of these facts is the factorization of the heat kernel;
by this we mean that the kernels
\beq K(\t\,;\bx,\by) := (e^{-\t \AA})(\bx, \by)~, \quad
K_a(\t\,;\bx_a,\by_a) := (e^{-\t \AA_a})(\bx_a,\by_a) ~~ (a\!\in\!\{1,2\}) \feq
are related by
\beq K(\t\,;\bx,\by) = K_1(\t\,;\bx_1,\by_1)\,K_2(\t\,;\bx_1,\bx_2)~, \label{prodK} \feq
a fact that is made apparent by the eigenfunction expansion \rref{eqheat}
and by Eq.s \rref{fk1k2} \rref{omk1k2}. \salto
In passing, let also mention that an analogous relation can be easily derived
for the heat trace; writing $K(\t)$, $K_a(\t)$ for the heat traces of $\AA$
and $\AA_a$ ($a \in \{1,2\}$), respectively, we obtain
({\footnote{In fact, Eq.s \rref{omfact} \rref{prodK} and the relations \rref{KTTr}
\rref{KTInt} allow us to infer the following chain of equalities:
$$ K(\t) = \int_\Om d\bx\;K(\t\,;\bx,\bx) = \int_{\Om_1} d\bx_1\;K_1(\t\,;\bx_1,\bx_1)
\int_{\Om_2} d\bx_2\;K_2(\t\,;\bx_2,\bx_2) = K_1(\t)\,K_2(\t) ~. $$}})
\beq K(\t) = K_1(\t)\,K_2(\t) ~. \label{prodKTr} \feq
In the present subsection we have analysed the case of a product
configuration with two factors; as a straightforward generalization,
one can consider a product with an arbitrary number of factors.
Examples of such multiple products will appear in Parts III and IV.
\subsection{The case of a slab: reduction to a lower-dimensional problem.} \label{slabSubsec}
By definition, we have a slab if
\beq \Om = \Om_1 \times \reali^{d_2}~, \qquad V = V(\bx_1) \feq
with $\Om_1$ a domain in $\reali^{d_1}$ ($d_1\!+\!d_2 = d$), and if
the boundary conditions prescribed for the field refer to its
behaviour on $\partial \Om_1 \times \reali^{d_2}$. Clearly, a slab
is a subcase of the general product case discussed in the previous
subsection, with $\Om_2 = \reali^{d_2}$ and $V_2 = 0$. In this subcase
the relevant operators are $\AA = - \Delta + V(\bx_1)$ acting in $L^2(\Om)$,
\beq \AA_1 := - \Delta_1 + V(\bx_1) \feq
acting in $L^2(\Om_1)$, and $\AA_2 := - \Delta_2$ acting in
$L^2(\reali^{d_2})$. \parn
The operator $\AA_1$ has its own eigenfunctions $F_{1, k_1}(\bx_1)
\equiv \Fkd(\bx_1)$ and eigenvalues $\om^2_{1,k_1} \equiv \oo^2_{k_1}$
$(k_1 \in \KK_1$); we assume $\AA_1$ to be strictly positive, so that
$\oo_{k_1} \geqs \eps$ for some $\eps >0$. Of course, $-\Delta_2$
is non-negative with eigenfunctions $F_{2, \bk_2}(\bx_2) =
(2 \pi)^{-d_2/2} e^{i \bk_2 \cdot \bx_2}$ and eigenvalues
$\om^2_{2, \bk_2} = |\bk_2|^2$, for $\bk_2 \in \reali^{d_2}$. \parn
In the sequel we write $\Dir_s(\bx_1,\bx_2 ; \by_1, \by_2)$ for the
Dirichlet kernel of $\AA$ at the points $\bx =(\bx_1,\bx_2)$ and
$\by = (\by_1,\by_2)$; $\Dir^{(1)}_s(\bx_1,\by_1)$ will be the
Dirichlet kernel of $\AA_1$. \parn
On the one hand, according to the general results
(\ref{Tidir00}-\ref{Tidirij}), the regularized VEV
$\la 0|\Tis_{\mu\nu}(\bx)|0\ra$ is determined by Dirichlet kernel
$\Dir_s(\bx,\by)$ and its derivatives evaluated on the diagonal
$\by = \bx$. The main intent of this subsection is to express $\Dir_s$
and its derivatives in terms of the reduced kernel
$\Dir^{(1)}_s$ at all points of the diagonal $\by = \bx$ (and, in fact,
on an even larger domain).
The starting point towards this goal is the identity
\beq \Dir_s(\bx_1, \bx_2; \by_1, \by_2) =
\hat{\Dir}_s(\bx_1,\by_1; |\bx_2 - \by_2|^2)~, \label{dirhats} \feq
involving a function $\hat{\Dir}_s : \Om_1 \times \Om_1 \times
[0,+\infty) \to \complessi$, $(\bx_1,\by_1,q) \mapsto
\hat{\Dir}_s(\bx_1,\by_1,q)$. This function and its partial
derivatives with respect to $q$, at $q=0$, are completely determined by
the kernel $\Dir^{(1)}_s$, according to the following rules:
\beq \hat{\Dir}_s(\bx_1, \by_1;0) = {\Ga(s\!-\! {d_1 \over 2}) \over
(4\pi)^{d_1/2}\,\Ga(s)} \; \Dir_{s-{d_1\over 2}}^{(1)}(\bx_1,\by_1)
\quad \mbox{for $s \in \complessi$, $\Re s\!>\!{d_1 \over 2}$} ~;
\label{DirGD} \feq
\beq {\partial^n\! \hat{\Dir}_{\!s} \over \partial q^n}(\bx_1, \by_1;0) =
{(-1)^n \Ga(s\!-\!{d_1 \over 2}\!-\!n) \over (4\pi)^{d_1/2}\,4^n\,\Ga(s)}\;
\Dir_{\!\!\!s-{d_1\over 2}-n}^{(1)}\!(\bx_1,\by_1) \quad\!\! \mbox{for
$s\!\in\!\complessi$, $\Re s\!>\!{d_1 \over 2}\!+\!n$}\,. \!\!
\label{DirGDder} \feq
We defer to Appendix \ref{appeslab} the derivation of Eq.s
(\ref{dirhats}-\ref{DirGDder}).
In particular, for the derivatives involved in Eq.s (\ref{Tidir00}-\ref{Tidirij})
on $\la 0 | \Tis_{\mu\nu}(\bx)|0\ra$ we obtain the following expressions
(where, for simplicity of notation, we write $(\bx,\by)$ for $(\bx_1,\bx_2; \by_1,\by_2)$):
\beq \Dir_{\s \pm 1 \over 2}(\bx, \by) \Big|_{\by = \bx} =
{\Ga({\s- d_2 \pm 1 \over 2}) \over (4\pi)^{d_2/2}\,\Ga({\s \pm 1 \over 2})} \;
\Dir_{\s - d_2 \pm 1 \over 2}^{(1)} (\bx_1,\by_1)\Big|_{\by_1 = \bx_1} ~;
\label{TmnDirRid1} \feq
\begin{equation}\begin{split}
& \l.\partial_{x_a^i y_b^j}\Dir_{\s+1\over 2}(\bx,\by)\r|_{\by = \bx}\!\! =
\l.\partial_{x_a^i x_b^j}\Dir_{\s+1\over 2}(\bx,\by)\r|_{\by = \bx}\!\! =
\l.\partial_{y_a^i y_b^j}\Dir_{\s+1\over 2}(\bx,\by)\r|_{\by = \bx}\!\! = 0 \\
& \; \mbox{for $(a,b)=(1,2)$ or $(a,b)=(2,1)$ and
$i \in \{1,...d_a\}$, $j \in \{1,...,d_b\}$}; \label{TmnDirRid2a}
\end{split}\end{equation}
\begin{equation}\begin{split}
& \l.\partial_{z_1^i w_1^j}\Dir_{\s+1 \over 2}(\bx, \by)\r|_{\by = \bx} =
{\Ga({\s- d_2 + 1 \over 2}) \over (4\pi)^{d_2/2}\,\Ga({\s+1 \over 2})} \;
\partial_{z_1^i w_1^j}\Dir_{\!\s - d_2 + 1 \over 2}^{(1)} (\bx_1,\by_1)
\Big|_{\by_1 = \bx_1} \\
& \hspace{2.5cm} \mbox{for $z,w \in \{x,y\}$ and $i,j \in \{1,...,d_1\}$} ~;
\label{TmnDirRid2}
\end{split}\end{equation}
\begin{equation}\begin{split}
& \l.\partial_{x_2^i y_2^j}\Dir_{\s+1\over 2}(\bx,\by)\r|_{\by = \bx}\! =
- \l.\partial_{x_2^i x_2^j}\Dir_{\s+1\over 2}(\bx,\by)\r|_{\by = \bx}\! =
- \l.\partial_{y_2^i y_2^j}\Dir_{\s+1\over 2}(\bx,\by)\r|_{\by = \bx}\! = \\
& \hspace{0.5cm} = \de_{ij}\;{\Ga({\s - d_2 -1 \over 2}) \over
(4\pi)^{d_2/2}\,2\,\Ga({\s+1 \over 2})}\,\Dir_{\s-d_2 -1 \over 2}^{(1)}
(\bx_1,\by_1)\Big|_{\by_1 = \bx_1} \quad \mbox{for $i,j \in \{1,...,d_2\}$} ~.
\label{TmnDirRid3}
\end{split}\end{equation}
Relations (\ref{TmnDirRid1}-\ref{TmnDirRid3}) are derived assuming
$\Re \s > d_2\!+\!1$, but it follows from them that the analytic
continuations in $\s$ of the Dirichlet kernel, of its reduced analogue
and of their derivatives fulfill the very same relations. \parn
Let us remark that the left-hand sides of the above equations depend
in principle on $\bx = (\bx_1,\bx_2)$, while the right-hand sides only
contain $\bx_1$; this confirms the expectation that the stress-energy
VEV ought to be independent of the variable $\bx_2$, due to the symmetry
of the slab configuration under translations regarding this variable alone.
Finally, using Eq.s (\ref{TmnDirRid1}-\ref{TmnDirRid3}) and
(\ref{Tidir00}-\ref{Tidirij}), it can be easily checked that the
components of the regularized stress-energy tensor VEV also fulfill
\begin{equation}\begin{split}
& \hspace{1.9cm} \la 0|\Tis_{ij}(\bx)|0\ra = 0 \qquad
\mbox{for $i,j\in \{d_1\!+\!1,...,d\}$, $i \neq j$} ~; \\
& \la 0|\Tis_{ij}(\bx)|0\ra = \la 0|\Tis_{ji}(\bx)|0\ra = 0
\qquad \mbox{for $i\!\in\!\{1,...,d_1\}$, $j\!\in\!\{d_1\!+\!1,...,d\}$} ~.
\end{split}\end{equation}
\section{Total energy and forces on the boundary} \label{SecEn}
We refer again to the general framework of Section \ref{back}, where
a quantized scalar field on a spatial domain $\Om$ and the associated
stress-energy tensor VEV are considered; we recall that $\AA$ indicates
the fundamental operator $-\Delta + V(\bx)$ acting in $L^2(\Om)$. \parn
From now on, we indicate with $da$ the area element on the boundary
$\partial\Om$, and use for $\Om$ the standard Lebesgue measure $d\bx$.
We also write $\textbf{n}(\bx) \equiv (n^{\ell}(\bx))_{\ell=1,...,d}$ for
the outer unit normal at a point $\bx \in \partial\Om$; this is assumed
to exist everywhere (which happens if $\partial \Om$ is globally smooth)
or almost everywhere (which happens if $\partial \Om$ has edges or corners).
\vspace{-0.4cm}
\subsection{The total energy.}
As anticipated in subsection \ref{TotEnSubsec}, the zeta-regularized
total energy can be defined as
\beq \EE^\s := \int_\Om d\bx\; \la 0|\Tis_{00}(\bx)|0\ra ~, \label{EEtot} \feq
provided that the above integral converges for $\s$ in a suitable complex domain;
using Eq. \rref{Tidir00}, the above definition yields
\begin{equation}\begin{split}
\EE^\s & = \mm^\s \l(\!{1 \over 4}+\xi\!\r)\!
\int_\Om d\bx\;\Dir_{{\s - 1\over 2}}(\bx,\by)\Big|_{\by = \bx}\; + \\
& \hspace{1.cm} + \mm^\s \l(\!{1 \over 4}-\xi\!\r)\! \int_\Om d\bx
\l[\Big(\partial^{x^\ell}\!\partial_{y^\ell}\!+\!V(\bx)\Big)
\Dir_{{\s + 1 \over 2}}(\bx,\by)\r]_{\by = \bx} \; . \label{eesse}
\end{split}\end{equation}
On the other hand, the eigenfunction expansion \rref{eqkerdi} for the
Dirichlet kernel (here used with $s = {\s + 1 \over 2}$) gives
\beq \int_\Om d\bx \l[\Big(\partial^{x^\ell}\!\partial_{y^\ell}\!+\!V(\bx)\Big)
\Dir_{{\s + 1 \over 2}}(\bx,\by)\r]_{\by = \bx}\! = \label{43} \feq
$$ = \int_\KK {dk \over \om_k^{\s + 1}} \int_\Om d\bx\;
\Big(\partial^\ell \Fk(\bx)\partial_\ell \Fkc(\bx)
+ V(\bx)\Fk(\bx)\Fkc(\bx)\Big) = $$
$$ = \!\int_\KK\! {dk \over \om_k^{\s + 1}}\!\l(\int_\Om d \bx\,
\Big(\!\Fk(\bx)(-\partial^\ell \partial_\ell\!+\!V(\bx))\Fkc(\bx)\Big)\!+
\!\int_{\partial \Om}\!da(\bx)\,\Fk(\bx)\,n^\ell(\bx)\partial_{\ell} \Fkc(\bx)\!\r)$$
where, in the last step, we have integrated by parts
({\footnote{\label{here} Here and in similar situations, whenever
we speak of an integration by parts we refer
to the identity
$$ \int_\Om d\bx\;(\partial_\ell f) g = \int_{\partial \Om} d a\;
f\,g \, n_\ell - \int_\Om d\bx\; f\,\partial_{\ell} g~, $$
holding for all sufficiently smooth functions $f,g$\,.}}). \parn
To go on we note that $(-\,\partial^\ell \partial_\ell\!+\!V)\Fkc =
\ov{\AA \Fk} = \om^2_k\,\Fkc$ and $\Fk(\bx) n^\ell(\bx)\partial_\ell\Fkc(\bx)=$
$\Fk(\bx) {\partial \Fkc \over \partial n}(\bx)$
$= \l.\Fk(\bx){\partial \Fkc(\by) \over \partial n_{\by}}\r|_{\by = \bx}$,
where we have introduced the normal derivative $\partial/\partial n
:= n^\ell \partial_{x^\ell}$\,. Substituting into Eq. \rref{43} and
summing over $k \in \KK$, we obtain
\begin{equation}\begin{split}
& \hspace{1.7cm}\int_\Om d\bx \l[\Big(\partial^{x^\ell}\!\partial_{y^\ell}\!+\!V(\bx)\Big)
\Dir_{{\s + 1 \over 2}}(\bx,\by)\r]_{\by = \bx} = \\
&  = \int_\Om d\bx\; \Dir_{{\s - 1 \over 2}}(\bx, \by)\Big|_{\by = \bx}
+ \int_{\partial\Om}\!da(\bx) \l.{\partial \over \partial n_{\by}}\,
\Dir_{{\s + 1 \over 2}}(\bx, \by) \r|_{\by = \bx} ~;
\end{split}\label{44} \end{equation}
inserting this result into Eq. \rref{eesse}, we conclude
\beq \EE^\s = E^\s + B^\s ~, \label{EtotEB}\feq
where we have introduced the \textsl{regularized bulk} and
\textsl{boundary energies}
\beq E^\s := {\mm^\s\! \over 2} \int_\Om d\bx\;\Dir_{{\s - 1\over 2}}(\bx,\bx)
= {\mm^\s\! \over 2}\; \Tr \AA^{{1 -\s \over 2}} ~, \label{defEs}\feq
\beq B^\s := \mm^\s \l(\!{1 \over 4}-\xi\!\r)\! \int_{\partial \Om}\!
d a(\bx)\,\l.{\partial \over \partial n_{\by}}\,
\Dir_{{\s + 1\over 2}}(\bx,\by)\r|_{\by = \bx} ~. \label{defBs} \feq
The derivation of the above result is a bit formal since, in general,
one cannot grant convergence of the integrals defining $E^\s$ and $B^\s$,
for suitable values $\s \in \complessi$\,. \parn
As for the bulk energy, it is easy to give an example in which
finiteness is granted for appropriate $\s$. To this purpose let us
consider the case \rref{iii}
(involving a bounded domain with Dirichlet boundary conditions); in this case,
recalling the Weyl estimates \rref{weyl}, we conclude
\beq \mbox{$E^\s$\; is finite if \;$\Re \s > d\!+\!1$} ~. \feq
As for the regularized boundary energy $B^\s$ let us mention that, for
$\Om$ bounded,
\beq B^\s = 0 \qquad \mbox{under Dirichlet or Neummann boundary
conditions on $\partial\Om$} \label{BDir} \feq
(since in the Dirichlet case we have $\Dir_s(\bx, \by)=0$ for
$\bx\!\in\!\partial \Om$ and all $\by$, while in the Neumann case
${\partial \over \partial n_{\by}} \Dir_{s}(\bx, \by) = 0$ for
$\by \in \partial \Om$ and all $\bx$). \parn
Applying the above considerations in the case of an unbounded domain
requires much caution. On the one hand, $E^\s$ can be infinite for all
$\s \in \complessi$\,; on the other hand, in the definition \rref{defBs}
of $B^\s$ it might be necessary to intend the integral $\int_{\partial\Om}
da$ as $\lim_{\ell \to + \infty} \int_{\partial \Om_\ell} d a$, where
$(\Om_\ell)_{\ell = 0,1,2,...}$ is a sequence of bounded subdomains such
that $\Om_\ell \subset \Om_{\ell+1}$ (for any $\ell \in \{0,1,2,..\}$) and
$\cup_{\ell=0}^{+\infty}\Om_\ell = \Om$ (note that this limit could either
be infinite or even fail to exist). \parn
If $E^\s$ exists finite for $\s$ belonging to a suitable open
subset of $\complessi$ and it is an analytic function of $\s$ on this
domain, a renormalization by analytic continuation can be implemented;
in general, following the extended version of the zeta approach, we
define the renormalized bulk energy as
\beq E^{ren} := RP \Big|_{\s=0} E^\s = RP \Big|_{\s=0}\!
\l({\mm^\s\! \over 2}\; \Tr \AA^{{1 -\s \over 2}}\r) \,. \label{eren} \feq
When the analytic continuation of $\Tr \AA^{{1 -\s \over 2}}$ is regular up
to $\s = 0$ the above prescription is reduced to
\beq E^{ren} := E^\s \Big|_{\s=0} = {1 \over 2}\; \Tr \AA^{1/2} \label{erenAC} \feq
(of course $\Tr \AA^{1/2}$ indicates the analytic continuation of
$\Tr \AA^{{1 -\s \over 2}}$ at $\s = 0$). \parn
In a similar way one can define the renormalized boundary and total energies as
\beq B^{ren} := RP\Big|_{\s = 0}\,B^\s ~; \feq
\beq \EE^{ren} := RP\Big|_{\s = 0}\,\EE^\s ~. \label{ETren} \feq
An alternative definition of the renormalized total energy could be
\beq \EE^{ren} := \int_\Om d\bx\; \la 0|\Ti_{00}(\bx)|0\ra_{ren} ~.
\label{ETrenAl} \feq
This possibility, which is considered rarely in this series of papers,
is not granted to be equivalent to \rref{ETren}; for example, it may
happen that the integral in the right-hand side of Eq. \rref{ETrenAl}
diverges, while the prescription \rref{ETren} always gives a finite
result by construction. For a comparison between the alternatives
\rref{ETren} \rref{ETrenAl}, see the final lines of subsection \ref{esSeg}
(dealing with a field on a segment, for several types of boundary conditions).
\vspace{-0.4cm}
\subsubsection{Reduced energy for a slab configuration.}
\label{totenslab}
Let us consider the slab configuration introduced in subsection \ref{slabSubsec},
so that $\Om := \Om_1 \times \reali^{d_2}$, the potential $V$ depends
only on $\bx_1\in\Om_1$, and the boundary conditions regard only
$\partial \Om_1 \times \Om_2$. We already observed in the mentioned
subsection that the regularized VEV $\la 0 |\Ti_{\mu\nu}|0\ra$ depends
only on $\bx_1 \in \Om_1$ (and not on $\bx_2 \in \reali^{d_2}$); so,
the integral in Eq. \rref{EEtot} defining the total energy diverges
due to an infinite volume factor. \parn
As a matter of fact, when dealing with a slab configuration one usually
considers in place of the total energy $\EE^\s$ the \textsl{reduced
total energy} $\EE^s_1$; this is the total energy per unit volume in
the ``free'' dimensions, i.e.,
\beq \EE_1^\s := \int_{\Om_1} d\bx_1\; \la 0|\Tis_{00}|0\ra ~. \feq
Recalling Eq.s (\ref{TmnDirRid1}-\ref{TmnDirRid3}) and using some
well-known identities regarding the gamma function, we infer
$$ \EE_1^\s = {\mm^\s\,\Ga({\s-d_2+1 \over 2})\over (4\pi)^{d_2/2}\,\Ga({\s+1\over 2})\!}
\l\{\!\l(\!{\s\!-\!1\!+\!d_2 \over 4({\s\!-\!1\!-\!d_2})}+\xi\r)\!\int_{\Om_1}\!
d\bx_1\,\Dir^{(1)}_{{\s -d_2- 1\over 2}}(\bx_1,\by_1)\Big|_{\by_1 = \bx_1}\! +
\hspace{-3.7cm}\phantom{\l[\!\Big(\partial^{x_1^\ell}\!\partial_{y_1^\ell}\Big)
\Dir^{(1)}_{{\s - d_2 + 1 \over 2}}\r]_{\by_1}} \r. $$
\beq \l. + \l(\!{1 \over 4}-\xi\!\r)\! \int_{\Om_1}\! d\bx_1\!
\l[\Big(\partial^{x_1^\ell}\partial_{y_1^\ell}\!+\!V(\bx_1)\Big)
\Dir^{(1)}_{{\s - d_2 + 1 \over 2}}(\bx_1,\by_1)\r]_{\by_1 = \bx_1}\r\}
~ .\!\!\! \label{eesseslab} \feq
Concerning the second term above, we can express the reduced
Dirichlet kernel $\Dir^{(1)}_{{\s - d_2 + 1 \over 2}}$ in terms
of the eigenfunctions $(\Fkd(\bx_1))_{k_1 \in \KK_1}$ and the eigenvalues
$(\oo_{k_1})_{k_1\in\KK_1}$ of the reduced operator $\AA_1 = -\Delta_1+V(\bx_1)$
and integrate by parts as in the general setting; working as in the
derivation of Eq.s \rref{43} \rref{44} and keeping in mind that
$\AA_1\Fkd = \oo_{k_1}^2 \Fkd$, we obtain
\begin{equation}\begin{split}
& \hspace{2.2cm}\int_{\Om_1}\! d\bx_1\!\l[\Big(\partial^{x_1^\ell}\partial_{y_1^\ell}\!+\!V(\bx_1)\Big)
\Dir^{(1)}_{{\s - d_2 + 1 \over 2}}(\bx_1,\by_1)\r]_{\by_1 = \bx_1} = \\
&  = \!\int_{\Om_1}\!d\bx_1\,\Dir^{(1)}_{{\s-d_2-1 \over 2}}(\bx_1,\by_1)\Big|_{\by_1 = \bx_1}
+ \int_{\partial\Om_1}\!\!da(\bx_1) \l.{\partial \over \partial n_{\by_1}}\,
\Dir^{(1)}_{{\s - d_2 + 1 \over 2}}(\bx_1, \by_1) \r|_{\by_1 = \bx_1} .
\end{split}\end{equation}
In conclusion, we have a result similar to Eq. \rref{EtotEB}:
\beq \EE_1^\s = E_1^\s + B_1^\s ~, \label{EtotEBslab}\feq
where we have introduced the \textsl{regularized reduced bulk} and
\textsl{boundary energies}
\begin{equation}\begin{split}
& E_1^\s := {\mm^\s\,\Ga({\s-d_2-1 \over 2})\over 2\,(4\pi)^{d_2/2}\,\Ga({\s-1\over 2})}
\int_{\Om_1}d\bx_1\;\Dir^{(1)}_{{\s -d_2- 1\over 2}}(\bx_1,\bx_1) = \\
& \hspace{1.6cm} = {\mm^\s\,\Ga({\s-d_2-1 \over 2})\over 2\,(4\pi)^{d_2/2}\,
\Ga({\s-1\over 2})}\; \Tr\AA_1^{{d_2 + 1 -\s \over 2}} ~,
\end{split}\end{equation}
\beq B_1^\s := {\mm^\s\,\Ga({\s-d_2+1 \over 2})\over (4\pi)^{d_2/2}\,\Ga({\s+1\over 2})\!}
\l(\!{1 \over 4}-\xi\!\r)\! \int_{\partial \Om_1}\!\!d a(\bx_1)
\l.{\partial \over \partial n_{\by_1}}\,
\Dir^{(1)}_{{\s - d_2 + 1\over 2}}(\bx_1,\by_1)\r|_{\by_1 = \bx_1} ~. \feq
The considerations of the previous subsection about convergence of the
bulk and boundary energies $E^\s$, $B^\s$ have obvious analogues for
the reduced energies $E^\s_1,B^\s_1$. Of course, the reduced bulk and
boundary energies are renormalized in terms of the analytic continuation
(or, possibly, of its regular part) at $u=0$.
\vspace{-0.4cm}
\subsection{Pressure on the boundary.} \label{pressuretmunu}
In this subsection we are interested in the \textsl{pressure}
$\textbf{p}(\bx) \equiv (p_i(\bx))_{i = 1,...,d}$, i.e., the force
per unit area produced by the field inside $\Om$ at a point $\bx$ on
the boundary $\partial\Om$. A possible characterization is the
following: we first introduce, for $\Re \s$ large, the
\textsl{regularized pressure} $\textbf{p}^\s(\bx)$ of components
\beq p^\s_i(\bx) := \la 0|\Tis_{i j}(\bx)|0\ra \, n^j(\bx) \qquad
\mbox{for $i \in \{1,...,d\}$} ~; \label{press1} \feq
then, we define the \textsl{renormalized pressure} at $\bx$ setting
\beq p^{ren}_i(\bx) := RP \Big|_{\s = 0}\, p^\s_i(\bx) \label{preren} \feq
where $RP|_{\s = 0}$ indicates the regular of the analytic continuation
evaluated at $\s = 0$ (of course, if the mentioned continuation is
regular up to $\s = 0$, the above prescription reduces to
$p^{ren}_i(\bx) := p^\s_i(\bx)|_{\s = 0}$, meaning that the analytic
continuation at $\s = 0$ has to be considered). \parn
It is important to point out that this is not the only reasonable
definition for the renormalized pressure at a point $\bx \in \Om$\,;
another possibility is
\beq p^{ren}_i(\bx) := \l(\lim_{\bx'\in\Om, \bx'\to\bx}
\la 0|\Ti_{ij}(\bx')|0\ra_{ren} \r) n^j(\bx) ~. \label{alt} \feq
In few words: in the approach (\ref{press1}-\ref{preren}), one
\textsl{stays at a point on the boundary}, and performs therein the
renormalization; in the approach \rref{alt}, one renormalizes
\textsl{at points inside} $\Om$, and then moves towards the boundary.
Notice that both approaches require the existence of the normal $\bn(\bx)$
(and thus lose meaning on edges and corner points of $\partial\Om$). \parn
As a matter of fact, the prescriptions (\ref{press1}-\ref{preren}) and
\rref{alt} do not always agree. The approach \rref{preren} (possibly,
in the restriced version) gives by construction a finite pressure;
on the contrary, this is not granted for the alternative prescription
\rref{alt}. As an example, in Part II of this series of papers we discuss
the case where the spatial domain $\Om$ is a wedge; in this case at
all boundary points not in the edge, where the normal is well defined,
the pressure defined according to Eq. \rref{alt} diverges. We
conjecture that, in general, at points $\bx \in \partial\Om$ where
the normal is well defined and the approach \rref{alt} gives a finite
pressure, the result obtained according to the latter prescription
agrees with the renormalized pressure defined by Eq. \rref{preren};
in fact, this happens in all the examples analysed in this series
of papers. \parn
In the rest of the present section, our analysis of the boundary forces
will mainly refer to the approach (\ref{press1}-\ref{preren}). \salto
In applications, one often considers a situation where a quantized field
is present both inside $\Om$ and in the complementary region $\Om^c :=
\reali^d \setminus \Om$\,. In this setting the force per unit area
acting on the boundary is the resultant of the pressure produced
by the field inside $\Om$, on the one hand, and by the field inside $\Om^c$,
on the other; the renormalized versions of both these observables
can be computed, separately, using either one of the two approaches
mentioned before.
\vspace{-0.4cm}
\subsection{Explicit expression for the (regularized) pressure.} \label{espli}
Let us stick to the viewpoint (\ref{press1}-\ref{preren}); in order
to implement it, we use Eq. \rref{Tidirij} for the regularized
stress-energy tensor that gives the following, for $\bx\in\partial\Om$ (and $\bn(\bx)$ well defined):
\begin{equation}\begin{split}
p^{\s}_i(\bx) & = \mm^\s \!\l[\!\Big({1\over 4} - \xi\Big) \de_{i j}
\Big(\!\Dir_{{\s - 1 \over 2}}(\bx,\by) -
(\partial^{\,x^\ell}\!\partial_{y^\ell}\!+\!V(\bx))
\Dir_{{\s + 1 \over 2}}(\bx,\by) \Big) \, + \r. \\
& \hspace{1.7cm} \l. + \l(\!\Big({1\over 2} - \xi\Big)
\partial_{x^i y^j} - \xi\, \partial_{x^i x^j} \!\r)\!
\Dir_{{\s + 1 \over 2}}(\bx,\by) \r]_{\by = \bx}\! n^j(\bx) ~.
\label{press1dir}
\end{split}\end{equation}
To go on, let us restrict the attention to the case of \textsl{Dirichlet
boundary conditions}; then, only the terms involving mixed derivatives
(both with respect to $\bx$ and $\by$) of the Dirichlet kernel yield
non-vanishing contributions on the boundary $\partial\Om$. Moreover,
the terms proportional to $\xi$ in Eq. \rref{press1dir} can be shown
to vanish, so that
\beq p^{\s}_i(\bx) = \mm^\s \l[\l(\! -{1 \over 4}\,
\de_{ij}\,\partial^{x^\ell}\!\partial_{y^\ell} + {1 \over 2}\,
\partial_{x^i y^j}\r) \Dir_{\s+1 \over 2}(\bx,\by)\r]_{\by=\bx}\!n^j(\bx) ~;
\label{pEeT} \feq
see Appendix \ref{apperell} for the proof. As a final step,
the analytic continuation of $\boma{p}^\s$ at $\s = 0$ must be considered.
\vspace{-0.4cm}
\subsection{An equivalent characterization of boundary forces.} \label{equiv}
In the literature, forces on $\partial \Om$ are often characterized
by a different approach, which does not require the knowledge of the
full stress-energy tensor; see, e.g., the monographies by Bordag et
al. \cite{BorNew,Bord}, Milton \cite{Mil} and Plunien et al. \cite{Plu}.
In this approach one considers a variation of the spatial domain $\Om$
controlled by a real parameter, and defines the pressure in terms of the
derivative of the bulk energy with respect to the mentioned parameter.
For example, if $\Om$ is a parallelepiped $(0,a) \times (0,b) \times (0,c)$
one could consider the variation of the length of any one of its sides,
say $a$; it is customary to define the total force on the face $\{x^1=a\}$
as the derivative of the bulk energy with respect to $a$, with the
sign changed. \parn
The idea that boundary forces are related to the variation of $\Om$
has been typically presented in simple examples like the previous one;
it can be of interest to propose a general formulation of this idea,
and to compare it with the characterization of boundary forces given
in subsection \ref{pressuretmunu} via the stress-energy tensor. \parn
For the sake of definiteness, let us consider the case where $\Om$
is a bounded domain in $\reali^d$ and Dirichlet boundary conditions
are prescribed on $\partial\Om$; besides, let $\dFI : \reali^d \to
\reali^d$ be a vector field. We assume $\Om$, its boundary $\partial\Om$
and $\dFI$ are regular enough to permit the subsequent calculations. \parn
First of all, consider the family of diffeomorphism
\beq \FI_\ee : \reali^d \to \reali^d~, \quad
\bx \mapsto \FI_\ee(\bx) := \bx + \ee\,\dFI(\bx) ~,  \label{def1} \feq
labelled by a small parameter $\ee >0$, that will be ultimately sent
to zero. For any $\ee$, the spatial domain
\beq \Ome := \FI_\ee(\Om) \quad(\subset \reali^d) ~ \label{def2} \feq
can be regarded as a deformation of the initial domain $\Om$. \parn
Of course, a relation analogous to \rref{defEs} holds  for the
regularized bulk energy $E_\ee^\s$ associated to the deformed domain
$\Ome$, i.e.,
\beq E_\ee^\s = {\mm^\s\! \over 2}\;
\sum_{k \in \KK}\;(\om_{\ee,k}^2)^{{1 -\s \over 2}} ~; \label{EOmee}\feq
in the above $(\om_{\ee,k}^2)_{k \in \KK}$ denote the eigenvalues
of the fundamental operator $\AA_\eps$, that is the operator
$- \Delta + V$ acting on the Hilbert space
$L^2(\Ome)$  (with Dirichlet boundary conditions on $\partial\Ome$)
instead of $L^2(\Om)$. \parn
Let us now consider the expansion of $E_\ee^\s$ to the first order
in $\ee$, which describes the variation of the regularized bulk
energy under the deformation \rref{def1} \rref{def2} of the space
domain. Due to Eq. \rref{EOmee}, the calculation of this expansion
can be reduced to the first order expansion of the eigenvalues
$\om_{\ee,k}$; this can be done by standard perturbation techniques,
as well known from the classic work of Rellich \cite{Rell}. As illustrated
in Appendix \ref{apperell}, the conclusion of this analysis is
\beq  E_\ee^\s = E^\s - \ee\,(1 - \s)\!\int_{\partial\Om}\!da(\bx)\;
\dFI^i(\bx)\,p^\s_i(\bx) + O(\ee^2) \label{eqpres} \feq
where $\boma{p}^\s \equiv (p_i^\s)$ is the regularized pressure,
given by Eq. \rref{EeDir}. Eq. \rref{eqpres} can be used for an
alternative, but equivalent definition of the regularised pressure;
from this viewpoint, the regularized pressure field $\boma{p}^\s$
is the unique vector field on $\partial \Om$ such that \rref{eqpres}
holds for each one-parameter deformation of the form (\ref{def1}-\ref{def2})
for the domain $\Om$. \parn
Let us now perform the analytic continuation up to $\s=0$,
\textsl{assuming that no pole occurs at this point}; $E_{\ee}^\s \Big|_{\s=0}$
and $\boma{p}^\s \Big|_{\s=0}$ are the renormalized bulk energy and
pressure, and Eq. \rref{eqpres} yields the relation
\beq  E_\ee^{ren} = E^{ren} - \ee \int_{\partial\Om}\!da(\bx)\;
\dFI^i(\bx)\,p^{ren}_i(\bx)  + O(\ee^2) ~. \label{eqpresren} \feq
This is the result anticipated at the beginning of this subsection:
a characterization of boundary forces in terms of the of the bulk energy
variation under deformations of the domain. We already mentioned the
frequent use of this idea in the literature, for particular choices of $\Om$.
\vspace{-0.4cm}
\subsection{Integrated forces on the boundary.} \label{force} Let us now
discuss the evaluation of the integrated force $\fo_\Op$ acting on an
arbitrary subset $\Op$ of the spatial boundary ($\Op \subset \partial \Om$;
possibly, $\Op = \partial \Om$). As in the case of the pressure considered
in subsection \ref{pressuretmunu}, we can give several alternative definitions
of this quantity. As a first possibility, we introduce the regularized
total force on $\Op$ (for large $\Re \s$)
\beq \fo_\Op^\s := \int_{\Op} da(\bx)\;\bp^\s(\bx) ~,
\label{Freg} \feq
where $\bp^\s \equiv (p^\s_i(\bx))$ indicates the regularized pressure
(see Eq. \rref{press1}); then, we define the \textsl{renormalized
total force} on $\Op$ as
\beq \fo_\Op^{ren} := RP\Big|_{\s = 0} \fo_\Op^\s \label{Fren} \feq
(clearly, when there is no pole in $\s = 0$, the above prescription
reduces to $\fo_\Op^{ren} := \fo_\Op^\s |_{\s = 0}$, meaning that the
analytic continuation in $\s = 0$ has to be considered). \parn
Another possibility is to put
\beq \fo_\Op^{ren} := \int_{\Op} da(\bx)\; \bp^{ren}(\bx) ~,
\label{FrenAlt} \feq
where $\bp^{ren}(\bx) = (p^{ren}_i(\bx))$ is the renormalized pressure,
defined according either to Eq. \rref{preren} or to Eq.
\rref{alt}. \parn
Similarly to what we pointed out in subsection \ref{pressuretmunu} for
the pressure, in general the two alternatives \rref{Fren} \rref{FrenAlt}
give different results; in fact, the prescription \rref{Fren} always
gives a finite result for $\fo_\Op^{ren}$, while \rref{FrenAlt}
can give an infinite result. \parn
In conclusion, let us stress that for the integrated force there hold
comments analogous to the ones at the end of subsection \ref{pressuretmunu},
when a quantized field is present both inside $\Om$ and in the complementary
region $\Om^c := \reali^d \setminus \Om$\,. In this case the total force
on any subset $\Op \subset \partial\Om$ is given by the resultant of the
forces corresponding, respectively, to the field inside and outside the
spatial domain $\Om$\,.
\vspace{-0.4cm}
\subsection{A comment on some previous ``anomalies''.}
In the previous subsections we have pointed out that the renormalized
versions of the total energy, of the pressure and of the integrated
forces on the boundary can be defined according to different prescriptions,
which in general are not equivalent (see Eq.s \rref{ETren} \rref{ETrenAl},
\rref{preren} \rref{alt}, \rref{Fren} \rref{FrenAlt} and the considerations
in the corresponding subsections). \parn
In consequence of this, there arise unavoidable ambiguities, or \textsl{anomalies},
when talking about the renormalized observables mentioned above.
For example, we have mentioned previously the possible non-equivalence of the
alternatives \rref{ETren} \rref{ETrenAl} for the total energy $\EE^{ren}$
and \rref{preren} \rref{alt} for the pressure $\bp^{ren}$; recall that it
may happen that the prescriptions \rref{ETrenAl} and \rref{alt} give infinite
results for $\EE^{ren}$ and $\bp^{ren}$, due to boundary singularities of the
stress-energy VEV which make divergent the integral $\int_\Om d\bx\,\la 0|\Ti_{0 0}(\bx)|0\ra_{ren}$
or the limit $\lim_{\bx' \in \Om,\,\bx' \to \bx} \la 0|\Ti_{ij}(\bx')|0\ra_{ren}\,n^j(\bx)$
($\bx \in \partial \Om$, $i \in \{1,...,d\}$). \parn
On the other hand, such boundary singularities of the renormalized stress-energy
VEV are not a specific consequence of the zeta regularization; indeed, they also
appear if one uses point-splitting, as indicated by the very systematic analysis
of Deutsch and Candelas \cite{DeuCan}. For the moment, the above mentioned anomalies
must be accepted as a problematic aspect of the main regularization schemes;
what we can do is just to record them when they appear, and hope that in the future
they can be better understood
({\footnote{One should probably look for their origin in some excessive
idealization of the physical model (for example, one could try
to describe in a more realistic manner the boundaries of the spatial domain; these
are ``hard'' and ``deterministic'' in the present formulation, but could
perhaps be replaced with ``soft'' or ``stochastic'' \cite{FordSt} boundaries).
For a different type of boundary anomaly, related
to vacuum effects for charged fermions, and for its cure
taking into account back reaction effects, see \cite{Joc}.
}}).
\section{Some variations of the previous schemes} \label{secVar}
The variations mentioned in the title are essentially of three kinds,
described hereafter in separate subsections. These variations will be
mainly used in the applications of Parts II, III and IV; however, the
first one will also be relevant for some subcases of the simple
application proposed at the end of the present Part I (see subsections
\ref{NNSeg} and \ref{segper}).
\subsection{The basic Hilbert space when $\boma{0}$ is an isolated point of $\boma{\si(\AA)}$; the case of Neumann and periodic boundary conditions.} \label{HiNPBC}
In this subsection we are going to consider a variation of the
framework developed in Sections \ref{back} and \ref{Secker} to deal
with the cases where the fundamental operator $\AA = -\Delta + V$
acting on $L^2(\Om)$ has its spectrum contained in $[0,+\infty)$,
with $0$ an isolated point; in other terms, $0 \in \si(\AA) \subset
\{0\} \cup [\eps^2,+\infty)$, for some $\eps > 0$\,. In this case $0$
is a proper eigenvalue, as it always occurs for isolated points of
the spectrum. \parn
A standard approach employed in the physical literature to treat
problems of this kind is to simply neglect the states of ``zero
energy''; see, e.g., \cite{Hawk,Mor1a,Wal}. According to the
formulation considered in the present paper, this amounts to the
following procedure: in place of $L^2(\Om)$, we define the basic
Hilbert space  as the orthogonal complement in $L^2(\Om)$ of the
null eigenspace, that is
\beq \L2m0(\Om) := (\ker \AA)^{\perp}
\;\quad(\,\subset L^2(\Om)\,) ~. \label{L2m0} \feq
It should be noted that the restriction of $\AA$ to $L^2_0(\Om)$ is
a selfadjoint, strictly positive operator in $L^2_0(\Om)$ with
spectrum contained in $[\eps^2, + \infty)$. In this situation,
$L^2_0(\Om)$ is the basic Hilbert space even from the viewpoint
of field quantization
({\footnote{By this, we mean that the Fock space $\Fock$ of the
quantized scalar field living in $\Om$ is the direct sum of all
symmetrized tensor powers of $\L2m0(\Om)$.}}). \parn
Let us recall the definition \rref{kerb}, giving the integral kernel
associated to a given operator on $L^2(\Om)$, and consider
Eq.s \rref{Dirdef}, \rref{deft} and \rref{SKer} for the Dirichlet,
heat, cylinder and modified cylinder kernels associated to $\AA$;
if the latter operator is redefined as the restriction of $-\Delta+V$
to $\L2m0(\Om)$, in the cited equations we should formally replace
$\de_\bx$, $\de_\by$ with $\E0 \de_\bx$, $\E0 \de_\by$ where $\E0$
is the orthogonal projection onto $\L2m0(\Om)$ (suitably extended
to distributions, so that it can be applied to $\de_\bx,\de_\by$).
With this modification, the expansions \rref{eqkerdi}, \rref{eqheat},
\rref{eqcyl} and \rref{SKer} for the kernels mentioned above hold
again, using the eigenfunctions of $\AA$ in $\L2m0(\Om)$
({\footnote{As an example, in the case described by Eq. \rref{mean0}
the projection $\E0$ onto $\L2m0(\Om)$ is given by
$\E0 f = f - {1 \over Vol(\Om)} \int_\Om d\bx\;f(\bx)$ ($Vol(\Om)$
is the volume of $\Om$); the previous prescription makes sense as
well for $f = \de_\bx$ and gives $\E0\de_\bx = \de_\bx - {1 \over Vol(\Om)}$\,.}). \parn
Typical configurations of the above type are those where $\AA =
\!-\Delta$\,, the spatial domain $\Om$ is bounded and the field fulfills
either Neumann or periodic boundary conditions on $\partial\Om$
({\footnote{As will be observed in subsection \ref{curvSubsec}, the
case of periodic boundaries would be more properly formulated in terms
of a free field on a torus, but this is cause of no concern for the
present considerations.}});
indeed, in such cases the spectrum of $\AA$ in $L^2(\Om)$ is purely
discrete, $0$ is an eigenvalue and $\ker\!\AA$ is formed by the
constant functions. Therefore $\L2m0(\Om)$\,, defined via Eq. \rref{L2m0},
is formed by the functions with mean zero:
\beq \L2m0(\Om) = \l\{f \in L^2(\Om) ~\Big|~
\int_\Om d\bx\, f(\bx) = 0 \r\} \,. \label{mean0} \feq
Let us mention that an analogous framework can be considered for
slab configurations where $\Om = \Om_1 \times \reali^{d_2}$ and
Neumann or periodic boundary conditions are prescribed on
$\partial\Om_1 \times \reali^{d_2}$. In these cases one works with
the reduced operator $\AA_1$ acting in $L^2(\Om_1)$; the latter
must then be replaced with the Hilbert space
\beq \L2m0(\Om_1) := (\ker \AA_1)^{\perp} = \l\{f \in L^2(\Om_1) ~\Big|~
\int_{\Om_1} d\bx_1\, f(\bx_1) = 0 \r\} \label{mean0slab} \feq
and the basic Hilbert space for the full theory on $\Om$ is
$\L2m0(\Om_1) \otimes L^2(\reali^{d_2})$\,. \parn
In the applications to be considered in the following, whenever $0$
is an isolated point of the spectrum we will always assume that the
fundamental operator $\AA$ (resp. $\AA_1$) has been redefined so that
it acts on the Hilbert space $\L2m0(\Om)$ of Eq. \rref{L2m0}
(resp. $\L2m0(\Om_1)$ of Eq. \rref{mean0slab}).
\vspace{-0.4cm}
\subsection{The case where $\boma{0}$ is in the continuous spectrum of $\boma{\AA}$.} \label{AAeps}
Let us pass to the case where the fundamental operator $\AA = - \Delta + V$
is non-negative ($\si(\AA) \subset [0,+\infty)$), and $0$ is in the
continuous spectrum of $\AA$; then $0$ has zero spectral measure and
is a non isolated point of the spectrum (otherwise, it would be a proper
eigenvalue). \parn
Here are two examples of this kind. To obtain them we consider the operator
$\AA := -\Delta$ in $L^2(\reali^d)$\,, or the operator $\AA := -\Delta$ in
$L^2(\Om)$ where $\Om$ is the half-space $\{ \bx \in \reali^d~|~ x^1 > 0 \}$\,,
and suitable boundary conditions, say Dirichlet, are specified on
$\partial \Om = \{ x^1 = 0 \}$\,. In these cases $\AA$ has a complete
orthonormal system of (improper) eigenfunctions $(F_\bk)_{\bk \in \KK}$
with corresponding eigenvalues $\om^2_\bk$, where: in the first case,
$\KK = \reali^d$ (with the Lebesgue measure $d \bk$), $\F_{\bk}(\bx)
:= (2 \pi)^{-d/2} e^{i \bk \sc \bx}$\,, $\om_\bk := |\bk|$\,; in the
second case, $\KK = (0,+\infty) \times \reali^{d-1}$ (again, with the
Lebesgue meaure $d \bk$, $\F_{\bk}(\bx) := \sqrt{2}(2\pi)^{-d/2}\sin(k^1 x^1)
e^{i k^2 x^2 + ... + k^d x^d}$ and, again, $\om_\bk := |\bk|$. In both
cases $\si(\AA) = [0,+\infty)$ and the spectrum is purely continuous. \parn
The case of $\AA$ non-negative, with $0$
in the continuous spectrum,
cannot be treated with the approach of the previous subsection: there
is no way to obtain a strictly positive operator by simply removing $0$
from the spectrum. In a more physical language, infrared divergences
cannot be simply ignored and we must devise a more sophisticated way
to treat them, as we are currently doing for ultraviolet divergences.
A natural approach to the problem is to represent $\AA$ as a limit
\beq \AA := ``{\lim_{\eps \to 0^+}}" \AA_{\eps} \feq
where $\AA_{\eps}$ is a selfadjoint operator, depending on a parameter
$\eps \in (0, \eps_0)$ and such that the spectrum of $\AA_{\eps}$ is
contained in $[\eps^2, +\infty)$; the deformed operator $\AA_\eps$ is
used everywhere in place of $\AA$, and the limit $\eps \to 0^+$ is
performed only at the end, after zeta renormalization has been carried
out. In particular, we define the \textsl{deformed smeared field operator}
\beq \Fiseps := (\mm^{-2} \AA_\eps)^{-\s/4} \Fi \feq
and the corresponding \textsl{deformed, regularized stress-energy tensor}
$\Tiseps_{\mu \nu}(x)$ whose VEV is given by
$$ \la 0 | \Tiseps_{\mu \nu}(x) | 0 \ra =  $$
$$ = \! \l. \l({1 \over 2}\!-\!\xi\!\r)\!(\partial_{x^\mu y^\nu}\!
+ \partial_{x^\nu y^\mu}\!)\! -\!\l({1\over 2}\!-\!2\xi\!\r)\!
\eta_{\mu\nu}\!\l(\partial^{x^\lam}\!\partial_{y^\lam}\!+\! V\r)\!
- \xi (\partial_{x^ \mu x^\nu}\!+ \partial_{y^\mu y^\nu}\!) \r|_{y=x}\!\!\cdot $$
\beq \cdot ~ \la 0|\Fiseps(x) \Fiseps(y) | 0 \ra ~; \label{tispropep} \feq
for the above VEV we have expression analogous to (\ref{Tidir00}-\ref{Tidirij})
in terms of the \textsl{deformed Dirichlet kernel}
\beq \Dir^{\eps}_s(\bx, \by) := \AA^{-s}_{\eps}(\bx,\by)
= \la \de_\bx | \AA^{-s}_{\eps}\, \de_\by \raL ~,  \feq
with $s = (u \pm 1)/2$\,. \parn
As mentioned above, in this generalized version of the local zeta
regularization the limit $\eps \to 0^+$ must be considered only
after the analytic continuation has been performed; in particular,
we define
\beq \la 0|\Ti_{\mu\nu}(x)|0\ra_{ren} := \lim_{\eps \to 0^+}
RP \Big|_{\s = 0} \la 0|\Tiseps_{\mu\nu}(x)|0\ra ~. \label{58} \feq
The above renormalized VEV can be expressed in terms of the renormalized
kernels $\Dik_{\pm 1/2}(\bx,\by)$ and $\partial_{z w} \Dik_{1/2}(\bx,\by)$,
where
\beq \Dik_{\!\pm {1 \over 2}}(\bx,\by) := \lim_{\eps \to 0^+}\!
RP \Big|_{\s=0}\Big(\mm^{\s} \Dir^{\eps}_{{\s \pm 1 \over 2}}(\bx,\by)\Big)\!
= \lim_{\eps \to 0^+}\! RP \Big|_{s= \pm {1 \over 2}}\Big(\mm^{2 s \mp 1}
\Dir^{\eps}_{s}(\bx,\by)\Big) \,,\! \label{DirRen} \feq
$$ \partial_{z w} \Dik_{{1 \over 2}}(\bx,\by)\! := \lim_{\eps \to 0^+}\!
RP \Big|_{\s=0} \Big(\mm^{\s} \partial_{z w} \Dir^{\eps}_{{\s + 1 \over 2}}(\bx,\by)\Big)\!
= \lim_{\eps \to 0^+}\!RP \Big|_{s={1 \over 2}}\Big(\mm^{2 s -1}
\partial_{z w} \Dir^{\eps}_{s}(\bx,\by)\Big)\,; $$
these functions play a role very similar to the ones introduced in
Eq. \rref{Dik} for a strictly positive $\AA$, and allow to express
the renormalized VEV \rref{58} as in Eq.s
(\ref{Tidir00R}-\ref{TidirijR}). \salto In the sequel we will
write $K^{\eps}$ and $T^{\eps}$ (or $\Tm^{\eps}$), respectively,
for the heat and cylinder (or modified cylinder) kernel associated
to $\AA_{\eps}$. Proceeding as in Section \ref{Secker} we obtain,
for $\Re s$ sufficiently large, \beq \Dir_s^{\eps}(\bx,\by) = {1
\over \Ga(s)} \int_0^{+\infty}
\!\!d\t\;\t^{s-1}\,K^{\eps}(\t\,;\bx,\by) ~; \label{DirHeatEp}\feq
\beq \Dir_s^{\eps}(\bx,\by) = {1 \over \Ga(2s)} \int_0^{+\infty}
\!\!d\t\;\t^{2s-1}\,T^{\eps}(\t\,;\bx,\by) ~; \label{DirCylEp}
\feq the above formulas are the starting point to discuss the
analytic continuation in $s$ of the Dirichlet kernel
$\Dir^{\eps}_s$, for any fixed $\eps \in (0,\eps_0)$. \parn
Let us associate to $\AA$ to the ``undeformed'' fundamental operator $\AA$
the heat and cylinder kernels
\beq K(t\,;\bx,\by) := e^{-t \AA}(\bx,\by)
~, \qquad T(t\,;\bx,\by) := e^{-t \sqrt{\AA}}(\bx,\by) ~, \feq
as well as the modified cylinder kernel
\beq \Tm(\t\,;\bx,\by) := (\sqrt{\AA\,}^{\;-1}e^{-t
\sqrt{\AA}})(\bx,\by)  \feq
(see subsection \ref{nonNeg}); these
can be represented
as in Eq.s \rref{eqheat}, \rref{eqcyl} and \rref{SKer} in terms of the
eigenfunctions $(\Fk)_{k\in \KK}$ and eigenvalues $(\om_k)_{k \in \KK}$
of $\AA$. We stress that the functions $\Dik_{\pm 1/2}$ of Eq. \rref{DirRen}
do \textsl{not} possess integral representations of the form
\rref{DirHeatEp} \rref{DirCylEp} with $K^{\eps}, T^{\eps}$ replaced
by $K,T$; in fact, the corresponding integrals for $K,T$ are
typically divergent.
In the sequel, we will present a more subtle way to obtain
$\Dik_{\pm 1/2}$ from $K$ or $T$ (and $\Tm$). \parn
Up to now we have not specified any particular form for $\AA_\eps$.
The following two choices seem to be natural:
\beq \AA_\eps := \AA + \eps^2 ~, \label{Aep1} \feq
\beq \AA_\eps := (\sqrt{\AA} +\eps)^2 ~. \label{Aep2} \feq
The first one corresponds to the idea, widespread in the physical
literature, to treat infrared divergences  adding a small mass $\eps$
\cite{MaS,Schwe}; the second one is less familiar and is justified by
the considerations that follow. \parn
Assuming $\AA_\eps$ to have either the form \rref{Aep1} or \rref{Aep2},
we readily obtain the following relations allowing to express the
deformed kernels $K^{\eps},T^{\eps}$ in terms of the analogous basic
kernels $K,T$:
\beq \AA_\eps := \AA +\eps^2 \quad \Rightarrow \quad
K^{\eps}(\t\,;\bx,\by) = e^{-\eps^2 \t}\; K(\t\,;\bx,\by) ~; \label{Kep}\feq
\beq \AA_\eps := (\sqrt{\AA} +\eps)^2 \quad \Rightarrow \quad
T^{\eps}(\t\,;\bx,\by) = e^{-\eps \t}\; T(\t\,;\bx,\by) ~. \label{Tep}\feq
In particular, assuming the kernels $K,T$ to be meromorphic functions
of $\t$, the above relations imply that the deformed kernels $K^{\eps},
T^{\eps}$ are meromorphic as well; in these cases, the deformed Dirichlet
kernel $\Dir^{\eps}_s$ admits integral representations analogous to
\rref{DirHankHeat} \rref{DirHankCyl}, involving the Hankel contour $\Hank$.
For example, one can write
\beq \Dir_s^{\eps}(\bx,\by) = {e^{-2i\pi s}\,\Ga(1\!-\!2s)\over 2\pi i}
\int_\Hank d\t\;\t^{2s-1}\,T^{\eps}(\t\,;\bx,\by) ~. \label{HankTep}\feq
Starting from the above representation we can derive explicit expressions for
the renormalized functions $\Dik_{\pm 1/2}(\bx,\by)$, and, more generally, for
\beq \Dik_{-{n \over 2}}(\bx,\by) := \lim_{\eps \to 0}
RP\Big|_{s = -{n \over 2}} \Big(\mm^{2 s + n} \Dir^{\eps}_s(\bx,\by)\Big)
\qquad (n\!\in\!\{-1,0,1,2,...\}) ~, \label{renor} \feq
that could be called ``renormalized Dirichlet kernels'' of order $-n/2$\,.
More precisely, let us assume the modified cylinder kernel $\Tm(\t\,;\bx,\by)$
associated to the fundamental operator $\AA$ to be
a meromorphic function of $\t$ in the
neighborhood of the positive real half-axis, fulfilling the bound
\beq |\Tm(\t\,;\bx,\by)| \leqs C\,|\t|^{-a - n +  1} \qquad
\mbox{for $\Re\t \to +\infty$ and some $C,a\!>\!0$} ~. \label{bout} \feq
Then, we obtain the following result, for $n = -1,0,1,2,...$ (see Appendix
\ref{appeEps}):
\beq \Dik_{-{n \over 2}}(\bx,\by) = (-1)^{n+1} \,\Ga(n\!+\!2)\,
\Res\Big(\t^{-(n +2)}\,\Tm(\t\,;\bx,\by)\,; 0\Big) ~. \label{Res1} \feq
Let us remark that Eq. \rref{Res1} has the same structure of
Eq. \rref{DirHankCylResTm}, dealing with the Dirichlet kernel when
$\AA$ is strictly positive. In the stricly positive case, the cited
result was derived rigorously (from Eq. \rref{DirHankS}), with no
need to introduce a regulating parameter $\eps$\,; in the present
framework, instead, it would be impossible to establish \rref{Res1}
without using the regulator $\eps$. \parn
One could derive results similar to Eq. \rref{Res1}, involving the
``renormalized derivatives'', e.g.,
\beq \partial_{z w} \Dik_{-{n \over 2}}(\bx,\by) :=
\lim_{\eps \to 0} RP\Big|_{s= -{n \over 2}} \Big(\mm^{2s+n}\,
\partial_{z w}\Dir^{\eps}_s(\bx,\by)\Big)\,, \label{renorde} \feq
where $z,w$ are spatial variables; indeed, for $n = -1,0,1,2,...$, we have
\beq \partial_{z w}\Dik_{-{n \over 2}}(\bx,\by) = (-1)^{n+1}\,\Ga(n\!+\!2)\,
\Res\Big(\t^{-(n+2)}\,\partial_{z w}\Tm(\t\,;\bx,\by)\,; 0\Big) \label{Res2} \feq
if $\partial_{z w} \Tm(\t\,;\bx,\by)$\,, as a function of $\t$,
fulfills conditions of the form stipulated previously for
$\Tm(\t\,;\bx,\by)$ (see, in particular, Eq. \rref{bout}). \salto
To conclude, let us discuss the pressure on the boundary in the present
framework; the main point is the fact that, as in the case of strictly
positive $\AA$, there are two possible prescriptions for the renormalized
pressure. The first alternative is to introduce, at each point $\bx \in
\partial\Om$, a \textsl{deformed, regularized pressure} with components
\beq p^{\eps\s}_i(\bx) := \la 0 | \Tiseps_{i j}(\bx) | 0 \ra n^{j}(\bx)
\qquad (i\!\in\!\{1,...,d\}) ~, \label{altmreg} \feq
where $\bn(\bx) \equiv(n^j(\bx))$ is the outer unit normal to the boundary;
we then define the renormalized pressure at $\bx$ as
\beq p^{ren}_i(\bx) := \lim_{\eps \to 0^+} RP \Big|_{\s=0} p^{\eps\s}_i(\bx) ~.
\label{altm} \feq
The second alternative is to put
\beq p^{ren}_i(\bx) := \l(\lim_{\bx'\in\Om, \bx'\to\bx}
\la 0|\Ti_{ij}(\bx')|0\ra_{ren} \r) n^j(\bx) \label{altt} \feq
where $\la 0|\Ti_{ij}(\bx')|0\ra_{ren}$ is defined according to Eq.
\rref{58} at all interior points $\bx'$ of the domain. The prescriptions
\rref{altm} \rref{altt} can, in general, give different results: this
happens, for example, in the case of a massless field on a wedge-shaped
domain, to be discussed in Section 5 of Part II.
\subsection{Some variations involving the spatial domain.} \label{curvSubsec}
In the literature, a scalar field fulfilling periodic boundary conditions
is often considered. To give a rigorous description of this configuration,
one should better give up to viewing $\Om$ as an open subset of $\reali^d$
and pass to a description in terms of tori.
For example, it is customary to speak of a field on the hypercube $(0,a)^d$
with periodic boundary conditions, where $a >0$ is some given length.
In the most precise description of this configuration, $\Om$ is not $(0,a)^d$
but rather the $d$-dimensional torus $\Toro^d_a := \reali^d/(a \interi)^d
\simeq (\reali/a \interi)^d$ (where $\interi$ is the set of integers, so
that $a \interi = \{ ...,-2a,-a,0,a,2a,...\}$)
({\footnote{\label{L20Tor}The considerations of subsection \ref{HiNPBC} for
the periodic case are easily rephrased in terms of the torus $\Toro^d_a$.
The operator $\AA := -\Delta$ acting in $L^2(\Toro^d_a)$ has $0$ as an
eigenvalue, with $\ker\AA$ formed by the constant functions; again, $0$
is eliminated viewing $\AA$ as an operator acting in
$$ \L2m0(\Toro^d_a) := (\ker\AA)^\perp = \l\{f \in L^2(\Toro^d_a) ~\Big|~
\int_{\Toro^d_a} d\bx\, f(\bx) = 0 \r\} \,. $$}}). \parn
In some applications to be considered in the subsequent Parts II and III,
the space domain $\Om$ is an open subset of $\reali^d$ but, in place of the
Cartesian coordinates $\bx \equiv (x^i)$, it is natural to use for it some
curvilinear coordinates $(q^i)_{i = 1,...,d} \equiv \bq$; in these coordinates,
the line element of $\reali^d$ will have the form
\beq d \ell^2 = a_{i j}(\bq)\, d q^i d q^j ~. \label{dlq} \feq
The above spatial coordinates induce a set of spacetime coordinates
$(q^\mu)_{\mu=0,...,d} \equiv q$ on $\reali \times \Om$ where
$q^0 := t$ and the $q^i$'s are as before; clearly, the spacetime line
element $ds^2 = - dt^2+d\ell^2$ will have the form
\begin{equation}\begin{split}
& \hspace{4.5cm} d s^2 = g_{\mu \nu}(q)\, dq^\mu dq^\nu ~, \\
& g_{00} := -1 ~, ~\quad g_{i 0} = g_{0 i} := 0~, ~\quad
g_{i j}(q) := a_{i j}(\bq) \qquad \mbox{for $i,j \in \{1,...,d\}$}~.
\label{dsq}
\end{split}\end{equation}
The analogue of Eq. \rref{tiquans} in the coordinate system $(q^\mu)$ is
\beq \Tis_{\mu \nu} := (1\!-\!2\xi)\partial_\mu \Fis\!\circ\partial_\nu \Fis\!
- \!\l({1\over 2}\!-\!2\xi\!\r)\!\eta_{\mu\nu}\!
\l(\!\partial^\lam\Fis\partial_\lam \Fis\!+\!V(\Fis)^2\r)
- 2 \xi \, \Fis\!\circ \nabla_{\!\mu \nu} \Fis \,, \label{tiquansq} \feq
where $\nabla_{\mu}$ is the covariant derivative induced by the (flat)
spacetime metric \rref{dsq}. In principle, the covariant derivative
$\nabla_{\mu}$ should appear in place of any derivative $\partial_{\mu}$;
however we are working with a \textsl{scalar} field and it is well-known that
\beq \nabla_{\!\mu} f = \partial_\mu f \qquad \mbox{if $f$ is
a scalar function} ~. \feq
The situation is different when we consider second order derivatives,
which explains the appearing of $\nabla_{\!\mu\nu}$ in \rref{tiquansq}.
Let us recall that
\beq \nabla_{\!\mu \nu} f = \partial_{\mu \nu} f -
\Ga^{\lam}_{\mu \nu} \partial_{\lam} f ~~( = \nabla_{\!\nu \mu} f)
\qquad \mbox{if $f$ is a scalar function} ~, \label{nabladue} \feq
\vfill \eject \noindent
{~}
\vskip -2.4cm \noindent
where we are using the spacetime Christoffel symbols
$\Ga^{\lam}_{\mu\nu} := {1 \over 2} g^{\lam \rho}(\partial_\mu g_{\rho \nu}
+ \partial_{\nu} g_{\mu \rho} - \partial_{\rho} g_{\mu \nu}$).
The above computational rule is more efficiently implemented recalling
that $q^0=t$ and using the space covariant derivatives $D_i$
corresponding to the line element \rref{dlq}; these rely on the
Christoffel symbols $\ga^k_{ij} := {1 \over 2}\,a^{kh}
(\partial_i a_{hj} + \partial_j a_{ih} - \partial_h a_{ij})$.
From Eq. \rref{dsq} one easily infers that $\Ga^k_{ij} = \ga^k_{ij}$
(for $i,j,k \in \{1,...,d\}$) are the only non-vanishing coefficients;
so, Eq. \rref{nabladue} for a scalar function $f$ on spacetime implies
\begin{equation}\begin{split}
& \hspace{2.2cm} \nabla_{ij} f = D_{ij}f = \partial_{ij}f
- \ga^{k}_{ij} \partial_k f ~,\\
& \nabla_{0 i} f = \partial_0(\partial_{i}f) =  \partial_i(\partial_0 f) =
\nabla_{i 0}f ~, \qquad \nabla_{0 0} f = \partial_{0 0} f~.
\label{compu}
\end{split}\end{equation}
As a further variation of our schemes, we can stipulate the spatial
domain $\Om$ to be an arbitrary $d$-dimensional Riemannian manifold,
possibly non flat; in any coordinate system $(q^i)_{i=1,...,d} \equiv \bq$
of $\Om$, the Riemannian line element $d \ell^2$ will have a representation
of the form \rref{dlq}. (Of course the position $\Om = \Toro^d_a$, considered
at the beginning of this paragraph in relation to periodic boundary conditions,
amounts to choosing for $\Om$ a very simple, flat Riemannian manifold).
Given any Riemennnian manifold $\Om$, we can associate to it the spacetime
$\reali \times \Om$ equipped with the line element $d s^2 = - dt^2 + d \ell^2$;
this takes the form \rref{dsq} in coordinates $(t,q^i) \equiv (q^0,q^i) \equiv
(q^\mu)_{\mu=0,...,d}$. \parn
As a final variation, we can assume the space domain $\Om$ to be an open subset
of a Riemannian manifold and prescribe boundary condtions on $\partial\Om$. \parn
Many results in Sections \ref{back}, \ref{Secker} and \ref{SecEn} are readily
adapted to the variations considered in this subsection for the space domain.
An essential point in making these adaptations is to remember that, when an
arbitrary coordinate system is employed, the second order derivatives of
scalar functions must be intended in a covariant sense and the computational
rules \rref{compu} must be applied.
\vskip -0.2cm \noindent
\section{The case of a massless field on the segment} \label{SegSec}
\vskip -0.4cm \noindent
\subsection{Introducing the problem for arbitrary boundary conditions.}
To conclude the present Part I we present a simple application of our
general formalism, namely a $1$-dimensional model describing a massless
scalar field living on a segment, with no background potential. This
means that
\beq d=1 ~, \qquad \Om = (0,a) \quad (a > 0)~, \qquad
\AA = -\partial_{x^1 x^1} \quad(V = 0) \label{seg} \feq
where $x^1 \in (0,a)$ is the standard Cartesian coordinate
({\footnote{\label{Foot14}We could have used, in place of
$x^1$, the rescaled spatial variable
$$ x^1_\star := x^1/a \in (0,1) ~, $$
in terms of which, we would have obtained simpler expressions for
the results reported in sequel.
Yet, we choose not to employ this rescaled coordinate in order to
make the comparison with kwnown results more straightforward.}}).
The field is assumed to fulfill Dirichlet, Neumann or periodic boundary
conditions at the boundary $\partial\Om = \{0\}\cup \{a\}$; we will deal
with each one of these alternatives separately, in the subsequent subsections
\ref{DirSeg}-\ref{segper}\,.
\vfill \eject \noindent
{~}
\vskip -2.2cm \noindent
In passing, we note that the setting described above is the $d = 1$
case both for the configuration with two parallel hyperplanes and for the
$d$-dimensional box (to be considered in Parts II and IV, respectively). \salto
Let us make some comparison with the previous literature about the
Casimir effect for a scalar field on a segment. First of all, we wish
to mention the book of Bordag et al. \cite{Bord} (see Chapter 2) and
the work by Fulling et al. \cite{FullAs}; these authors derive the total
bulk energy, for several boundary conditions, using regularization methods
different from zeta approach. More precisely, \cite{Bord} uses an
exponential cut-off regularization followed by Abel-Plana resummation,
while \cite{FullAs} employs essentially a point-splitting procedure.
These authors also obtain the force acting on the end-points of the
segment by differentiating the expression for the total energy with
respect to the lenght of the segment (see the comments at the beginning
of subsection \ref{equiv}). Let us also mention the paper \cite{MaTru1}
by Mamaev and Trunov deriving the stress-energy VEV for a massless
scalar field on a segment in the case of periodic boundary conditions,
via point-splitting regularization. \parn
In all cases where the present section has an intersection with
\cite{Bord,FullAs,MaTru1}, our results are in agreement with these
references
(\footnote{\textsl{Note added}. After this manuscript was written, we became
aware of a very recent paper by Mera and Fulling \cite{MeFu} who
consider a \textsl{massive} scalar field on a segment,
regularized via an exponential cutoff, and compute
the stress-energy VEV by the method of images, i.e., as
a sum over infinitely many optical paths.
(Even though different from zeta regularization,
this approach is related to the cylinder kernel
due to the cutoff structure.) The authors
of \cite{MeFu} consider the zero cutoff limit,
whose treatment can be understood as a renormalization,
and also give a number of results on the zero mass limit.
With a small addition to these results, presented
hereafter, our renormalized stress-energy VEV
can be shown to agree with the zero mass limit of
the calculations in \cite{MeFu}. As an example
let us consider the $(0,0)$ component of
the stress-energy VEV, i.e., the energy density,
in the case of Neumann boundary conditions; in \cite{MeFu} this is expressed as the sum
of a term diverging quadratically in the cutoff,
and of other terms which have a finite
limit when the cutoff is sent to zero. If one
renormalizes removing the divergent term, one
obtains from \cite{MeFu} that
$$ \langle 0 | T_{0 0}(x^1) | 0 \rangle_{ren} = $$
$$ = - {m \over 2 \pi a}
\sum_{n=1}^{+\infty} {K_1(2 m n a) \over n}
- {m \xi \over \pi} \sum_{n=-\infty}^{+\infty} {K_1(2 m |x^1 + n a|) \over |x^1 + n a|}
- {2 m^2 \over \pi} (\xi - {1 \over 4})
\sum_{n=-\infty}^{+\infty} K_0(2 m |x^1 + n a|)
$$
where $m$ is the field mass and $K_0, K_1$ are modified Bessel functions of
the second kind (see Eq.s (10) (15-17) and (51) of \cite{MeFu}, here
re-written with $L=a$, $\beta = \xi - 1/4$ and putting $r=l=0$
to indicate the choice of Neumann boundary conditions).
It is known that $K_0(z) = - \ln z + O(1)$ and $K_1(z)
= 1/z + O(z \ln z)$ for $z \vain 0^{+}$. So one
expects that, for \hbox{$m \vain 0^{+}$},
\hbox{$m  \sum_{n=1}^{+\infty} K_1(2 m n a)/n$}
\hbox{$\vain (\sum_{n=1}^{+\infty} 1/n^2)/(2 a) = \pi^2/(12 a)$}
and $m^2 \sum_{n=-\infty}^{+\infty} K_0(2 m |x^1 + n a|) \vain 0$;
these two facts are in fact established in \cite{MeFu}.
Let us add to these results the remark that, for \hbox{$m \vain 0^{+}$},
\hbox{$m \sum_{n=-\infty}^{+\infty}
K_1(2 m |x^1 + n a|)/|x^1 + n a|$} $\vain (1/2) \sum_{n=-\infty}^{+\infty}
1/(x^1 + n a)^2 = \pi^2/(2 a^2 \sin^2 ({\pi \over a} x^1))$
(for the computation of the last series by contour integral methods
see \cite{Hen}, p. 268, Eq. 4.9-4).
In view of these facts
$$ \lim_{m \vain 0^{+}} \langle 0 | T_{0 0}(x^1) | 0 \rangle_{ren} =
- {\pi \over 24 a^2} - \xi \,{\pi \over 2 a^2 \sin^2({\pi \over a} x^1)}~, $$
in agreement with Eq. \rref{TmnNN1} of the present work. Let us emphasize
that, differently from \cite{MeFu}, our zeta approach gives the renormalized
stress-energy VEV by mere analytic continuation, with no need to rimove
divergent terms. }}).
\vfill \eject \noindent
{~}
\vskip-0.6cm
\noindent
\subsection{Cylinder and Dirichlet kernels.}
For any one of the previously mentioned boundary conditions, we
perform our analysis in the manner explained hereafter. First of all,
we determine explicitly the cylinder kernel $T(\t\,;x^1,y^1)$
associated to the fundamental operator $\AA$\,; to this purpose
we consider a complete orthonormal set of eigenfunctions
$(\Fk)_{k \in \KK}$ for $\AA$ with eigenvalues $(\om^2_k)_{k\in \KK}$.
The label set $\KK$ is countable and $\int_\KK dk$ means
$\sum_{k\in \KK}$; so, the eigenfunction expansion \rref{eqcyl}
for the cylinder kernel reads
\beq T(\t\,;x^1,y^1) = \sum_{k\in \KK} e^{-\om_k\t}\Fk(x^1)\Fkc(y^1)
~. \label{TSum1} \feq
Once the cylinder kernel has been computed explicitly by evaluating
the above sum, we can proceed to determine the modified cylinder
kernel $\Tm(\t\,;x^1,y^1)$ as the primitive of $-T(\t\,;x^1,y^1)$
vanishing for $\t \to + \infty$ (see Eq. \rref{STKerPrim}). \parn
In the next subsections $T$ and $\Tm$ will be computed explicitly,
for several kinds of boundary conditions. In all cases, the cylinder
kernel $T$ and the \textsl{spatial derivatives} of the modified cylinder
kernel $\Tm$ will be found to have meromorphic extensions in the $\t$
variable to an open complex neighborhood of $[0,+\infty)$,
with possible poles only at $\t = 0$, vanishing
exponentially for $\Re\t\to +\infty$; thus, the framework of subsection
\ref{anacont} can be applied straightforwardly. \parn
To be more precise: for $y^1 \neq x^1$, $T, \Tm$ and their derivatives
have analytic extensions in $\t$ to a neighborhood of $[0,+\infty)$.
When evaluated on the diagonal $y^1 = x^1$, the
modified cylinder kernel $\Tm$ is found to have a logarithmic singularity
in $\t = 0$, while the cylinder kernel $T$ and the spatial derivatives
of both $T$ and $\Tm$ are meromorphic in a neighborhood of $[0,+\infty)$
with only a pole singularity in $\t = 0$\,. Because of
this, one can resort to Eq.s \rref{DirHankCyl} and \rref{DirHankS} to
obtain the analytic continuation of the Dirichlet kernel and of its
derivatives, required in order to determine the regularized VEV of the
stress-energy tensor; explicitly, we have
\beq \Dir_{\s - 1 \over 2}(x^1,y^1)\Big|_{y^1 = x^1} = {e^{- i \pi(\s-1)}
\,\Ga(2\!-\!\s) \over 2 \pi i} \int_\Hank d\t\;\t^{\s-2}\, T(\t\,;x^1,y^1)
\Big|_{y^1 = x^1} ~; \label{DirRPP11} \feq
\begin{equation}\begin{split}
& \partial_{z w}\Dir_{\!{\s+1 \over 2}}(x^1\!,y^1)\Big|_{y^1 = x^1} =
-{e^{-i \pi (\s+1)}\,\Ga(1\!-\!\s) \over 2 \pi i} \int_\Hank d\t\;
\t^{\s-1}\,\partial_{z w}\Tm(\t\,;x^1,y^1)\Big|_{y^1 = x^1} \\
& \hspace{5.cm} \mbox{for $z,w \in \{x^1,y^1\}$} ~. \label{DirRPP21}
\end{split}\end{equation}
In order to obtain the analytic continuations at $\s = 0$ of the above
functions, one can simply set $\s = 0$ in the expressions on the right-hand
sides of Eq.s \rref{DirRPP11} \rref{DirRPP21} and explicitly evaluate the
remaining integrals along the Hankel contour via the residue theorem
({\footnote{Notice that the analogoue of Eq. \rref{DirRPP11} in terms
of $\Tm$ is not so simple and straightforward to employ, due to the
logarithmic behaviour of the modified cylinder kernel near $\t = 0$;
on the other hand, the analogue of \rref{DirRPP21} in terms of $T$
has a singularity in the gamma function for $\s=0$. Thus, for the
computations in which we are interested, there is no better strategy
than using Eq. \rref{DirRPP11} with $T$ and Eq. \rref{DirRPP21} with
$\Tm$; this also explains why, in the sequel, we will frequently
refer to both kernels.}}); as indicated in subsection \ref{anacont}, this gives
\beq \Dir_{- {1 \over 2}}(x^1,y^1)\Big|_{y^1 = x^1} =
- \,\Res\Big(\t^{-2}\,T(\t\,;x^1,y^1)\Big|_{y^1 = x^1}\,; 0\Big)~; \label{DirRPP11Res} \feq
\beq \partial_{z w}\Dir_{{1 \over 2}}(x^1\!,y^1)\Big|_{y^1 = x^1}\!=
\Res\Big(\t^{-1}\partial_{z w}\Tm(\t\,;x^1,y^1)\Big|_{y^1 = x^1}; 0\Big)
\quad \mbox{for $z,w\!\in\!\{x^1,y^1\}$} \! \label{DirRPP21Res} \feq
(use Eq. \rref{DirHankCylRes} and the analogue of Eq. \rref{DirHankCylResTm}
for the derivatives of $\Dir_s$). \salto
Before moving on, let us mention that analogous considerations
can be made concerning the traces $\Tr\AA^{-s}$, $T(\t)$ (see
Eq. \rref{TrAs} and Eq.s \rref{KTTr} \rref{KTInt}, respectively).
Indeed, we can compute the cylinder trace $T(\t)$ according to
Eq. \rref{case2KT} that in the present case reads
\beq T(\t) = \sum_{k\in \KK} e^{-\om_k\t} ~. \feq
By explicit evaluation of the above sum for the boundary conditions
considered in the sequel, it becomes apparent that $T(\t)$ possesses
the same features of its local counterpart. Thus, we can resort again
to the general framework of subsection \ref{anacont} to obtain the
analytic continuation of $\Tr\AA^{-s}$; in particular, due to
Eq. \rref{DirHankCylTr}, the continuation at $s = -1/2$ is
\beq \Tr \AA^{1/2} = - \,\Res\Big(\t^{-2}\,T(\t)\,; 0\Big) ~.
\label{SegTrA} \feq
\subsection{The stress-energy tensor.}
We can now determine explicitly the renormalized VEV of the stress-energy
tensor; in fact, since no singularity arises, Eq.s (\ref{Tidir00R}-\ref{TidirijR})
and \rref{DikAC} imply
\beq {~}\hspace{-0.5cm} \la 0 | \Ti_{0 0}(x^1) | 0 \ra_{ren}  =
\!\l[\!\l(\!\frac{1}{4}\!+\!\xi\!\r)\!\Dir_{- {1\over 2}}(x^1,y^1) +
\l(\!\frac{1}{4}\!-\!\xi\!\r)\! \partial_{x^1 y^1}\Dir_{{1 \over 2}}(x^1,y^1)
\r]_{y^1 = x^1} , \label{T100} \feq
\beq \la 0 | \Ti_{0 1}(x^1) | 0 \ra_{ren} =
\la 0 | \Ti_{1 0}(x^1) | 0 \ra_{ren} = 0 ~, \label{T1i0} \feq
\begin{equation}\begin{split}
& \hspace{5.cm} \la 0 | \Ti_{11}(x^1) | 0 \ra_{ren}  = \\
& = \l[\!\Big({1\over 4} - \xi\Big)\Dir_{-{1 \over 2}}(x^1,y^1)
+ {1\over 4}\; \partial_{\,x^1 y^1} \Dir_{{1 \over 2}}(x^1,y^1)
- \xi\,\partial_{x^1 x^1}\Dir_{{1 \over 2}}(x^1,y^1) \r]_{y^1 = x^1}. \label{T1ij}
\end{split}\end{equation}
In the following, the scheme outlined above will be illustrated in
detail, as an example, for the case of Dirichlet boundary conditions.
For the other boundary conditions we will be more synthetic but, in
any case, we will always report the expressions for $T(\t\,;\bx,\by)$,
$\Tm(\t\,;\bx,\by)$ and $\la 0|\Ti_{\mu\nu}|0\ra_{ren}$; in particular
recall the considerations of subsection \ref{ConfSubsec} and note that
in this case Eq. \rref{xic} gives
\beq \xi_1 = 0 ~. \label{xi1}\feq
\vspace{-0.9cm}
\subsection{The total energy.}\label{esSeg} Since no singularity appears,
we can use the general prescription \rref{erenAC}; the latter, along with
Eq. \rref{SegTrA}, allow us to derive an explicit expression for the bulk
energy, for any one of the several boundary conditions to be considered
in the following. More precisely, we have
\beq E^{ren} = -\,{1 \over 2}\;\Res\Big(\t^{-2}\,T(\t)\,; 0\Big)
\label{SegERen} \feq
where $T(\t)$ is the cylinder trace of Eq. \rref{KTInt}. \parn
In passing, let us remark that the renormalized boundary energy
$B^{ren}$ always vanishes identically in the cases considered,
due to the prescribed boundary conditions (indeed, the same
statement can be made for the regularized version $B^\s$;
see Eq. \rref{BDir}). \parn
We also mention the following fact: by direct comparison of the
results reported in subsections \ref{DirSeg}-\ref{segper} it appears
that the results derived using Eq. \rref{SegERen} could as well be
deduced integrating over $(0,a)$ the conformal part of the renormalized
energy density $\la 0|\Ti_{00}|0 \ra_{ren}$\,. On the contrary, the
non-conformal part of the latter appears to diverge in a non-integrable
manner near the end-points $x = 0$ and $x = a$.
\subsection{The boundary forces.} Let us remark that, since the boundary
is zero-dimensional, the nominal ``pressure'' on the boundary points
$x^1 = 0$, $x^1=a$ does in fact coincide with the force on these points;
because of this, we adopt the notation
\beq F_{ren}(x^1) \equiv p^{ren}(x^1) \qquad \mbox{for $x^1 = 0,a$} ~. \feq
For all the (non periodic) boundary conditions to be analysed in the
following subsections, there are in principle two definitions of the
renormalized boundary forces; these descend from the two alternatives
pointed out in the general discussion on pressure of subsection
\ref{pressuretmunu}. \parn
Let us indicate with $n^1(x^1)$ the unit ``outer normal'' at the points
on the boundary, so that $n^1(0) = -1$ and $n^1(a) = 1$. The first
definition reads
\beq F_{ren}(x^1) := \l. \la 0|\Tis_{11}(x^1)|0\ra \r|_{\s=0}\,n^1(x^1)
\label{alt1} \feq
(see Eq. \rref{preren} and notice that the prescription of taking the
regular part is superfluous, since no sigularity arises; namely, we
first compute the regularized stress-energy tensor at the boundary
point $x^1$, and then we analytically continue at $\s=0$).
The second alternative is to define (see Eq. \rref{alt})
\beq F_{ren}(x^1) := \l(\lim_{{x'}^1 \in (0,a),\,{x'}^{1} \to x^1}
\la 0 | \Ti_{1 1}({x'}^1) | 0 \ra_{ren} \r) n^1(x^1) \label{alt2} \feq
(i.e., we first renormalize at inner points of the interval $(0,a)$,
and then move towards the boundary). As a matter of fact, for all boundary
conditions considered in the next subsections, the equivalence between
\rref{alt1} and \rref{alt2} will be checked by direct computation.
\vspace{-0.4cm}
\subsection{Dirichlet boundary conditions.}\label{DirSeg} As a first example,
let us consider the case where the field fulfills Dirichlet conditions at
both the end points of the segment $(0,a)$, that is
\beq \Fi(t,x^1) = 0 \qquad \mbox{for $t \in \reali$, and
$x^1 = 0$ or $x^1 = a$} ~. \feq
A complete orthonormal set of eigenfunctions $(\Fk)_{k \in \KK}$ for
$\AA$ and the related eigenvalues $(\om_k^2)_{k \in \KK}$ are
\beq \Fk(x^1) := \sqrt{2 \over a}\,\sin(k\,x^1)\,, \quad\!\om_k^2 := k^2\!
\qquad \mbox{for }\, k\in \KK \equiv \l\{{n \pi\over a} ~\Big|~
n = 1,2,3,... \r\} . \label{speDD}\feq
The expansion \rref{TSum1} for the cylinder kernel associated to $\AA$ reads
\beq T(\t\,;x^1,y^1) = {2 \over a} \, \sum_{n = 1}^{+\infty}
e^{- {n \pi \over a}\,\t} \sin\!\l({n \pi \over a}\,x^1\r)\!
\sin\!\l({n \pi \over a}\,y^1\r) ; \label{TDD1}\feq
re-writing the trigonometric functions in terms of complex exponentials,
the right-hand side of the above equation reduces to a sum of four geometric
series, which can be explicitly evaluated. The final result is
\beq T(\t\,;x^1,y^1) =
{1 \over 2a}\!\l[{\cos({\pi \over a}(x^1\!-\!y^1)) - e^{-{\pi \over a}\t}
\over \cosh({\pi \over a}\t)\!-\!\cos({\pi \over a}(x^1\!-\!y^1))}
- {\cos({\pi \over a}(x^1\!+\!y^1)) - e^{-{\pi \over a}\t} \over
\cosh({\pi \over a}\t)\!-\!\cos({\pi \over a}(x^1\!+\!y^1))} \r]; \label{TTDD1} \feq
the same expression is also reported, e.g., in \cite{FulGus,FullAs}, but
therein it is not used to compute the full, renormalized stress-energy VEV. \parn
Expressing the hyperbolic functions in terms of exponentials,
we easily obtain the primitive of $T$ which vanishes exponentially
for $\t \to +\infty$, that is $-\Tm$; in conclusion
\begin{equation}\begin{split}
& \Tm(\t\,;x^1,y^1) = -{1 \over 2\pi}\bigg[\ln\!\bigg(1 -2 e^{-{\pi \over a}\t}
\cos\Big({\pi \over a}(x^1\!-\!y^1)\Big) + e^{-{2\pi \over a}\t} \bigg) \; + \\
& \hspace{5cm} - \ln\!\bigg(1 -2 e^{-{\pi \over a}\t}
\cos\Big({\pi \over a}(x^1\!+\!y^1)\Big) + e^{-{2\pi \over a}\t}\bigg)\bigg]\,.
\end{split}\end{equation}
Both $T$ and the space derivatives of $\Tm$
have meromorphic extensions in $\t$ to a
complex neighborhood of $[0,+\infty)$,
with poles only at $\t =0$;
so, we can employ Eq.s
\rref{DirRPP11Res} \rref{DirRPP21Res}
to obtain from them the
renormalized Dirichlet kernel and its spatial derivatives.
For example, since
\begin{equation}\begin{split}
& \hspace{5cm} T(\t\,;x^1,y^1)\Big|_{y^1 = x^1} = \\
& {1\over \pi\t} - {\pi(3\!-\!\sin^2({\pi \over a}\,x^1))
\over 12a^2 \sin^2({\pi \over a}\,x^1)}\;\t\,+
{\pi^3(15(2\!+\!\cos({2\pi \over a}\,x^1))-\sin^4({\pi \over a}\,x^1))
\over 720 a^4 \sin^4({\pi \over a}\,x^1)}\;\t^3 + O(\t^5) ~,
\end{split}\end{equation}
for $\t \to 0$, evaluating explicitly the residue in Eq. \rref{DirRPP11Res},
it follows
\beq \Dir_{- {1 \over 2}}(x^1,y^1)\Big|_{y^1 = x^1} =
{\pi \over 12a^2}\;{3 - \sin^2({\pi \over a}\,x^1) \over
\sin^2({\pi \over a}\,x^1)} ~. \feq
Proceeding similarly for the derivatives of the Dirichlet kernel,
and then using Eq.s (\ref{T100}-\ref{T1ij}), one obtains the following
expression for the renormalized VEV of the stress-energy tensor:
\begin{equation}\begin{split}
& \la 0 | \Ti_{\mu\nu}(x^1) | 0 \ra_{ren} \Big|_{\mu,\nu = 0,1}
\hspace{-0.1cm} = A \l(\!\!\barray{cc}
-1\!& 0  \\
0\! & -1 \farray \!\!\r)\! + \xi\; B (x^1) \l(\!\barray{cc}
1   &   0 \\
0   &   0 \farray\!\r) , \\
& \hspace{0.12cm} A := {\pi \over 24 a^2}~,\qquad B(x^1) := {\pi \over 2 a^2}\,
{1 \over \sin^2 ({\pi \over a}\,x^1)} \quad \mbox{for $x^1 \!\in\! (0,a)$} ~.
\label{TmnDD1}
\end{split}\end{equation}
Let us now discuss the renormalized bulk energy. To this purpose, we
first note that the expansion \rref{case2KT} for the cylinder trace
gives
\beq T(\t) = \sum_{n = 1}^{+\infty} e^{- {n \pi \over a}\,\t}
= {1 \over e^{{\pi \over a}\t} - 1} ~; \label{TDD1Tr}\feq
then, using prescription \rref{SegERen}, we readily infer
\beq E^{ren} = -\,{\pi \over 24 a} ~. \label{En1DD} \feq
In conclusion, let us consider the boundary forces; it is easily
seen that both definitions \rref{alt1} and \rref{alt2} give
(with $A$ as in Eq. \rref{TmnDD1})
\beq F_{ren}(0) = A ~, \qquad  F_{ren}(a) = - A ~. \label{FDir} \feq
\subsection{Dirichlet-Neumann boundary conditions.} Let us now consider
the case where Dirichlet and Neumann boundary conditions are respectively
prescribed at the two end points of the segment $(0,a)$:
we assume
\beq \Fi(t,0) = 0 ~, \quad \partial_{x^1}\Fi(t,a) = 0
\qquad \mbox{for $t \in \reali$} ~. \feq
In this case, a complete orthonormal set of eigenfunctions $(\Fk)_{k \in \KK}$
for $\AA$ and the related eigenvalues $(\om_k^2)_{k \in \KK}$ are described by
\beq \Fk(x^1)\!:= \sqrt{2 \over a}\,\sin(k\,x^1), ~~
\om_k^2 := k^2 \!\quad\! \mbox{for }\,k\!\in\!\KK \equiv\!
\l\{\!\Big(n\!+\!{1\over 2}\Big){\pi\over a}
\,\Big|\, n = 0,1,2,...\r\} .\! \feq
Using the expansion \rref{TSum1}, we can determine the cylinder kernel $T$
and then obtain the modified kernel $\Tm$ as minus the primitive of $T$,
vanishing for $\t \to +\infty$; the final results are
\beq T(\t\,;x^1\!,y^1)\!= {1 \over a}\!\l[{\sinh({\pi \over 2a}\,\t)
\cos({\pi \over 2a}(x^1\!-\!y^1)) \over \cosh({\pi \over a}\,\t)
- \cos({\pi \over a}(x^1\!-\!y^1))} - {\sinh({\pi \over 2a}\,\t)
\cos({\pi \over 2a}(x^1\!+\!y^1)) \over \cosh({\pi \over a}\,\t)
- \cos({\pi \over a}(x^1\!+\!y^1))}\r] , \! \label{TTDN1}\feq
\begin{equation}\begin{split}
& \hspace{5.5cm} \Tm(\t\,;x^1,y^1) = \\
& {1 \over 2\pi}\l[\ln\!\l({\cos({\pi \over 2a}(x^1\!-\!y^1))\!+\!\cosh({\pi \over 2a}\t)
\over \cos({\pi \over 2a}(x^1\!-\!y^1))\!-\!\cosh({\pi \over 2a}\t)}\r)\!
- \ln\!\l({\cos({\pi \over 2a}(x^1\!+\!y^1))\!+\!\cosh({\pi \over 2a}\t)
\over \cos({\pi \over 2a}(x^1\!+\!y^1))\!-\!\cosh({\pi \over 2a}\t)}\r)\r] .
\end{split}\end{equation}
Using the above expressions along with Eq.s \rref{DirRPP11Res},
\rref{DirRPP21Res} and (\ref{T100}-\ref{T1ij}), one obtains the
renormalized VEV of the stress-energy tensor:
\begin{equation}\begin{split}
& \hspace{0.2cm} \la 0 | \Ti_{\mu\nu}(x^1) | 0 \ra_{ren} \Big|_{\mu,\nu = 0,1}
\hspace{-0.1cm} = A \l(\!\barray{cc} 1 & 0  \\
0 & 1 \farray \!\r)\! + \xi\; B (x^1) \l(\!\barray{cc}
1   &   0 \\ 0  &   0 \farray\!\r) , \\
& A := {\pi \over 48 a^2} ~, \qquad B(x^1) := {\pi \over 2 a^2}\,
{\cos({\pi \over a}\,x^1) \over \sin^2 ({\pi \over a}\,x^1)} \quad
\mbox{for $x^1 \!\in\! (0,a)$} ~. \label{TmnDN1}
\end{split}\end{equation}
Next, we derive the cylinder trace using again the expansion \rref{case2KT}:
\beq T(\t) = \sum_{n = 0}^{+\infty} e^{-(n+{1 \over 2}){\pi \over a}\,\t}
= {e^{{\pi \over 2a}\t} \over e^{{\pi \over a}\t} - 1} ~. \label{TDN1Tr}\feq
Now prescription \rref{SegERen} allows us to obtain the renormalized
total bulk energy:
\beq E^{ren} = {\pi \over 48 a} ~. \feq
Concerning the boundary forces, also in this case definitions
\rref{alt1} \rref{alt2} agree and give
\beq F_{ren}(0) = - A ~, \qquad  F_{ren}(a) = A \label{F1DN}\feq
where $A$ is as in Eq. \rref{TmnDN1}; notice, in particular, that the
above expressions have the opposite sign with respect to the ones of
Eq. \rref{FDir}, corresponding the case of Dirichlet boundary conditions.
\vspace{-0.4cm}
\subsection{Neumann boundary conditions.}\label{NNSeg}
We are now going to study the case where
\beq \partial_{x^1}\Fi(t,x^1) = 0 \qquad \mbox{for $t \in \reali$,
and $x^1 = 0$ or $x^1 = \pi_a$} ~. \feq
In this case, according to the considerations of subsection \ref{HiNPBC}, the
Hilbert space $L^2(0,a)$ has to be replaced with the space $\L2m0(0,a)$ of
square integrable functions on $(0,a)$ with mean  zero (see Eq. \rref{mean0});
in this space, a complete orthonormal set of eigenfunctions $(\Fk)_{k \in \KK}$
for $\AA = -\partial_{x^1 x^1}$ and the corresponding eigenvalues
$(\om_k^2)_{k \in \KK}$ are given by
\beq \Fk(x^1)\! := \sqrt{2 \over a}\,\cos(k\,x^1)\,, \quad\! \om_k^2 := k^2
\qquad\!\! \mbox{for }\, k\in \KK \equiv \l\{{n \pi\over a} ~\Big|~
n = 1,2,3,... \r\} . \label{speNN} \feq
The cylinder kernel associated to $\AA$ can be evaluated according to
Eq. \rref{TSum1} to obtain
\beq T(\t\,;x^1,y^1) = {1 \over 2a} \l[{\cos({\pi \over a}(x^1\!-\!y^1)) - e^{-\t}\over
\cosh\t - \cos({\pi \over a}(x^1\!-\!y^1))} + {\cos({\pi \over a}(x^1\!+\!y^1))
- e^{-\t} \over \cosh\t - \cos({\pi \over a}(x^1\!+\!y^1))} \r] ; \label{TTNN1}\feq
while for the modified cylinder kernel, computed as minus the primitive
of $T$, we obtain
\begin{equation}\begin{split}
& \Tm(\t\,;x^1,y^1) = -{1 \over 2\pi}\bigg[\ln\!\bigg(1 -2 e^{-{\pi \over a}\t}
\cos\Big({\pi \over a}(x^1\!-\!y^1)\!\Big) + e^{-{2\pi \over a}\t} \bigg) \; + \\
& \hspace{5cm} + \ln\!\bigg(1 -2 e^{-{\pi \over a}\t} \cos\Big({\pi \over a}(x^1\!+\!y^1)\!\Big)
+ e^{-{2\pi \over a}\t}\bigg)\bigg] \,.
\end{split}\end{equation}
Resorting once more to Eq.s \rref{DirRPP11Res}, \rref{DirRPP21Res}
and (\ref{T100}-\ref{T1ij}), the renormalized VEV of the stress-energy
tensor is found to be
\beq \la 0 | \Ti_{\mu\nu}(x^1) | 0 \ra_{ren} \Big|_{\mu,\nu = 0,1}
\hspace{-0.1cm} = A \l(\!\!\barray{cc} -1\!& 0  \\
0\! & -1 \farray \!\!\r)\! - \xi\; B (x^1) \l(\!\barray{cc}
1   &   0 \\ 0  &   0 \farray\!\r) ,
\label{TmnNN1} \feq
where $A$ and $B(x^1)$ are defined as in Eq. \rref{TmnDD1}. \parn
In the case under analysis the spectrum of $\AA$ coincides with the one
obtained for Dirichlet boundary conditions (compare Eq.s \rref{speDD}
\rref{speNN}); therefore, the cylinder trace is again given by Eq.
\rref{TDD1Tr}, and we derive the same renormalized bulk energy as
in Eq. \rref{En1DD}:
$$ E^{ren} = -\,{\pi \over 24 a} ~. $$
Finally, both definitions \rref{alt1} and \rref{alt2} give for the
boundary forces the same results as in the case of Dirichlet
boundary conditions (see Eq. \rref{FDir}):
\beq F_{ren}(0) = A ~, \qquad  F_{ren}(a) = - A \feq
(again, $A$ is as in Eq. \rref{TmnDD1}).
\vspace{-0.4cm}
\subsection{Periodic boundary conditions.}\label{segper} The last case we
consider for the segment configuration is the one where the field satisfies
periodic boundary conditions:
\beq \Fi(t,0) = \, \Fi(t,a)~, \quad \partial_{x^1}\Fi(t,0) = \,\partial_{x^1}\Fi(t,a)
\qquad \mbox{for $t \in \reali$} ~. \feq
As explained in subsection \ref{curvSubsec}, this case would be more
properly formulated in terms of a free scalar field on the $1$-dimensional
torus $\Toro^1_a := \reali/(a\interi)$\,. Besides, similarly to the case of
Neumann boundary conditions, recall that the basic Hilbert space is
$\L2m0(\Toro^1_a) = \{f\!\in\!L^2(\Toro^1_a) \,|\, \int_0^a dx^1 f(x^1) = 0\}$
(see subsection \ref{HiNPBC} and the footnote \ref{L20Tor} of page \pageref{L20Tor}).
In this space a complete orthonormal set of eigenfunctions $(\Fk)_{k \in \KK}$
for $\AA$, with the corresponding eigenvalues $(\om_k^2)_{k \in \KK}$, is
\beq \Fk(x^1) := \sqrt{1 \over a}\,e^{i k\,x^1}, \quad\!
\om_k^2 := k^2 \qquad\! \mbox{for }\,k\!\in\! \KK \equiv
\l\{\pm\,{2 n \pi\over a} ~\Big|~ n = 1,2,3,...\r\} . \feq
Let us pass to determine the cylinder and modified cylinder kernel associated
to $\AA$; using the same methods of the previous subsections, we obtain
\beq T(\t\,;x^1,y^1) = {\cos({2\pi\over a}(x^1\!-\!y^1)) - e^{-{2\pi\over a} \t}
\over a \l[\cosh({2\pi\over a}\t) - \cos({2\pi\over a}(x^1\!-\!y^1))\r]} ~,
\label{TSegPer} \feq
\beq \Tm(\t\,;x^1,y^1) = -\,{1 \over 2\pi} \ln\!\bigg(1 -2\,e^{-{2\pi \over a}\t}
\cos\Big({2\pi \over a}(x^1\!-\!y^1)\!\Big) + e^{-{4\pi \over a}\t} \bigg) \feq
(the same expression for $T$ is also reported, e.g., in \cite{FulGus},
again for other purposes). \parn
Eq.s \rref{DirRPP11Res}, \rref{DirRPP21Res} and (\ref{T100}-\ref{T1ij}),
yield the following expression for the renormalized VEV of the
stress-energy tensor:
\beq \la 0 | \Ti_{\mu\nu}(x^1) | 0 \ra_{ren} \Big|_{\mu,\nu = 0,1}
= {\pi \over 6 a^2} \l(\!\barray{cc} -1\!& \!0  \\
0\! & \!-1 \farray \!\r) \,. \label{TP1} \feq
Let us stress that the above results respects the invariance under
translations $x^1 \mapsto x^1\!+\!\al$ (for any $\al \in \reali$) of
the given configuration, since it does not depend explicitly on the
spatial coordinate $x^1$\,. \parn
To conclude, we discuss the renormalized bulk energy. We first note
that expansion \rref{case2KT} for the cylinder trace yields, in the
present case,
\beq T(\t) = \l(\sum_{n = -\infty}^{-1}\! +\, \sum_{n = 1}^{+\infty}\;\r)
e^{- {2|n|\pi \over a}\,\t} = 2 \sum_{n = 1}^{+\infty} e^{- {2 n \pi \over a}\,\t}
= {2 \over e^{{2\pi \over a}\t} - 1} ~; \label{TP1Tr}\feq
then, using once more prescription \rref{SegERen}, we obtain for the
bulk energy
\beq E^{ren} = -\,{\pi \over 6 a} ~. \label{En1P} \feq
\vskip 0.5cm \noindent
\textbf{Acknowledgments.}
This work was partly supported by INdAM, INFN and by MIUR, PRIN 2010
Research Project  ``Geometric and analytic theory of Hamiltonian systems in finite and infinite dimensions''.
\vfill \eject \noindent
\appendix
\section{Appendix. On the form \rref{tiquan} for the stress-energy tensor}
\label{appetmunu}
Eq. \rref{tiquan} is the quantized version of a classical formula
for the stress-energy tensor \cite{BirDav,MorBo,Call,Oha}, which
we review here for completeness. Following Section \ref{back}, we
refer to Minkowski spacetime; after an inertial frame has been
chosen, the latter is identified with $\reali^{d+1} = \reali \times \reali^d
\ni x \equiv (x^\mu) \equiv (t,\bx)$. We confine the attention to
a subset of the form $\reali\times\Om$, where $\Om \subset \reali^d$
is a spatial domain; let us consider a classical scalar field $\phi$
on $\reali \times \Om$ described by an arbitrary Lagrangian density
$\LL = \LL(\phi,\partial\phi, x)$. The associated canonical stress-energy
tensor is
\beq T^{can}_{\mu \nu} := -\dd{\partial \LL \over
\partial(\partial^\mu \phi)} \, \partial_\nu \phi
+ \eta_{\mu\nu}\LL ~, \feq
and fulfills
\beq \partial^\mu T^{can}_{\mu\nu}= -\,\partial_\nu\LL \feq
along the solutions of the field equations. If $\Si$ is any spacelike
hypersurface with normal unit vector $N^\mu$ and volume element $d v$,
we define the canonical momentum
\beq \quad P^{can}_\nu(\Si) := \int_\Si d v \,
N^{\mu}\, T^{can}_{\mu \nu} ~. \label{psigma} \feq
For simplicity we rescrict the attention to the case
\beq \Si = \{t\} \times \Om ~, \feq
for a fixed $t \in \reali$, writing $P^{can}_{\nu}(t)$ for the
corresponding canonical momentum. In this case we can take $(N^\mu) =
(1,0,\ldots,0)$ and $\int_\Si dv$ corresponds to integration on $\Om$
with respect to the usual volume element $d\bx$, so
\beq P^{can}_\nu (t) := \int_\Om d \bx \;
T^{can}_{0 \nu}(t, \bx) ~. \label{pt} \feq
Writing $x$ for $(t, \bx)$, we have ${d P^{can}_\nu/d t}(t) = \int_\Om
d\bx \, \partial_0 T^{can}_{0 \nu}(x)$ $ = \int_\Om d\bx \,
(-\partial^0 T^{can}_{0 \nu})(x) = \int_\Om d \bx \, (\partial^i T^{can}_{i \nu} -
\partial^\mu T^{can}_{\mu \nu})(x)$ $ = \int_\Om d \bx \,
[\partial^i T^{can}_{i \nu}(x) + (\partial_\nu \LL)(\phi(x),\partial\phi(x),x)]$,
i.e., by the $d$-dimensional divergence theorem,
\beq {d P^{can}_\nu \over d t}(t) = \int_\Om d \bx \;
(\partial_\nu \LL)(\phi(x), \partial \phi(x), x)
+ \int_{\partial \Om} d a(\bx) \; n^i(\bx) \,
T^{can}_{i \nu}(x) ~. \label{evol} \feq
Here (and in the sequel) $\boma{n}(\bx) = (n^i(\bx))$ is the outer
unit vector in $\reali^d$ normal to the boundary $\partial \Om$ at
$\bx$, and $d a$ is the $(d\!-\!1)$-dimensional area element
({\footnote{Obviously enough, if $\Om$ is unbounded we intend
$\int_{\partial \Om}\! da(\bx)\, n^i(\bx) := \lim_{\ell \to +\infty}
\int_{\partial \Om_\ell}\! da_{\ell}(\bx) \,n^i_{\ell}(\bx)$ where $\Om_1
\subset \Om_2 \subset \Om_3 \subset \ldots \, $ are bounded domains such
that $\cup_{\ell = 1}^{+\infty} \Om_\ell = \Om$.}}).
For a number of reasons briefly reviewed in the sequel, it is customary
(see, e.g., \cite{Call,Sch}) to consider an ``improved stress-energy
tensor'' of the form
\beq T_{\mu \nu} := T^{can}_{\mu \nu} + \partial^\lam F_{\lam\mu\nu} ~, \label{Timp} \feq
where $F_{\lam \mu \nu}$ is a covariant tensor of rank $3$ such that
\beq F_{\lam \mu \nu} = - F_{\mu \lam \nu} ~. \label{such} \feq
Condition \rref{such} implies $\partial^\mu (\partial^\lam F_{\lam \mu \nu})
= 0$, thus ensuring
\beq \partial^\mu T_{\mu \nu} = \partial^\mu T^{can}_{\mu \nu} ~. \label{ensure} \feq
We can give definitions similar to \rref{psigma} and \rref{pt} using
the improved stress-energy tensor; in particular, for each $t \in
\reali$, we define the ``improved momentum''
\beq P_\nu(t) := \int_\Om d\bx\; T_{0\nu}(t,\bx) ~. \feq
We claim that
\beq P_\nu (t) = P^{can}_\nu (t) + \int_{\partial \Om} da(\bx)
\; n^i(\bx) \, F_{i 0 \nu}(t, \bx) ~, \label{claim} \feq
where $d a$ and $\boma{n}(\bx)$ have the same meaning as before. To
prove this, note that
\beq T_{0 \nu} - T^{can}_{0 \nu} = \partial^\lam F_{\lam 0 \mu} =
\partial^0 F_{00 \nu} + \partial^i F_{i 0 \nu} = \partial^i F_{i 0 \nu} \feq
($F_{0 0 \nu} = 0$ due to Eq. \rref{such}); thus $P_\nu (t) =
P^{can}_\nu (t) + \int_\Om d \bx \, \partial^i F_{i 0 \nu}
(t, \bx)$, and the $d$-dimensional divergence theorem yields
Eq. \rref{claim}. \parn
In many cases of interest, the boundary term in Eq. \rref{claim} is
zero. In particular, this happens if $\Om = \reali^d$ and
$F_{i 0 \nu}(t,\bx)$ vanishes rapidly for $\bx\to\infty$. In the
case of a bounded domain, the boundary term can be zero if $F_{i 0 \nu}$
depends suitably on the field $\phi$ and the latter fulfills
appropriate conditions on $\partial\Om$. \parn
Whether or not the boundary term in Eq.\! \rref{claim} vanishes, using
Eq.\! \rref{ensure} we prove that the improved momentum evolves
according to the analogue of Eq.\! \rref{evol}, i.e.,
\beq {d P_\nu \over d t}(t) = \int_\Om d \bx\; (\partial_\nu \LL)
(\phi(x), \partial \phi(x), x) + \int_{\partial \Om} d a(\bx) \;
n^i(\bx) \, T_{i \nu}(x) ~. \label{evoll} \feq
The improved stress-energy tensor is symmetric if and only if
\beq \partial^\lam(F_{\lam \mu \nu} - F_{\lam \nu \mu}) =
- \, (T^{can}_{\mu \nu} - T^{can}_{\nu \mu}) ~. \label{sim} \feq
When $T^{can}_{\mu \nu}$ is not symmetric and it can be found a rank
$3$ tensor $F_{\lam \mu \nu}$ fulfilling conditions \rref{such} and
\rref{sim}, the symmetry of the improved stress-energy tensor \rref{Timp}
is itself a good reason to consider this object. \parn
There are reasons to consider the improved stress-energy tensor even
in the case when $T^{can}_{\mu \nu}$ is itself symmetric (of course,
in this case Eq. \rref{sim} requires that $\partial^\lam F_{\lam\mu\nu}$
be symmetric in $\mu$ and $\nu$). One of these reasons has been pointed
out by Callan et al. \cite{Call}; in few words, after quantization the
divergences of the improved tensor can happen to be softer than the
divergences of the canonical one, especially in perturbative renormalization.
\salto
In this paper we are interested in a field theory governed by the equation
$0 = (-\partial_{tt} +\Delta - V)\phi = (\partial_\mu\partial^\mu - V)\phi$
which arises from the Lagrangian
\beq \LL := - {1 \over 2}\,\partial^\mu \phi\, \partial_\mu \phi
- {1\over 2}\,V \phi^2 ~. \feq
The corresponding canonical stress-energy tensor is
\beq T^{can}_{\mu \nu} = \partial_\mu \phi\,\partial_\nu \phi
- {1\over 2} \, \eta_{\mu \nu}(\partial^\lam \phi \, \partial_\lam \phi
+ V \phi^2) ~; \feq
this is symmetric and fulfills (along solutions of the field equations)
\beq \partial^\mu T^{can}_{\mu \nu} = - {1\over 2}\,(\partial_\nu V) \, \phi^2 ~. \feq
Condition \rref{such} is satisfied by the tensor
\beq F_{\lam \mu \nu} := - \xi (\eta_{\lam \nu} \partial_\mu
- \eta_{\mu \nu} \partial_\lam) \phi^2 ~, \label{fla} \feq
where $\xi$ is a real parameter. In this case
\beq \partial^\lam F_{\lam \mu \nu} = - \xi(\partial_{\mu \nu}
- \eta_{\mu\nu} \partial^\lam \partial_\lam) \phi^2 ~; \label{defe} \feq
this tensor is symmetric in $\mu$ and $\nu$, so $T_{\mu \nu}$ is
symmetric as well. The derivatives of $\phi^2$ in Eq. \rref{defe}
can be re-expressed using the field equation $\partial_\mu \partial^\mu
\phi = V \phi$, yielding
\beq \partial^\lam F_{\lam \mu \nu} = - 2 \xi\,\partial_\mu \phi\,\partial_\nu \phi
+ 2 \xi\,\eta_{\mu \nu} (\partial^\lam \phi \,\partial_\lam \phi + V \phi^2)
- 2 \xi \, \phi \,\partial_{\mu \nu} \phi ~. \feq
Thus the improved stress-energy tensor \rref{Timp} takes the form
\beq T_{\mu \nu} = (1\!-\!2\xi)\partial_\mu \phi \, \partial_\nu \phi
- \l({1 \over 2} - 2 \xi \r) \eta_{\mu \nu}(\partial^\lam \phi \partial_\lam \phi + V \phi^2)
- 2 \xi \, \phi \, \partial_{\mu \nu} \phi ~, \label{ticlas} \feq
of which \rref{tiquan} is a natural quantization. \parn
Let us recall that the momentum $P_\nu$ corresponding to the
improved tensor $T_{\mu \nu}$ is related to the canonical one
via Eq. \rref{claim}; in the present framework where
$F_{\lam \mu \nu}$ is given by Eq. \rref{fla}, the boundary term
in Eq. \rref{claim} vanishes under Dirichlet boundary conditions
($\phi(t, \bx) =0$ for all $\bx \in \partial \Om$); this term
vanishes as well if $\Om = \reali^d$ and $\phi(t, \bx)$,
$\partial_{\lam} \phi(t, \bx)$ vanish rapidly for $\bx \to \infty$. \parn
In the case $V = 0$, the above improved tensor \rref{ticlas}
allows another interpretation: this is the functional derivative
with respect to the metric of the action describing a scalar field
that interacts with a gravitational field via the scalar curvature,
in the limit where $\phi$ is small and the metric is the Minkowski
metric $\eta_{\mu \nu}$ plus a perturbation of the second order in
$\phi$. Concerning this, see \cite{Call,Oha,Par} and \cite{ptp,tes}. \parn
Again for $V = 0$, the action of the field coupled to gravity is
conformally invariant for $\xi = (d-1)/(4d)$ \cite{Wal}.
\vskip -0.5cm
\parn
\section{Appendix. Smoothness properties of some integral kernels}\label{AppKer}
We refer to an operator $\AA$ in $L^2(\Om)$ with features described by
Eq.s \rref{i}, \rref{ii} or \rref{iii} (recall that the situation \rref{i}
is more general than \rref{ii}, and \rref{ii} is more general than \rref{iii}.
In the present appendix we present a number of results which are useful in
relation to several integral kernels associated to $\AA$\,. Proving these
results would require a heavy use of functional analysis, which is not among
the purposes of the present work; therefore, hereafter we just sketch the
basic ideas that will be presented with more details elsewhere \cite{InPre}.
\salto
\textbf{General results for case \rref{i}.} As in the cited equation
we assume $\AA$ to be any strictly positive, selfadjoint operator
in $L^2(\Om)$. A basic step towards our goal consists in associating a scale
of Hilbert spaces $\HH^r$ to the powers $\AA^r$ for $r \in \reali$
(see \cite{AlbSpr} for a similar construction); speaking somehow loosely, we
can describe $\HH^r$ as the Hilbert space of generalized functions $f : \Om \to \complessi$
such that $\AA^r f \in L^2(\Om)$, equipped with the inner product
\beq \la f|g\ra_r := \la\AA^{r/2}f|\AA^{r/2}g \ra \label{eqinner} \feq
(as usual, $\la~|~\ra$ denotes the inner product of $L^2(\Om)$) which induces the norm
\beq \|f\|_r := \sqrt{\la f|f\ra_r} = \| \AA^{r/2} f \| \feq
({\footnote{Let us sketch the precise definition of $\HH^r$, given in \cite{InPre}.
For any real $r$ we consider in $L^2(\Om)$ the dense linear subspace $\DD^r$ on
which $\AA^{r/2}$ is well defined according to the general framework for functional
calculus of selfadjoint operators in Hilbert spaces (see, e.g., \cite{ReSi});
if $(\Fk)_{k in \KK}$ is a (generalized) complete orthonornmal set of eigenfunctions
of $\AA$ with eigenvalues $\om_k^2$ ($\om_k > \eps > 0$), then $\DD^r$
is formed by the functions $f \in L^2(\Om)$ such that
$\int_{\KK} d k \,\om^{2 r}_k| \la \Fk|f \ra |^2 < + \infty$.
We introduce on $\DD^r$ an inner product $\la~|~\ra_r$ following Eq. \rref{eqinner},
and define $\HH^r$ to be the completion of $\DD^r$ with respect to $\la~|~\ra_r$\,;
for $r_2 \geqs r_1$ one finds $\DD^{r_2} \subset \DD^{r_1}$, a fact implying
$\HH^{r_2} \subset \HH^{r_1}$. Of course $\AA^0$ coincides with the identity
operator $\textbf{1}$ of $L^2(\Om)$; thus $\DD^0 = \HH^0 = L^2(\Om)$ and
$\la~|~\ra_0$ is the usual $L^2$ inner product $\la~|~\raL$.
For any $r \geqs 0$, $\DD^r$ is itself complete, so $\HH^r = \DD^r \subset L^2(\Om)$;
for $r < 0$ the completion $\HH^r$
is larger than $\DD^r$, and even of $L^2(\Om) = \HH^0$. If $r$ is sufficiently
large, $\HH^{-r}$ contains elements which cannot even be interpreted as ordinary
functions on $\Om$; for example, as shown later in this appendix, the Dirac
delta $\de_\bx$ at any point $\bx \in \Om$ can be interpreted as an element
of $\HH^{-r}$ for all $r > d/2$\,.}}).
Let $(\Fk)_{k \in \KK}$ be any complete orthonormal set of (generalized)
eigenfunctions of $\AA$ with eigenvalues $(\om_k^2)_{k \in \KK}$
($\om_k \geqs \eps > 0$ for all $k \in \KK$); then $\HH^r$
is made of the generalized function $f$ on $\Om$ such that
\beq
\int_K dk\;\om_k^{2r}\,|\la \Fk | f\ra|^2 < + \infty ~; \label{char} \feq
moreover, for any pair of functions $f,g \in \HH^r$, there holds
\beq \la f | g\ra_r = \int_K dk\;\om_k^{2r}\;\ov{\la \Fk |
f\ra}\, \la \Fk | g \ra ~. \feq
Of course, $\HH^0$ is the usual space $L^2(\Om)$, and it can be proved
that $\HH^{r_2} \cemb \HH^{r_1}$ for $r_2 \geqs r_1$
({\footnote{Given any pair of topological vector spaces $\XX,\YY$, we say
that $\XX$ is continuously embedded in $\YY$, and we write $\XX \cemb \YY$,
if $\XX$ is a linear subspace of $\YY$ and the identity map from $\XX$ into
$\YY$ is continuous. So, for example, $\HH^{r_4} \cemb \HH^{r_3} \cemb L^2(\Om)
\cemb \HH^{r_2} \cemb \HH^{r_1}$ if $r_4 \geqs r_3 \geqs 0 \geqs r_2 \geqs r_1$\,.}}).
Moreover, from the characterization \rref{char} of $\HH^r$ and from the identity
$\AA^s \Fk = \om_k^{2s} \Fk$ it follows that
\beq \mbox{$\AA^{-s}$ maps continuously $\HH^{r}$ into $\HH^{r'}$
for $r, r' \in \reali$ such that $r' - r < 2 \Re s$} ~; \label{AAsc} \feq
similarly, one proves that
\beq \mbox{$e^{-\t\AA}, e^{-\t\sqrt{\AA}}$ map continuously $\HH^{r}$
into $\HH^{r'}$ for all $r,r' \in \reali$} ~. \label{Expsc} \feq
Next, consider the inner product of $L^2(\Om)$; for all $r \in \reali$ this
can be extended to a continuous, sesquilinear map
\beq \la ~|~\ra : \HH^{-r} \times \HH^r \to \complessi ~, \label{ExtCon} \feq
which allows, in particular, to infer the isomorphic identification
$(\HH^r)' = \HH^{-r}$\,. \salto
\textbf{Results for case \rref{ii}.} In accordance with the cited equation,
we now assume $\AA = -\Delta + V$ on an open subset $\Om$ of $\reali^d$ with
given boundary conditions, $V : \Om \to \reali$ a $C^\infty$ potential; all
these ingredients are chosen so as to ensure the strict positivity of $\AA$.
In this case it can be proved that
\beq \HH^r \cemb C^j(\Om) \qquad \mbox{for $j \in \naturali$, $r\in\reali$
with $\dd{r > {d \over 2} + j}$} ~, \label{CEmb} \feq
where $C^j(\Om)$ carries the topology of uniform convergence of the derivatives
up to order $j$ on compact subsets of $\Om$; this is induced by the family of seminorms
\beq p_{j K}(f) := \max_{|\al| \leqs j,\,\bx \in K} |\partial^{\al} f(\bx)|
\qquad (\mbox{$K \subset \Om$ compact}) \label{pjk} \feq
({\footnote{\label{FootImb}Let us sketch the derivation of Eq. \rref{CEmb}. Due to
well-known results on elliptic operators \cite{Chaz}, one has
$$ \HH^r \cemb H^r_{loc}(\Om) \qquad \mbox{for all $r \geqs 0$} ~, $$
where $H^r_{loc}(\Om)$ is the standard local Sobolev space of order $r$ (that is,
the space of functions $f:\Om \to \complessi$ such that $(1-\Delta)^{r/2}(\varphi f)
\in L^2(\Om)$ for all smooth, compactly supported functions $\varphi: \Om \to \complessi$).
On the other hand, the usual Sobolev imbedding theorems \cite{Adams} give
$$ H^r_{loc}(\Om) \cemb C^j(\Om) \qquad \mbox{for all $r\in\reali$,
$j \in \naturali$ with $\dd{r > {d \over 2} + j}$} ~. $$
Summing up, the embeddings discussed in this footnote yield the thesis \rref{CEmb}.}}).
By duality, Eq. \rref{CEmb} implies (under the same assumptions on $j$ and $r$)
\beq (C^j(\Om))' \cemb (\HH^r)' = \HH^{-r} ~. \label{CEmbDu} \feq
Now, let $j \in \naturali$ and $\bx \in \Om$; the prescription
\beq \la \de_\bx , f \ra := f(\bx) \feq
makes sense for all $f \in C^j(\Om)$ and allows to interpret the Dirac delta function
$\de_\bx$ as an element of the dual space $(C^j(\Om))'$. Moreover, it  can be shown
that the mapping $\delta: \bx \in \Om \mapsto \de_\bx \in (C^j(\Om))'$ is itself of class $C^j$
({\footnote{For example, setting $j = 1$, one finds that the map $\bx \mapsto \de_\bx
\in (C^1(\Om))'$ is $C^1$ with derivatives $(\partial_i \de)_\bx$ such that
$$ \la (\partial_i \de)_\bx , f \ra = \partial_i  f(\bx) \qquad (i \in \{1,...,d\}) $$
for all $f \in C^1(\Om)$\,.}}); of course, this fact and Eq.s \rref{CEmbDu} imply that
\beq \de : \Om \to \HH^{-r} ~, \quad \bx \mapsto \de_\bx \quad
\mbox{is $C^j$ if $r \in \reali, j \in \naturali$ and $\dd{r > {d \over 2} + j}$}~.
\label{delta} \feq
Finally, let $j \in \naturali$ and suppose that
\beq \barray{c} \mbox{$\BB : \HH^{-(d/2 + j_2 + \epsii)} \to \HH^{d/2 + j_1 + \epsii}$
is linear and continuous} \\
\mbox{for some $\epsii > 0$ and all $j_1, j_2 \in \naturali$ with $j_1 + j_2 \leqs j$}~;
\farray \label{BBC} \feq
we claim that, in this case,
\beq \barray{c} \mbox{the kernel $\Om \times \Om \to \complessi$,
$(\bx,\by) \mapsto \BB(\bx,\by) := \la \de_\bx |\BB\,\de_{\by}\raL$} \\
\mbox{is (well defined and) of class $C^j(\Om \times \Om)$}\,. \farray
\label{KerCj} \feq
In fact, each derivative $\partial\BB(~,~)$ of order $\leqs j$ involves
(in an arbitrary order) $\al_i$ operations of derivation with respect to
$x^i$ and $\be_i$ operations of derivation with respect to $y^i$ ($i=1,...,d$)
where $j_1 := \al_1 + ... + \al_d$ and $j_2 := \be_1 + ... + \be_d$ are such that
$j_1 + j_2 \leqs j$; any such derivative exists and is continuous on $\Om \times
\Om$, with the explicit expression $\partial \BB(\bx,\by) = \la (\partial^\al \de)_\bx
|\BB\,(\partial^\be \de)_{\by}\raL$
({\footnote{note that, due to \rref{delta}, the map $\by \mapsto (\partial^\be\de)_{\by}$
is continuous from $\Om$ to $\HH^{-(d/2 + j_2 + \epsii)}$; due to \rref{BBC},
$\by \mapsto \BB (\partial^\beta \de)_{\by}$ is continuous from $\Om$ to $\HH^{d/2 + j_1 + \epsii}$;
finally, due to \rref{delta} the map $\bx \mapsto (\partial^\al \de)_\bx$ is
continuous from $\Om$ to $\HH^{-(d/2 + j_2 + \epsii)}$, and the sesquilinear form
$\la~|~\raL$ is continuous on $\HH^{-(d/2 + j_1 + \epsii)} \times \HH^{d/2 + j_1 + \epsii}$.}}).
\saltino
\textsl{Applications to the Dirichlet, heat and cylinder kernels: regularity results.}
Let $s \in \complessi$, $j \in \naturali$; if $\Re s > d/2 + j/2$, using
Eq. \rref{AAsc} one easily infers that the operator $\BB := \AA^{-s}$
fulfills the condition \rref{BBC} (with $\epsii =
\Re s - d/2 - j/2$); thus \rref{KerCj} holds. In conclusion, we have the
following result for the Dirichlet kernel $\Dir_s(\bx,\by) := \la \de_\bx|\AA^{-s}\de_\by\raL$:
\beq \Dir_s \in C^j(\Om \times \Om) \qquad \mbox{for $s \in \complessi$,
$j \in \naturali$ and $\dd{\Re s > {d \over 2} + {j \over 2}}$} ~. \label{B15} \feq
Similarly, for any $\t > 0$ and any $j \in \naturali$, due to \rref{Expsc} the
operators $\BB = e^{-\t\AA}$ or $\BB = e^{-\t\sqrt{\AA}}$ fulfill
the condition \rref{BBC}, implying Eq \rref{KerCj}. So, the heat and cylinder kernels
$K(\t\,;\bx,\by) := \la \de_\bx|e^{-\t\AA}\de_\by\raL$, $T(\t\,;\bx,\by) :=
\la \de_\bx|e^{-\t\sqrt{\AA}}\de_\by\raL$ are of class $C^j$ in $\bx,\by$, for
all $j \in \naturali$; in conclusion, these kernels are $C^\infty$ in $\bx,\by$:
\beq K(\t\,;~,~), T(\t \,;~,~) \in C^\infty{(\Om \times \Om)} \qquad
\mbox{for each $\t > 0$} ~. \label{B16} \feq
\textbf{Results for case \rref{iii}.} We now consider the case described by
the cited equation. Thus $\AA = - \Delta + V$ on bounded domain $\Om$
with $C^\infty$ boundary $\partial\Om$, on which Dirichlet conditions are imposed;
the potential $V$ is in $C^\infty(\ov{\Om})$ and $V(\bx) \geqs 0$ for
all $\bx \in \ov{\Om}$ (recall that, due to these assumptions, $\AA$ is selfadjoint
and strictly positive).
\saltino
\textsl{Smoothness up to the boundary.}
In the setting described above, one can strengthen the results discussed in the
previous paragraph of this appendix replacing systematically the space $C^j(\Om)$
with $C^j(\ov{\Om})$; this space consists of the functions which are continuous with
their derivatives up to order $j$ on the closure $\ov{\Om} = \Om \cup \partial\Om$,
and it is equipped with the norm
\beq \|f\|_{C^j} := \max_{|\al| \leqs j,\,\bx \in \ov{\Om}}|\partial^\al\!f(\bx)| ~. \feq
Similarly, the space $C^j(\Om \times \Om)$ can be replaced with $C^j(\ov{\Om}\times\ov{\Om})$,
endowed with the norm
\beq \|f\|_{C^j} := \max_{|\al| + |\be| \leqs j,\;\bx,\by \in \ov{\Om}}
|\partial^\al_\bx \partial^\be_\by f(\bx,\by)| ~. \label{normajj} \feq
In the present case Eq. \rref{CEmb} has the stronger version
\beq \HH^r \cemb C^j(\overline{\Om}) \qquad \mbox{for $j \in \naturali$, $r\in\reali$
with $\dd{r > {d \over 2} + j}$} ~; \label{CEmbb} \feq
similarly, Eq. \rref{KerCj} (with the assumptions \rref{BBC} on $\BB$) holds with
$C^j(\Om \times \Om)$ replaced by $C^j(\ov{\Om}\times\ov{\Om})$
({\footnote{In order to derive Eq. \rref{CEmbb} one proceeds similary to
footnote \ref{FootImb} using the embeddings $\HH^r \cemb H^r(\Om)$ and $H^r(\Om)
\cemb C^j(\ov{\Om})$; these follow from Theorem 3 on p. 155 of \cite{Mikh} and some
standard interpolation theory (see, e.g., \cite{Int2,Int3}).}});
due to these facts, Eq.s \rref{B15} and \rref{B16} for
the Dirichlet, heat and cylinder kernels hold
in the stronger versions
\beq \Dir_s \in C^j(\ov{\Om} \times \ov{\Om}) \qquad \mbox{for $s \in \complessi$,
$j \in \naturali$ with $\dd{\Re s > {d \over 2} + {j \over 2}}$} ~; \label{B155} \feq
\beq K(\t\,;~,~), T(\t \,;~,~) \in C^\infty{(\ov{\Om} \times \ov{\Om})} \qquad
\mbox{for each $\t > 0$} ~. \label{B166} \feq
\saltino
\textsl{Estimates for the eigenfunctions and eigenvalues of $\AA$.}
Due to the boundedness of $\Om$, $\AA$ has purely point spectrum; in fact,
one can build for this operator a complete orthonormal set of proper eigenfunctions
$(\Fk)_{k \in \KK}$ with eigenvalues $(\om^2_k)_{k \in \KK}$, where $\KK = \{1,2,3,...\}$
and the labels are chosen so that $0 < \om_1 \leqs \om_2 \leqs \om_3 \leqs ...$
(with the possibility that some of these inequalities are equalities, to deal with
the case of degenerate eigenvalues). \parn
Let us discuss the smoothness properties of the eigenfunctions and derive some
norm bounds for them. Clearly, for each $k \in \{1,2,3,..\}$ and $r \in \reali$,
we have $\Fk \in \HH^r$ and $\|\Fk\|_r = \|\AA^{r/2}\Fk\| = \|\om^r_k\Fk \| = \om^r_k$\,.
If $r > {d \over 2}+j$ these facts and the imbedding \rref{CEmbb} imply
$\Fk \in C^j(\ov{\Om})$; since this holds for each $j \in \naturali$, we have
\beq \Fk \in C^\infty(\ov{\Om}) ~. \feq
To go on, let us note that \rref{CEmbb} means the following: for each $j \in \naturali$
and $r > j + {d \over 2}$, there exists a constant $\Lambda_{j, r} \in (0,+\infty)$ such that
\beq \|f\|_{C^j} \leqs \Lambda_{j, r} \, \| f\|_{r}
\qquad \mbox{for all $f \in \HH^r$}\,. \label{due} \feq
This statement with $f = \Fk$ gives $\|\Fk\|_{C^j} \leqs \Lambda_{j r} \om_k^{r}$
for $j \in \naturali$, $r > j + {d \over 2}$ or, equivalently,
\beq \|\Fk\|_{C^j} \leqs \Lambda_{j \epsii}\;\om_k^{j + d/2 + \epsii} \quad
\mbox{for all $j \in \naturali$, $\epsii > 0$} \label{esteig} \feq
(where $\Lambda_{j \epsii}$ stands for $\Lambda_{j, j + d/2 + \epsii}$).
These results should be kept in mind in the sequel, together with the already mentioned
Weyl asymptotics \rref{weyl}
$$ \om_k \sim C \,k^{1/d} \qquad \mbox{for $k \to + \infty$} ~, $$
where $C := 2 \sqrt{\pi}\, \Ga(d/2\!+\!1)^{1/d}\,\mbox{Vol}(\Om)^{-1/d}$
({\footnote{Concerning \rref{weyl}, we have already given references
\cite{Shu} \cite{Mikh}.
Eq. \rref{esteig} is known as well in the literature, see \cite{Shis1,Shis2,Yak};
here we have proposed an alternative derivation of this result just because it arose
naturally from the general framework of the present appendix.}}).
\saltino
\textsl{On the eigenfunction expansion for the Dirichlet kernel.}
In the case we are considering, the eigenfunction expansion \rref{eqkerdi}
for the Dirichlet kernel takes the form
\beq \Dir_s(\bx,\by) = \sum_{k=1}^{+\infty} {1 \over \om_k^{2s}}\;\Fk(\bx)\Fkc(\by)~;
\label{takeform} \feq
hereafter we discuss the absolute convergence of this expansion with respect
to the norm \rref{normajj} of $C^j(\ov{\Om} \times \ov{\Om})$, showing that
\beq \sum_{k=1}^{+\infty} {1 \over \,|\om_k^{2s}|\,}\;\|\Fk(\cdot)\Fkc(\cdot\cdot) \|_{C^j}
< + \infty \qquad \mbox{if $\Re s > d + {j \over 2}$} ~. \label{absconv} \feq
To this purpose, we first note that
\begin{equation}\begin{split}
& \sum_{k=1}^{+\infty} {1 \over \,|\om_k^{2s}|\,}\;\|\Fk(\cdot)\Fkc(\cdot\cdot) \|_{C^j}
= \sum_{k=1}^{+\infty} {1 \over \,\om_k^{2 \Re s}}\;\max_{|\al| + |\be| \leqs j}
\max_{\bx \in \ov{\Om}, \by \in \ov{\Om}} |\partial^\al \Fk(\bx) \partial^\be \Fkc(\by)|
\leqs \label{sos1} \\
& \hspace{3.5cm} \leqs \sum_{k = 1}^{+\infty}{1 \over \om_k^{2\Re s}}\;
\max_{j_1 + j_2 \leqs j} \|\Fk\|_{j_1} \|\Fk\|_{j_2} ~.
\end{split}\end{equation}
Let $k \in \{1,2,3,...\}$, $j_1 + j_2 \leqs j$ and $\epsii > 0$.
Due to the estimate \rref{esteig} we have
$$ \|\Fk\|_{j_1} \| \Fk \|_{j_2} \leqs \Lambda_{j_1 \epsii}\,\Lambda_{j_2 \epsii}\;
\om_k^{d + j_1 + j_2 + 2 \epsii} ~; $$
but (recalling that $\om_k \geqs \om_1$)
$$ \om_k^{j_1 + j_2}  = \left({\om_k \over \om_1}\right)^{\!j_1 + j_2} \om_1^{j_1 + j_2}
\leqs \left({\om_k \over \om_1}\right)^{\!j} \om_1^{j_1 + j_2} =
{\om_k^{j} \over \om_1^{j - j_1 - j_2}} ~, $$
whence
\beq \|\Fk\|_{j_1} \|\Fk\|_{j_2} \leqs {\Lambda_{j_1 \epsii}\,\Lambda_{j_2 \epsii} \over
\om_1^{j - j_1 - j_2}}\;\om_k^{d + j + 2 \epsii} ~. \label{sos2} \feq
Inserting Eq. \rref{sos2} into Eq. \rref{sos1}  we obtain the following, for all $\epsii > 0$:
\beq \sum_{k=1}^{+\infty} {1 \over \,|\om_k^{2s}|\,}\;\|\Fk(\cdot)\Fkc(\cdot\cdot)\|_{C^j}
\leqs \l(\max_{j_1 + j_2 \leqs j} {\Lambda_{j_1 \epsii}\,\Lambda_{j_2 \epsii} \over
\om_1^{j - j_1 - j_2}}\r) \sum_{k = 1}^{+\infty} {1 \over \om_k^{2\Re s - d - j - 2 \epsii}} ~.
\label{sos3} \feq
From here to the end of the paragraph we assume
\beq \Re s > d + {j \over 2} ~; \label{assums} \feq
hereafter we show that the series on the left-hand side of Eq. \rref{sos3} converges
for a suitable $\epsii > 0$, a fact yielding the thesis \rref{absconv}. Indeed, due to
Eq. \rref{assums} there is $\epsii > 0$ such that
\beq \Re s = d + {j \over 2} + \l({d \over 2}+1\r) \epsii ~; \feq
expressing $\Re s$ in this way, and using the Weyl asymptotics \rref{weyl} we get
\beq {1 \over \om_k^{2\Re s - d - j - 2 \epsii}} = {1 \over \om_k^{(1 + \epsii) d}} \sim
{1 \over C^{(1 + \epsii) d} k^{1 + \epsii}} \qquad \mbox{for $k \to + \infty$}~, \feq
which implies convergence for the series on the left-hand side of Eq. \rref{sos3}.
\saltino
\textsl{On the eigenfunction expansions for the heat and cylinder kernels.}
In the type of configuration under analysis, the expansions \rref{eqheat} and
\rref{eqcyl} can be re-expressed, respectively, as
\beq K(\t\,;\bx,\by) = \sum_{k=1}^{+\infty} e^{-\t \om^2_k}\,\Fk(\bx)\Fkc(\by) ~,
\label{takeHeat} \feq
\beq T(\t\,;\bx,\by) = \sum_{k=1}^{+\infty} e^{-\t \om_k}\,\Fk(\bx)\Fkc(\by) ~.
\label{takeCyl} \feq
Using considerations similar to the ones of paragraph c), one shows the absolute
covergence of these expasions in the norm \rref{normajj} of $C^j(\ov{\Om} \times \ov{\Om})$,
for all $\t > 0$ and for each $j \in \naturali$:
\beq \sum_{k=1}^{+\infty} {e^{-\t \om^2_k}} \| \Fk(\cdot ) \Fkc(\cdot \cdot) \|_{C^j}
< + \infty ~, \qquad \sum_{k=1}^{+\infty} {e^{-\t \om_k}} \| \Fk(\cdot ) \Fkc(\cdot \cdot) \|_{C^j}
< + \infty ~. \label{absconvheat} \feq
\saltino
\textsl{Another result.} In subsection \ref{slKT} it is stated
that, under the assumptions \rref{iii} for $\AA$ and considering
a complete orthonormal set $(\Fk)_{k=1,2,3,...}$ with the usual
features for this case, the function
\beq \hat{T}(\t\,;\bx,\by) := \sum_{k=1}^{+\infty} e^{-\om_k \t}
|\Fk(\bx)| |\Fk(\by)| \feq
admits a uniform bound
\beq \hat{T}(\t\,;\bx,\by) \leqs \check{T}(\t) < + \infty \qquad
\mbox{for all $\bx,\by \in \ov{\Om}$ and $\t > 0$} ~. \label{ub} \feq
The proof of this statements starts from the inequality \rref{esteig} with $j=0$
and any $\epsii >0$, this implies $|\Fk(\bx)|, |\Fk(\by)| \leqs
\Lambda_{0 \epsii}\;\om_k^{d/2 + \epsii}$, whence
\beq \hat{T}(\t\,;\bx,\by) \leqs \check{T}(\t)~\mbox{for all $\bx,\by \in \ov{\Om}$
and $\t > 0$}~, \quad \check{T}(\t) := \Lambda^2_{0 \epsii}
\sum_{k=1}^{+\infty} e^{-\om_k \t} \om_k^{d + 2 \epsii}\,. \feq
Finally, the Weyl estimates \rref{weyl} ensure $\check{T}(\t) < + \infty$ for
each $\t >0$.
\section{Appendix. A $\boma{(d\!+\!1)}$-dimensional Green function and its relation with the cylinder kernel}
\label{appeg}
Let $\Om \subset \reali^d$ be an open set, and let $\AA$ be a strictly
positive selfadjoint operator in $L^2(\Om)$ (keeping into account suitable
boundary conditions on $\partial\Om$). We consider the cylinder kernel
$T(\t\,;\bx, \by) := (e^{-\t \sqrt{\AA}})(\bx,\by)$
($\bx,\by \in \Om$, $\t \in (0,+\infty)$). \parn
The aim of the present appendix is to illustrate the fact mentioned in
subsection \ref{ment}, namely, the possibility to relate $T$ to a $(d+1)$
dimensional Green function. To this purpose we consider in $\reali^{d+1}$
the domain $\OO := (0,+\infty) \times \Om \ni (\t,\bx)$ and the Hilbert
space $L^2(\OO)$; we introduce therein the operator
\beq \PP := - \partial_{\t\t} + \AA ~, \feq
with suitable boundary conditions on $\partial\OO := (\{0\}\!\times\!\Om)
\cup((0,+\infty)\!\times\!\partial\Om)$\,. More precisely, we assume
Dirichlet boundary conditions on $\{0\} \times \Om$ and the previously
given boundary conditions for $\AA$ on $(0,+\infty) \times \partial\Om$.
The operator $\PP$ is selfadjoint; in the sequel we will prove that it
is strictly positive. \parn
Let us introduce the Green function
\beq G(\t, \bx; \t', \by) := \PP^{-1}((\t,\bx),(\t',\by))\equiv
\la \de_\t \, \de_\bx | \PP^{-1}\de_{\t'} \, \de_\by \ra ~; \label{gren} \feq
this is characterized by the equation
\beq (-\partial_{\t\t} + \AA_\bx)G(\t, \bx; \t', \by) = \de(\t - \t') \de(\bx - \by) ~, \feq
and by the boundary conditions prescribed on $\partial \OO$.
We claim that the cylinder kernel $T$ is related to $G$ by
\beq \l.\partial_{\t'} G(\t,\bx;\t',\by) \r|_{\t'=0}
= T(\t\,; \bx, \by) ~. \label{gt} \feq
To prove this (and the previous statement on the strict positivity of
$\PP$), let $(\Fk)_{k\in \KK}$ be a complete orthonormal system of
eigenfunctions of $\AA$ with corresponding eigenvalues $(\om_k^2)_{k \in \KK}$;
clearly, the functions $\t \in (0,+\infty) \mapsto \sqrt{\!{2 \over \pi}}\,
\sin(\lam \t)$ ($\lam \in (0,+\infty)$) are a complete orthonormal system
in $L^2((0,+\infty))$ and are eigenfunctions of $-\partial_{\t \t}$
vanishing for $\t =0$. These facts ensure that the family of functions
\beq Y_{(\lam,k)}(\t,\bx) := \sqrt{\!{2 \over \pi}}\,\sin(\lam \t)\,\Fk(\bx)
\qquad \mbox{for $(\lam,k) \in (0,+\infty) \times \KK$} \label{corton} \feq
is a complete orthonormal system of eigenfunctions of $\PP$, with
\beq \PP\,Y_{(\lam,k)} =  (\lam^2 + \om^2_k)\,Y_{(\lam,k)} ~. \feq
We recall that we are assuming $\om^2_k \geqs \eps^2$ for some $\eps > 0$;
the eigenvalues $(\lam^2 + \om^2_k)$ also have $\eps^2$ as a lower bound,
so $\PP$ is strictly positive. \parn
The Green function of Eq. \rref{gren} can be expressed via the equation
\beq G(\t, \bx; \t', \by) = \!\int_{(0,+\infty) \times \KK}
{d \lam \; d k \over \lam^2 \!+\! \om_k^2}
\l(\!\sqrt{2 \over \pi} \, \sin(\lam \t)\Fk(\bx)\!\r)\!\!
\l(\!\sqrt{2 \over \pi} \, \sin(\lam \t')\Fkc(\by)\!\r), \feq
which, evaluating explicitly the integral in $\lam$
({\footnote{Just observe that, by symmetry arguments and the residue
theorem, we have
$$ \int_0^{+\infty} d\lam\;{\sin(\lam \t)\sin(\lam \t') \over \lam^2 + \om_k^2}
= {1 \over 2} \int_{-\infty}^{+\infty} d\lam \;{\sin(\lam \t) \sin(\lam \t')
\over \lam^2 + \om_k^2} = {\pi \over 4 \om_k} \; \Big( e^{-\om_k |\t-\t'|}
- e^{-\om_k (\t+\t')} \Big) ~. $$}}),
reduces to
\beq G(\t, \bx; \t', \by) = \int_\KK d k \;{e^{-\om_k |\t-\t'|}
- e^{-\om_k (\t+\t')}\over 2 \om_k}\,\Fk(\bx)\,\Fkc(\by) ~. \label{Gfk}\feq
To go on, let us differentiate both sides of Eq. \rref{Gfk} with respect
to $\t'$; this gives
\beq \partial_{\t'} G(\t,\bx;\t',\by) = \int_\KK d k
\l[{\mbox{sgn}(\t\!-\!\t') \over 2}\; e^{-\om_k |\t-\t'|} +
{1 \over 2}\;e^{-\om_k (\t+\t')} \r]\! \Fk(\bx) \Fkc(\by) ~. \feq
Setting $\t'= 0$ (and recalling that $\t>0$), the last equation yields
\beq \l.\partial_{\t'} G(\t,\bx;\t',\by) \r|_{\t'=0} =
\int_\KK d k \; e^{-\om_k \t}\,\Fk(\bx) \Fkc(\by) ~. \feq
The right-hand side of the above equality is just the representation
of the cylinder kernel $T$ given by Eq. \eqref{eqcyl}; thus we have
proved Eq. \rref{gt}. \salto
\textbf{An example.} Hereafter, the approach based on \rref{gt} is used to
derive the expression \rref{eq2} for the cylinder kernel $T$ in the case where
\beq \Om := \reali^d~, \qquad \AA := - \Delta ~. \feq
The associated $(d+1)$ dimensional domain and the operator $\PP$ are
\beq \OO := (0,+\infty) \times \reali^d ~, \qquad
\PP := -\partial_{\t \t} - \Delta = - \Delta_{d+1} ~, \feq
with Dirichlet boundary conditions on $\partial \OO$, which coincides
with the hyperplane $\{0\} \times \reali^d$; of course, in the above
$\Delta_{d+1}$ indicates the $(d+1)$-dimensional Laplacian. The Green
function $\GG$ of $-\Delta_{d+1}$ on the half-space $\OO$ can be obtained
by the familiar method of images from the Green function $\GG_0$ of
$-\Delta_{d+1}$ on the full space $\reali^{d+1}$; thus
\beq \GG(\t, \bx; \t', \by) = \GG_0(\t, \bx; \t',\by)
- \GG_0(\t, \bx; -\t', \by)~, \label{gg} \feq
\beq \GG_0(\t,\bx; \t', \by) := \l\{\! \barray{cc}
{1 \over 2 \pi} \ln((\t - \t')^2 + (\bx - \by)^2) & \mbox{if $d=1$}, \\
\dd{\Ga({d + 1 \over 2}) \over (d-1) 2 \pi^{d + 1 \over 2}
((\t - \t')^2 + |\bx - \by|^2)^{d-1 \over 2}} & \mbox{if $d \geqs 2$.}
\farray \r. \label{gg0} \feq
Inserting Eq.s \rref{gg} \rref{gg0} into Eq. \rref{gt} we obtain for
the cylinder kernel $T$ the expression \rref{eq2}
$$ T(\t\,;\bx,\by) = {\Ga({d+1 \over 2})\,\t \over
\pi^{d+1 \over 2}(\t^2 + |\bx-\by|^2)^{{d+1 \over 2}}} $$
(in all dimensions, including $d=1$).
\section{Appendix. Derivation of Eq. \rref{HankCont}}
\label{apperes}
Let $\t \mapsto h(\t)$ be a complex-valued function, analytic in
a neighborhood of $[0,+\infty)$ and exponentially vanishing for
$\Re \t \to + \infty$. For any given $s \in \complessi$ with
$\Re s > 0$\,, consider the integral
\beq I(s,h) := \int_\Hank d\t \; \t^{s - 1}\, h(\t) ~, \label{Irho}\feq
where $\Hank$ is a Hankel contour (see below Eq. \rref{HankIden} and
Fig. \ref{fig:HankFig} on page \pageref{fig:HankFig} for the description
of this path); the complex power $\t^{s-1}$ in the above equation is
defined following Eq.s (\ref{power}) (\ref{arg}). Hor any $\de > 0$,
$\Hank$ is homotopic to the path $\Hank_\de$ described as follows:
\beq \Hank_\de = \Hank_\de^{+} \cup \Hank_\de^0 \cup \Hank_\de^{-} ~,
\qquad \mbox{with} \label{ISum}\feq
$$ \Hank_\de^{\pm} := \{\t \in \complessi ~|~
\t = \vt \pm i \de,~ \vt \in [0,+\infty) \} ~, $$
$$ \Hank_\de^0 := \{\t \in \complessi ~|~
\t = \de\,e^{i\te},~ \te \in (\pi/2,3\pi/2)\} ~. $$
Due to this remark and to the analyticity of $h$ we can replace $\Hank$
with $\Hank_{\delta}$ in Eq. \rref{Irho}; so, for any $\de > 0$ we have
\beq I(s,h) = I^{+}_\de(s,h) + I^0_\de(s,h) + I^{-}_\de(s,h) ~, \label{noet} \feq
$$ I_\de^{\pm}(s,h) := \mp\!\int_0^{+\infty}\!\!\!d\vt\,(\vt \pm i \de)^{s-1}
h(\vt \pm i \de) ~, \quad\! I_\de^{0}(s,h) := i\!\int_{\pi/2}^{3 \pi/2}\!\!
d\te\,(\de\,e^{i\te})^s h(\de\,e^{i\te}) ~. $$
We are now going to consider the limit $\de \to 0^+$. Notice that in this
limit $(\vt + i\de)^{s - 1} \to \vt^{s - 1}$ while $(\vt - i\de)^{s -1} \to
e^{2 i \pi(s -1)} \vt^{s - 1} = e^{2 i \pi s} \vt^{s - 1}$; moreover,
$h(\vt \pm i\de) \to h(\vt)$. Due to these results, we easily infer
\beq \lim_{\de \to 0^+} I_\de^{+}(s,h) =
- \int_0^{+\infty} d\vt\; \vt^{s - 1} h(\vt) ~, \feq
\beq \lim_{\de \to 0^+} I_\de^{-}(s,h) = e^{2i \pi s}
\int_0^{+\infty} d\vt\;\vt^{s - 1} h(\vt) ~. \feq
Passing to the integral $I_\de^{0}(s,h)$, noting that $|h(\de\,e^{i\te})|
\leqs C$ for small $\de$ and recalling that $\Re s > 0$ by hypothesis,
we obtain
\beq |I_\de^{0}(s,h)| \leqs C \de^{\Re s} \int_{\pi/2}^{3 \pi/2}
d\te\; e^{-(\Im s)\te} \to 0 \qquad \mbox{for $\de \to 0^+$} ~. \feq
Summing up, in the limit $\de \to 0^+$ one obtains from Eq. \rref{noet} that
\beq I(s,h) = (e^{2 i \pi s} - 1) \int_0^{+\infty} d\vt\;\vt^{s - 1}h(\vt) ~; \feq
noting that $e^{2i\pi s} - 1 = 2i e^{i\pi s} \sin(\pi s)$, the above relation yields
\beq I(s,h) = 2 i e^{i\pi s} \sin(\pi s)
\int_0^{+\infty} d\vt\;\vt^{s - 1} h(\vt) ~, \feq
which is equivalent to Eq. \rref{HankCont} for $\Re s >0$, $s \notin \{1,2,3,...\}$\,.
For more information concerning the Mellin transform and its contour integral
representations see, e.g., \cite{Mell,Mell2}.
\section{Appendix. Derivation of Eq.s (\ref{dirhats}-\ref{DirGDder})}
\label{appeslab}
We refer to the framework of subsection \ref{slabSubsec} about the slab
$\Om = \Om_1 \times \reali^{d_2}$; we retain all the assumptions and
notations of the cited subsection. In particular, $(\Fkd)_{k_1 \in \KK_1}$
is a complete orthonormal system of eigenfunctions of $\AA_1$ with related
eigenvalues $\oo_{k_1}^2$. A complete orthonormal set of eigenfunctions
$(\Fk)_{k \in \KK}$ with eigenvalues $(\om_k^2)_{k \in \KK}$ for the
operator $\AA$ in $L^2(\Om)$ is given by
\begin{equation}\begin{split}
& \Fk(\bx) = \Fkd(\bx_1)\,{e^{i\bk_2 \cdot \bx_2} \over (2\pi)^{d_2/2}}~,
\qquad \om_k^2 = \oo_{k_1}^2 + |\bk_2|^2 \, \\
& \hspace{0.5cm} \mbox{for $\bx = (\bx_1,\bx_2)$ and
$k = (k_1,\bk_2)\!\in\!\KK_1\!\times\!\reali^{d_2}$}
\end{split}\end{equation}
In the present case, the eigenfunction expansion \rref{eqkerdi} of the
Dirichlet kernel at any two points $\bx = (\bx_1,\bx_2)$ and $\by = (\by_1,\by_2)$
can be re-expressed as follows:
\begin{equation}\begin{split}
& \Dir_s(\bx_1,\bx_2;\by_1,\by_2) = \int_{\KK_1 \times \reali^{d_2}}
{dk_1\;d\bk_2 \over (\oo_{k_1}^2\!+\!|\bk_2|^2)^s}\;\Fkd(\bx_1)\,\ov{\Fkd}(\by_1) \;
{e^{i\bk_2\cdot (\bx_2 - \by_2)} \over (2\pi)^{d_2}} = \\
&\hspace{1.3cm} = \int_{\KK_1 \times \reali^{d_2}} {dk_1\;d \bh \over \oo_{k_1}^{2 s - d_2}
(|\bh|^2\!+\!1)^s } \;\Fkd(\bx_1) \, \ov{\Fkd}(\by_1) \;
{e^{i \oo_{k_1} \bh \cdot (\bx_2 - \by_2)} \over (2\pi)^{d_2}} \label{Dird1}
\end{split}\end{equation}
where, in the last passage, we performed the change of variables
$\bk_2 = \oo_{k_1} \bh$. On the other hand, it is known that, for any
$\bz \in \reali^{d_2}$,
\beq \int_{\reali^{d_2}}\!{d\bh \over (2\pi)^{d_2}}\;
{e^{i \bh \cdot \bz} \over (|\bh|^2\!+\!1)^s} =
{|\bz|^{s - {d_2\over 2}} \over (2\pi)^{d_2/2}\,2^{s-1} \Ga(s)}\;
K_{s-{d_2 \over 2}}(|\bz|) \quad \mbox{if $s\!\in\!\complessi$,
$\Re s\!>\!{d_2 \over 2}$} ~, \feq
with $K_\nu$ denoting the modified Bessel function of the second kind of
order $\nu\in\complessi$ (see, e.g., \cite{AroSmi,NIST,Wat}). Thus
\begin{equation}\begin{split}
& \hspace{4.5cm} \Dir_s(\bx_1, \bx_2; \by_1, \by_2) = \\
& \int_{\KK_1} {dk_1 \over \oo_{k_1}^{2s-d_1}}\;\Fkd(\bx_1)\,\ov{\Fkd}(\by_1)\;
{(\oo_{k_1} |\bx_2\!-\!\by_2|)^{s - {d_2\over 2}} \over
(2\pi)^{d_2/2}\,2^{s-1} \Ga(s)}\; K_{s-{d_2 \over 2}}(\oo_{k_1}|\bx_2\!-\!\by_2|)
~. \label{DirGD1}
\end{split}\end{equation}
For the sake of brevity, for any $\nu \in \complessi$, we put
\beq \GD_\nu:(0,+\infty) \to \complessi ~, \qquad
\vi \mapsto \GD_\nu(\vi) := \vi^{\nu/2} K_\nu(\sqrt{\vi}) ~; \feq
due to the asymptotic behaviour of the Bessel function $K_\nu$ near
zero (see \cite{NIST}, p.252, Eq.10.30.2), for $\Re \nu > 0$ this
function has continuous extention to $\vi = 0$, given by
\beq \GD_\nu(0) = 2^{\nu - 1} \Ga(\nu) ~. \label{Gnu0} \feq
To proceed, note that Eq. \rref{DirGD1} can then be rephrased in terms of the
function $\GD_\nu$ as follows:
\begin{equation}\begin{split}
& \hspace{2.7cm} \Dir_s(\bx_1, \bx_2; \by_1, \by_2) =
\hat{\Dir}_{\!s}(\bx_1,\by_1;|\bx_2\!-\!\by_2|^2) ~ , \\
& \hat{\Dir}_s (\bx_1, \by_1;q) = {2^{1-s} \over (2\pi)^{d_2/2}\Ga(s)}
\int_{\KK_1} {dk_1 \over \oo_{k_1}^{2s-d_2}}\;\Fkd(\bx_1)\,\ov{\Fkd}(\by_1)\;
\GD_{\!s - {d_1 \over 2}}(\oo_{k_1}^2\,q) ~. \label{DirGD2}
\end{split}\end{equation}
This proves Eq. \rref{dirhats}, also giving an explicit expression for
the function $\hat{\Dir}_s$. To proceed note that, due to the well-known
facts on the derivatives of the Bessel function $K_\nu$ for any $\nu \in
\complessi$ (see \cite{NIST}, p.252, Eq.10.29.4), we have
\begin{equation}\begin{split}
{d\GD_\nu \over d\vi}\,(\vi) & = \l.{d \over d\zi}\Big(\zi^\nu K_\nu(\zi)\Big)
\r|_{\zi=\sqrt{\vi}}\; {1 \over 2\sqrt{\vi}} =
\Big(\!\!-\zi^\nu K_{\nu-1}(\zi)\Big)_{\!\zi = \sqrt{\vi}}\;{1 \over 2\sqrt{\vi}} = \\
& \qquad = -\,{1 \over 2}\;\vi^{\nu-1 \over 2} K_{\nu-1}(\sqrt{\vi}) =
-\,{1 \over 2}\;\GD_{\nu-1}(\vi) ~.
\end{split}\end{equation}
Using the above identity it can be proved by induction that
\beq {d^n \GD_\nu \over d\vi^n}\,(\vi) = \l(\!-\,{1 \over 2}\r)^{\!\!n}
\GD_{\nu-n}(\vi) \qquad \mbox{for $n \in \{1,2,3,...\}$} ~; \feq
this fact, along with Eq. \rref{DirGD2}, implies
\beq {\partial^n\! \hat{\Dir}_s \over \partial q^n}(\bx_1,\by_1;q)\! =
{(-1)^n 2^{1-s-n} \over (2\pi)^{d_2/2}\Ga(s)}\!\int_{\!\KK_1}\!\!
{dk_1 \over \oo_{k_1}^{2s-d_2-2n}}\,\Fkd(\bx_1)\ov{\Fkd}(\by_1)\,
\GD_{\!\!s - {d_2 \over 2}-n}(\oo_{k_1}^2q) \,.\! \label{DirGD3} \feq
Now, recalling Eq. \rref{Gnu0} we conclude that, for any $n \in
\{1,2,3,...\}$ and any $s\in\complessi$ with $\Re s > {d \over 2}\!+\!n$,
\beq {\partial^n \hat{\Dir}_s \over \partial q^n}(\bx_1, \by_1;0) =
{(-1)^n \Ga(s\!-\!{d_2 \over 2}\!-\!n) \over (4\pi)^{d_2/2}\,4^n\,\Ga(s)}
\int_{\!\KK_1} {dk_1 \over \oo_{k_1}^{2s-d_2-2n}}\;
\Fkd(\bx_1)\ov{\Fkd}(\by_1) \,. \label{DirGD4} \feq
Due to a representation analogous to \rref{eqkerdi} holding for the
reduced Dirichlet kernel $\Dir_s^{(1)}$, Eq. \rref{DirGD4} implies
Eq. \rref{DirGDder}. Finally, Eq. \rref{DirGD} is just the case
$n = 0$ of Eq. \rref{DirGDder}.
\section{Appendix. Some results on boundary forces}\label{apperell}
As in the final part of subsections \ref{espli} and \ref{equiv},
we work on a domain $\Om$ with Dirichlet boundary conditions.
\vspace{-0.4cm}
\subsection{Derivation of Eq. \rref{pEeT} for the pressure.}
We start from Eq. \rref{press1dir}, holding for general boundary conditions.
In the Dirichlet case that we are considering, only the terms involving
mixed derivatives (with respect to both $\bx$ and $\by$) of the Dirichlet
kernel yield non-vanishing contributions on the boundary $\partial\Om$
(\footnote{One can easily infer this statement using the eigenfunction
expansion \rref{eqkerdi}.});
thus, for any $\bx\!\in\!\partial\Om$, Eq. \rref{press1dir} reduces to
\begin{equation}
p^{\s}_i(\bx)  = \mm^\s\!\l[\l(\!-\!\l(\!{1 \over 4}\!-\!\xi\!\r)\!
\de_{ij}\,\partial^{x^\ell}\!\partial_{y^\ell}\! + \!\l(\!{1 \over 2}\!-\! \xi\r)\!
\partial_{x^i y^j} \r)\!\Dir_{\s+1 \over 2}(\bx,\by)\r]_{\by=\bx}\!n^j(\bx) ~. \label{pEeTT}
\end{equation}
We now claim that the terms proportional to $\xi$ in Eq. \rref{pEeTT}
vanish, i.e., that
\begin{equation}
\l[\l(\! \de_{ij}\,\partial^{x^\ell}\!\partial_{y^\ell}
- \partial_{x^i y^j}\r) \Dir_{\s+1 \over 2}(\bx,\by)\r]_{\by=\bx}\!n^j(\bx) =0
\qquad \mbox{for all $\bx \in \partial \Om$} ~; \label{vanish}
\end{equation}
this will yield Eq. \rref{pEeT}. Using the eigenfunction expansion \rref{eqkerdi}
for the Dirichlet kernel, we see that Eq. \rref{vanish} holds if we are able
to prove that
\beq \l(\! \de_{i j}\,\partial^\ell \Fk \partial_\ell \Fkc - {1 \over 2}\,
\partial_i \Fk \partial_j \Fkc  - {1 \over 2}\,\partial_j \Fk \partial_i \Fkc\r)\!\!(\bx)\;
n^j(\bx) = 0 \quad\! \mbox{for $k\!\in\!\KK$, $\bx\!\in\!\partial \Om$}\,. \label{direct} \feq
The simplest way to prove Eq. \rref{direct} is to derive the following,
equivalent statement: for all $k \in \KK$ and all (sufficiently smooth)
vector field $\dFI \equiv (\dFI^i) : \partial \Om \to \reali^d$,
\beq \int_{\partial \Om} d a \, \dFI^i\!\l(\! \de_{i j}\,\partial^\ell \Fk \partial_\ell \Fkc
- {1 \over 2}\,\partial_i \Fk \partial_j \Fkc - {1 \over 2}\,\partial_j \Fk \partial_i \Fkc\r)
\!n^j = 0 ~. \label{toske} \feq
Let us sketch a derivation of Eq. \rref{toske}, for given $k \in \KK$
and $\dFI$ $: \partial \Om \to \reali^d$. To this purpose we consider
a smooth extension of $\dFI$ to a vector field $\boma{\dFI} :
\partial \Om \cup \Om \to \reali^d$ and fix the attention on the integral
\beq {1 \over 2} \int_\Om d \bx\;(\partial^j \partial_j)(\partial_i \dFI^i)\,|\Fk|^2 =
{1 \over 2} \int_\Om d\bx\; \partial_i(\partial^j \partial_j \dFI^i)\,|\Fk|^2
\label{f5} \feq
(note that $\partial^j \partial_j=\Delta$). We re-express both
sides in the above identity integrating by parts with respect to all the
derivatives appearing therein ({\footnote{See the footnote on page \pageref{here}.}),
considering them in the two orders proposed in the two sides; some of
the boundary terms arising in this way vanish since $\Fk$ is zero on
$\partial \Om$\,. The difference between the two expressions thus
obtained, which is obviously zero, is found to coincide with the
left-hand side of Eq. \rref{toske}.
\vspace{-0.4cm}
\subsection{Derivation of Eq. \rref{eqpres}.} Let us stick to the framework
of subsection \ref{equiv} in which the domain $\Om$ is bounded, and Dirichlet
boundary conditions are prescribed; the
operator $\AA = -\Delta + V$, acting in
$L^2(\Om)$, has a complete orthonormal system of eigenfunctions $\Fk$ with
eigenvlaues $\om^2_k$, labelled by a countable set $\KK$. We consider, for
small $\ee >0$, a deformation of the domain $\Om$ of the form
(\ref{def1}-\ref{def2}), controlled by a vector field $\dFI$ on $\reali^d$.
The operator $\AA_{\ee} := - \Delta + V$ acting in $L^2(\Om_{\ee})$ has a
complete orthonormal system of eigenfunctions $F_{\ee, k}$ with eigenvalues
$\om^2_{\ee, k}$. \parn
In subsection \ref{equiv} we have already considered the regularized bulk
energy corresponding to $\Om_{\ee}$; this is (see Eq. \rref{EOmee})
$$ E_\ee^\s = {\mm^\s\! \over 2}\;\sum_{k \in \KK}\;(\om_{\ee,k}^2)^{{1 -\s \over 2}} ~. $$
We now consider the limit $\ee \to 0$, and expand everything to the first
order in $\ee$. Eq. \rref{eqpres} that we want to derive concerns the
expansion in $\ee$ of the bulk energy $E_\ee^\s$; as already mentioned,
Eq. \rref{EOmee} can be used to make contact with the expansion of
the eigenvalues $\om^2_{\ee, k}$, on which we now fix our attention. \parn
The variation of the eigenvalues under a deformation of the spatial domain for
the Dirichlet Laplacian (or similar operators) has been the subject of
classical investigations. Here we refer to the book of Rellich \cite{Rell}
(see Chapter II, at the end of $\S 6$), whose results can be expressed in
this way with our notations:
\beq \om_{\ee,k}^2 = \om_k^2 + \eps\,\oo_k^2 + O(\ee^2) \qquad
\mbox{with} \qquad \oo_k^2 := \la \Fk | \BB \Fk \ra ~, \label{ooke}\feq
where $\BB$ is the selfadjoint operator in $L^2(\Om)$ defined by
\beq \BB f := \partial_i\Big(\!(\partial^i\dFI^j\!+\partial^j\dFI^i)\,\partial_i f\Big)\!
+ \!\l({1 \over 2}\,\Delta \partial_\ell \dFI^\ell\!+ \dFI^\ell \partial_\ell V\r)\!f \label{defBe}\feq
(as matter of fact, \cite{Rell} gives the expression of $\BB$ for $V=0$,
but the extension to a nonzero $V$ is straightforward). Keeping in mind
these facts, we return to Eq. \rref{EOmee} for the regularized bulk energy;
this implies
\beq E_\ee^\s = E^\s + \ee\,\Ee^\s + O(\ee^2)~, \qquad
\Ee^\s := \l({1\!-\!\s \over 2}\r){\mm^\s\! \over 2}\;
\sum_{k \in \KK}\;\om_k^{-1-\s}\,\oo_k^2 ~. \label{defEe}  \feq
To go on, we note that the definition of $\oo_k^2$ in Eq. \rref{ooke},
with $\BB$ as in Eq. \rref{defBe}, yields
\beq \oo_k^2 = \int_\Om d\bx \;\Fkc
\l[\partial_i\Big(\!(\partial^i\dFI^j\!+\partial^j\dFI^i)\,\partial_i \Fk\Big)\! +
\!\Big({1 \over 2}\,\Delta \partial_\ell \dFI^\ell\!+ \dFI^\ell \partial_\ell V\Big) \Fk\r]. \feq
The above result can be re-expressed in terms of surface integrals
on $\partial\Om$ via suitable integrations by parts
({\footnote{See again the footnote \ref{here} on page \pageref{here}.}});
while making these computations, one must use the identity $\Delta \Fk
= (V\!-\!\,\om_k^2)\Fk$ (and its complex conjugate), and recall that
$\Fk$ vanishes on $\partial \Om$. In this way we obtain
\beq \oo_k^2 = \!\int_{\partial\Om}\! da(\bx)\,n^j \dFI^i \Big[\de_{ij}\,
\partial^\ell\Fkc \partial_\ell\Fk -(\partial_i\Fkc\partial_j\Fk\!
+\!\partial_j\Fkc\partial_i\Fk) \Big] \, . \feq
We plug this relation into Eq. \rref{defEe}, exchange the summation
with the integration and use the expansion \rref{eqkerdi}, which in
this case reads $\Dir_s(\bx,\by)\!=\!\sum_{k \in \KK}{1 \over \om_k^{2s}}\,
\Fk(\bx)\Fkc(\by)$; in this way we infer
\begin{equation}\begin{split}
\Ee^\s & = -(1\!-\!\s)\,\mm^\s\!\int_{\partial\Om}\!da(\bx)
\,n^j(\bx)\,\dFI^i(\bx) ~ \cdot \\
& \hspace{1cm} \cdot \l[\l(\!-{1 \over 4}\,\de_{ij}\,
\partial^{x^\ell}\!\partial_{y^\ell} +  {1 \over 2}\,\partial_{x^i y^j} \!\r)\!
\Dir_{\s+1 \over 2}(\bx,\by)\r]_{\by=\bx} \,. \label{EeDir}
\end{split}\end{equation}
Now, comparing the above result with Eq. \rref{pEeT} for the
regularized pressure we see that
\beq  \Ee^\s = - (1 - \s) \int_{\partial\Om}\!da(\bx)\;\dFI^i(\bx)\,p^\s_i(\bx)
\label{eqees}~; \feq
the first equality in \rref{defEe} and Eq. \rref{eqees} give
the thesis \rref{eqpres}.
\section{Appendix. Derivation of Eq. \rref{Res1}} \label{appeEps}
Consider the framework developed in subsection \ref{AAeps} for a
fundamental operator $\AA$, such that $\si(\AA) \subset [0,+\infty)$
and $0$ is
in the continuous spectrum of $\AA$.
Herefter we are
using the deformed fundamental operator
\beq \AA_\eps := (\sqrt{\AA}+\eps)^2 ~. \feq
We already observed in subsection \ref{AAeps} (see Eq. \rref{Tep})
that the cylinder kernels $T$ and $T^\eps$, respectively
associated to $\AA$ and $\AA_\eps$, are related by
\beq T^{\eps}(\t\,;\bx,\by) = e^{-\eps\t}\, T(\t\,;\bx,\by) ~; \label{TepA}\feq
we also showed (see Eq. \rref{HankTep}) that, assuming the map
$\t \mapsto T(\t\,;\bx,\by)$ (for fixed $\bx,\by \in \Om$) to admit
an extension in a neighborhood of the real half-axis $[0,+\infty)$
to a meromorphic function of $\t$, we have
\beq \Dir_s^{\eps}(\bx,\by) = {e^{-2i\pi s}\,\Ga(1\!-\!2s)\over 2\pi i}
\int_\Hank d\t\;\t^{2s-1}\,T^{\eps}(\t\,;\bx,\by) ~. \label{HankTepA}\feq
The integral in the right-hand side of the above equation is an
analytic function of $s$ on the whole complex plane, while the
gamma function is meromorphic with simple poles at positive
half-integer values of $s$. Taking into account these facts,
hereafter we show how to evaluate the renormalized kernels
\beq \Dik_{s_0}(\bx, \by) := \lim_{\eps \to 0^{+}}
RP \Big|_{s = s_0} \Big(k^{2(s-s_0)}\Dir^{\eps}_{s}(\bx,\by)\Big) \feq
considering, separately, the cases $s_0 = - n/2$ ($n \in \{0,1,2,...\}$) and
$s_0  = n/2$ ($n\in \{1,2,3...\}$). Putting together the results obtained for
$s_0 = -n/2$ ($n \in \{0,1,2,...\}$) and for $s_0 = 1/2$, we will finally obtain
the proof of Eq. \rref{Res1}. \parn
We remark that, for $s_0 = \pm 1/2$, the above renormalized kernels
coincide with the functions introduced in Eq. \rref{DirRen}.
\vskip 0.6cm \noindent
\textbf{Case 1: $\boma{s_0 = -{n \over 2}}$\,, $\boma{n \in \{0,1,2,...\}}$\,.}
The right-hand side of Eq. \rref{HankTepA} is clearly an analytic
function of $s$ for $\Re s < {1 \over 2}$\,; thus, for $n \in \{0,1,2,...\}$
$\Dir^{\eps}_{s}(\bx, \by)$ has an analytic continuation at $s=-n/2$,
hereafter indicated with $\Dir^{\eps}_{-n/2}(\bx, \by)$, that is simply
obtained substituting this value of $s$ in the integral representation
\rref{HankTepA}. The resulting integrand is meromorphic so that, by the
residue theorem,
\beq \Dir_{-{n \over 2}}^{\eps}(\bx,\by) = (-1)^n\,\Ga(n + 1)\,
\Res\Big(\t^{-(n+1)}\,e^{-\eps t}\,T(\t\,;\bx,\by)\,;0 \Big) ~.
\label{DirResEp} \feq
Of course, in the present case the prescription of taking the regular
part in Eq. \rref{DirRen} is pleonastic, and the cited equation is
reduced to
\beq \Dik_{-{n \over 2}}(\bx, \by) = \lim_{\eps \to 0^{+}}
\Dir^{\eps}_{-{n \over 2}}(\bx, \by) ~. \feq
\vfill \eject \noindent
On the other hand, computation of the previous limit gives
({\footnote{To prove this, one can proceed as follows. In the present
case the cylinder kernel is assumed to be a meromorphic function of
$\t$, so there is an expansion $T(\t\,;\bx,\by) = {1\over\t^q}
\sum_{k = 0}^{+\infty}e_k(\bx,\by)\,\t^k$ for some $q \in \interi$
(converging, at least, for small $\t$; note that under the assumptions
for the validity of \rref{asinTD} we have $q=d$); of course
$e^{-\eps \t} = \sum_{k=0}^{+\infty} {(-\eps)^k \over k!} \t^k$,
so the Cauchy formula for the product of two series yields
$$ \Res\Big(\t^{-(n+1)}e^{-\eps t} T(\t\,;\bx,\by);0 \Big)\!
= \sum_{k = 0}^{q+n}\!{(-\eps)^k\! \over k!}\,e_{q+n-k}(\bx,\by)\,
\stackrel{\eps \to 0}{\longrightarrow}\, e_{d+n}(\bx,\by) =
\Res\Big(\t^{-(n+1)}T(\t\,;\bx,\by);0 \Big)\,. $$
This proves Eq. \rref{DirRes}.}})
\beq \Dik_{-{n \over 2}}(\bx,\by) = (-1)^n\,\Ga(n + 1)\,
\Res\Big(\t^{-(n+1)}\,T(\t\,;\bx,\by)\,; 0 \Big) ~.
\label{DirRes} \feq
The above result can be reformulated in terms of the modified
cylinder kernel $\Tm$ associated to $\AA$\,. Indeed, recall that
$T(\t\,;\bx,\by) = - \partial_\t \Tm(\t\,;\bx,\by)$ (see Eq. \rref{STKer})
and note that, for any pair of functions $f,g$ meromorphic near a
point $\t_0$, we have $\Res(f g';\t_0) = -\Res(f'g;\t_0)$; these facts
(and the standard identity $z\,\Ga(z) = \Ga(1\!+\!z)$) give
\beq \Dik_{-{n \over 2}}(\bx,\by) = (-1)^{n+1}\,\Ga(n + 2)\,
\Res\Big(\t^{-(n+2)}\,\Tm(\t\,;\bx,\by)\,; 0 \Big) ~.
\label{DirResTm} \feq
\vskip 0.4cm \noindent
\textbf{Case 2: $\boma{s_0 = +{n \over 2}}$\,, $\boma{n \in \{1,2,3,...\}}$
(with a special attention for the subcase $\boma{n=1}$).} A substantial
difference occurs with respect to Case 1. In fact, due to the gamma
function appearing in Eq. \rref{HankTepA}, the function $\Dir_s^{\eps}(\bx,\by)$
described by this equation has a genuine singularity at $s = n/2$\,;
in order to remove this singularity, it is essential to retain only
the regular part in Eq. \rref{DirRen}. To this purpose, for $s$ in
a neighborhood of $n/2$, we introduce the variable $u := 2s-n$ and
note that, for $u \to 0$,
\begin{equation}\begin{split}
& \hspace{2.6cm} \mm^{2s-n}e^{-2i\pi s}\,\Ga(1\!-\!2s)\,\t^{2s-1} = \\
& = {\t^{n-1} \over (n\!-\!1)!} \l[{1 \over u} +
\Big(\ln(\mm\t) + \ga_{EM} - i\pi - H_{n-1} \Big) + O(v)\r] , \label{Sersp}
\end{split}\end{equation}
where $\ga_{EM} \simeq 0.577216$ is the Euler-Mascheroni constant and,
for $m \in \{0,1,2,...\}$, $H_m := \sum_{k=1}^m {1\over k}$ ($H_0 := 0$)
denotes the $m$-th harmonic number
({\footnote{\label{Foot2} To obtain Eq. \rref{Sersp}, one uses the following
well known facts \cite{NIST}: for $m \in \{0,1,2,...\}$,
$$ \Ga(-u-m) = {(-1)^m\,\Ga(-u) \over (u\!+\!1)...(u\!+\!m)} ~; \qquad
(u+1)...(u+m) = m!\,\Big[1 + H_m\,u + O(u^2)\Big] ~ , $$
$$ \Ga(-u) = -\,{1 \over u} - \ga + O(u) ~, \qquad
e^{-i\pi u}\, (\mm\t)^u = e^{u (\ln(\mm\t)-i\pi)}\! =
1 +(\ln(\mm\t)-i\pi) u + O(u^2) \qquad \!\mbox{for $u\!\to\! 0$}\,. $$}}). \parn
It follows that
\begin{equation}\begin{split}
& \hspace{2.75cm} RP\Big|_{s = {n \over 2}} \Big(\mm^{2s-n}\Dir_{s}^{\eps}(\bx,\by)\Big)= \\
& = {1 \over 2\pi i}\!\int_\Hank d\t\; {\t^{n-1}\,e^{-\eps\t}
\over (n\!-\!1)!}\,\Big(\ln(\mm\t) + \ga - i\pi - H_{n-1}\Big)\,T(\t\,;\bx,\by) ~.
\label{HankSeru}
\end{split}\end{equation}
According to Eq. \rref{DirRen}, the renormalized function $\Dik_{n\over 2}(\bx,\by)$
is the limit $\eps \to 0^+$ of the above expression. Under suitable
hypotheses on the behaviour of $T$ for $\Re\t \to +\infty$ (namely,
$|T(\t\,;\bx,\by)| \leq C\, |\t|^{-a-n}$ for some $C,a\!>\!0$), we
can exchange the limit and the integral to obtain
\beq \Dik_{n\over 2}(\bx,\by) = {1 \over 2\pi i}\int_\Hank d\t\;
{\t^{n-1}\over (n\!-\!1)!}\,\Big(\ln(\mm\t) + \ga - i\pi - H_{n-1}\Big)\,
T(\t\,;\bx,\by) ~; \label{HankRen}\feq
the term $\ln(\mm\t)$ prevents us from using the residue theorem,
so we must find alternative ways to evaluate explicitly the above
integral. In the special case $n = 1$,
we can proceed as follows. First we recall that $T(\t\,;\bx,\by) =
- \partial_\t \Tm(\t\,;\bx,\by)$ and integrate by parts
Eq. \rref{HankRen} to obtain
\beq \Dik_{{1\over 2}}(\bx,\by) = {1 \over 2\pi i} \int_\Hank d\t\;
\t^{-1}\; \Tm(\t\,;\bx,\by) ~; \label{Hankp1}\feq
the resulting integrand is meromorphic in $\t$ so that we can resort
to the residue theorem to obtain
\beq \Dik_{{1\over 2}}(\bx,\by) = \Res \Big(\t^{-1}\,\Tm(\t\,;\bx,\by)\,;0\Big) ~. \label{eqqu2} \feq
\salto
\textbf{Conclusion.} Eq. \rref{DirResTm} for $\Dik_{-{n \over 2}}$
with $n \in \{0,1,2,...\}$ and Eq. \rref{eqqu2} for $\Dik_{{1 \over 2}}$
prove Eq. \rref{Res1} for  $\Dik_{-{n \over 2}}$ with $n \in\{-1,0,1,2,...\}$\,.
\vfill \eject \noindent

\end{document}